\documentclass[opre,nonblindrev]{informs3} % arXiv preprint

\OneAndAHalfSpacedXI
\usepackage{graphicx}
\usepackage{amsfonts}
\usepackage{subfigure}
\usepackage{longtable}
\usepackage{rotating}
\usepackage{epstopdf}
\usepackage{color}
\usepackage{multirow}
\usepackage{booktabs}
\usepackage{bbm}
\usepackage{setspace}
\usepackage{enumerate}
\usepackage{diagbox}
\usepackage{threeparttable} 

\usepackage{xurl}
\usepackage{natbib}
\usepackage[hypertexnames=false]{hyperref}
\hypersetup{
    colorlinks=true,
    linkcolor=black,
    filecolor=black,
    urlcolor=black,
    citecolor=blue,
}
\ifdefined\pdfsuppressptexinfo\pdfsuppressptexinfo=-1\relax\fi % omit file names of included figures from the PDF

\usepackage{tabularx,tikz,microtype,enumitem}
\usetikzlibrary{arrows.meta,positioning}

\usepackage{endnotes}
\makeatletter
\def\enoteheading{\section*{\notesname
		\@mkboth{\MakeUppercase{\notesname}}{\MakeUppercase{\notesname}}}%
	\noindent
}
\makeatother

\usepackage{lipsum}
\makeatletter
\renewcommand\footnoterule{%
  \kern-3\p@
  \hrule\@width 6.5in
  \kern2.6\p@}
\makeatother

 \bibpunct[, ]{(}{)}{,}{a}{}{,}%
 \def\bibfont{\fontsize{9}{11}\selectfont}%

 \def\argmax{{\rm argmax}\,}
 \def\argmin{{\rm argmin}\,}

 \def\co{{\rm co }\,}

 \def\ph{\phi}

 \def\rh{\rho}

 \def\today{\ifcase\month\or January\or February\or  March\or  April\or
 May\or June\or July\or August\or  September\or October\or November\or
 December\fi  \space\number\day, \number\year}

 \def\qed{\hspace*{10pt}\hfill{$\square$}\hfilneg\par}
 \def\sqr#1#2{{\vcenter{\hrule height .#2pt
       \hbox{\vrule width .#2pt height#1pt \kern#1pt\vrule width.#2pt}
                        \hrule height.#2pt}}}
 \def\square{\mathchoice\sqr54\sqr54\sqr{2.1}3\sqr{1.5}3}

 \theoremstyle{TH}
 \newtheorem{thm}{Theorem}%[section]
 \newtheorem{lem}{Lemma}
 \newtheorem{cor}{Corollary}%[thm]
 \newtheorem{prop}{Proposition}

 \newtheorem{defi}{Definition}
 \newtheorem{exap}{Example}

 \newcommand{\enu}{\begin{enumerate}}
 \newcommand{\eenu}{\end{enumerate}}
 \newcommand{\des}{\begin{description}}
 \newcommand{\edes}{\end{description}}
 \newcommand{\iit}{\begin{itemize}}
 \newcommand{\eiit}{\end{itemize}}
 \newcommand{\tab}{\begin{tabular}}
 \newcommand{\etab}{\end{tabular}}
 \newcommand{\ben}{\begin{equation}}   \newcommand{\een}{\end{equation}}
 \newcommand{\bea}{\begin{eqnarray}}   \newcommand{\eea}{\end{eqnarray}}
 \newcommand{\beq}{\begin{eqnarray*}}  \newcommand{\eeq}{\end{eqnarray*}}

\newcommand{\E}{\mathbb E}
\newcommand{\ind}{\mathbf 1}
\newcommand{\cK}{\mathcal K}
\newcommand{\cP}{\mathcal P}
\newcommand{\cD}{\mathcal D}
\newcommand{\Rp}{\mathcal R}
\newcommand{\Rstar}{\mathcal R^*}
\newcommand{\RE}{\mathcal R_E}
\newcommand{\RR}{\mathcal R_{\rm res}}
\newcommand{\Shat}{\widehat U}
\newcommand{\pos}[1]{\left[#1\right]_+}

\renewcommand{\argmax}{\mathop{\rm arg\,max}}
\renewcommand{\argmin}{\mathop{\rm arg\,min}}
\allowdisplaybreaks[2]
\graphicspath{{Optimal auctions with projection bias/fig/}}
\TheoremsNumberedThrough     % Preferred (Theorem 1, Lemma 1, Theorem 2)
\EquationsNumberedThrough    % Default: (1), (2), ...
\newcommand{\PaperTitle}{Mechanism Design under Belief Disagreement: Allocation Flexibility and Rent Capacity}
\begin{document}
\RUNAUTHOR{Lu and Zhang}
 \RUNTITLE{Mechanism Design under Belief Disagreement}
\TITLE{\PaperTitle}
\ARTICLEAUTHORS{%
    \AUTHOR{Lijian Lu\thanks{The authors used generative AI tools (ChatGPT and Claude Code) to assist with editing, derivations, proof validation, and numerical verification. Authors retain sole responsibility for the paper's content.}}
    \AFF{HKUST Business School, Hong Kong University of Science and Technology, \EMAIL{lijianlu@ust.hk}}
    \AUTHOR{Ce ZHANG}
    \AFF{HKUST Business School, Hong Kong University of Science and Technology, \EMAIL{czhangdp@connect.ust.hk}}
}

\ABSTRACT{%
A seller jointly chooses allocation and payments when buyers misperceive competition. Under independent private values, statewise participation, and payment restrictions, allocation determines both required rents and their delivery capacity. We separate these effects under a common subjective law. For three values and shared-top beliefs, explicit score and capacity conditions characterize globally optimal lower-value lotteries, including top-rent overflow and welfare differences among revenue optima. A two-buyer construction establishes a strict allocation advantage over every reserve under the same fixed reporting certificate and positive margin. A transport test and exact band-boundary incentive equalities identify separate conditions restoring objective-law revenue ceilings. For three bands and a common affine belief interval, an explicitly certified reserve attains the ceiling under stated score conditions. A case scaled to public collectible-auction records yields a nominal shared-top allocation gain of 24.72 cents per auction. A one-cent report-by-report margin yields matched gains of 1.48 cents under projection and 3.32 cents under shared-top beliefs; the three-band point-belief gain of 2.50 cents disappears with common-interval protection. These comparisons bound the additional recurring costs an allocation redesign can cover at the specified reporting requirement.
}
\SUBJECTCLASS{Games/group decisions: bidding/auctions; Programming: linear.}
\AREAOFREVIEW{Markets, Platforms \& Revenue Management.}
\maketitle
%============================ MAIN TEXT ============================
\section{Introduction}\label{sec:intro}
An auction operator must distinguish actual competition from the competition buyers perceive. The distinction is particularly relevant for collectibles, where individual tastes shape both willingness to pay and expectations of rivals' tastes, affecting bidding and optimal reserves \citep{gagnon-bartsch2021projection}. When buyers overweight a rival configuration, a benefit delivered there supplies perceived utility at low objective cost. Payment restrictions tie this opportunity to allocation: a seller who cannot pay a losing buyer must give her a chance to win. The central design question is when creating this rent capacity justifies serving a lower-value buyer.

Consider two independent buyers with values \$30, \$20, or \$12, each equally likely. Both believe a rival has the highest value with probability two-thirds and each other value with probability one-sixth. The best reserve allocation with optimized payments earns \$22.72. Joint allocation and payment design earns \$23.17 by giving a \$20 report a 5\% chance of a free object against a \$30 report. This lottery replaces a \$4 middle-type discount against a low rival with one dollar of expected utility against a high rival. The benefits deliver the same perceived rent because buyers regard the latter state as four times as likely. After adjusting the other payments to preserve incentives, revenue rises by approximately 44 cents and allocated surplus falls by 11 cents. Appendix~\ref{app:examples} gives the complete schedules.

A second example shows why revenue at a truthful equilibrium is insufficient for evaluating the design. Two independent buyers value the object at \$10 with probability 20\% and \$1 otherwise. Projection leads high and low types to assign probabilities 44\% and 14\% to a high rival. The optimal truthful rule with nonnegative payments earns \$3.76 but awards a free object to a uniformly selected buyer after two high reports. Always reporting high is also a strict equilibrium: a high report receives a half chance of a free object, whereas a unilateral low report loses. Revenue is then zero. Allocation flexibility must therefore be evaluated against a specified reporting guarantee.

These examples motivate two linked design questions. When does allocation create sufficiently valuable rent capacity to change the optimal service rule? How much is allocation flexibility worth when the operator requires a specified certificate of truthful reporting? We study joint allocation and payment design with independent private values, known subjective laws, fixed entry, and participation constraints in every rival configuration. Disagreement is between buyers' beliefs and the seller's objective law, while all buyer types may share the same subjective law. Each reserve allocation receives optimized payments under the same beliefs, payment restrictions, and reporting requirement as joint design. These matched comparisons isolate the value of allocation flexibility separately from payment optimization.

Our first contribution characterizes the interaction between required rents and rent-delivery capacity. Under a common subjective law, incentive constraints reduce to allocation-dependent interim rent floors, and capped rent delivery is a fractional-knapsack problem. For three values and a shared tilt toward the highest category, explicit score and capacity conditions characterize globally optimal lower-value lotteries. The analysis covers both rents that fit in the cheapest state and rents that overflow to costlier tiers. A lower-value lottery can displace capacity already used to deliver top-type rents; pricing this displacement determines whether the lottery is worthwhile and explains welfare differences among revenue optima. The closed-form allocation results apply under their stated conditions and fully classify the calibrated belief path. The exact common-law program remains available outside those conditions.

Our second contribution establishes a strict allocation advantage under a reporting guarantee and identifies restrictions that eliminate the advantage. A two-buyer construction strictly outperforms every reserve under the same fixed report-by-report deletion schedule and positive utility margin. Two separate restoration arguments then delimit the opportunity. Under subjective independence, a transport test identifies when admissible rival reports reproduce the objective law at each deletion step, transferring incentive inequalities to that law and imposing its revenue ceiling. For three reporting bands, exact incentives at every value throughout a common affine belief interval imply boundary equalities that also hold under the objective law, even when that law lies outside the protected interval. Under explicit score conditions, a reserve with a positive interior certificate attains the resulting ceiling for the given bands.

A calibrated numerical case using public collectible-auction records quantifies these tradeoffs. Nominal allocation gains under shared-top beliefs are 24.72 cents per auction. A one-cent report-by-report margin yields matched gains of 1.48 cents under projection and 3.32 cents under shared-top beliefs. With three reporting bands, the 2.50-cent gain at point beliefs becomes zero under common-interval protection. The records discipline value scale and heterogeneity under the maintained observation model, where beliefs, entry, and reporting requirements remain scenario inputs. Paired refits assess reoptimized gains, while fixed-contract checks assess incentive sensitivity. These gains quantify the additional recurring costs a redesign can cover at the chosen reporting requirement.

\noindent\textbf{Outline.} The paper is organized as follows. Section~\ref{sec:related} reviews related literature. Section~\ref{sec:model} defines comparison standards. Section~\ref{sec:allocation} characterizes optimal allocation and rent delivery. Section~\ref{sec:implementation} studies reporting costs and revenue ceilings. Section~\ref{sec:structure} collects supporting reductions, and Section~\ref{sec:data} evaluates the matched case. Detailed proofs and computational methods are in the online appendix.

\section{Related Literature}\label{sec:related}
Table~\ref{tab:contribution_comparison} distinguishes the central contributions from their methodological foundations. The closest antecedent, \citet{LuZhang2026}, optimizes payments under fixed efficient or highest-band allocation. Its alignment, rent-concentration, fixed-point, and capped-delivery results support our payment reductions. We endogenize allocation within a specified reporting interface, so service changes both the rents required by incentives and the capacity to deliver them. The distinctive allocation results characterize this interaction, including displacement of top-type rent capacity and welfare differences among revenue optima. 

Virtual values \citep{Myerson1981}, linear programming \citep{Vohra2012}, and Bayesian duality \citep{Cai2021} provide the methodological foundations. In our setting, subjective laws determine incentive benefits, whereas the objective law determines their expected cost. Payment restrictions connect this difference to service by limiting where rents can be delivered. Objective independence and statewise participation also distinguish our setting from correlated-information surplus extraction \citep{Cremer1988}. Related operational studies examine participation in sequential selling \citep{Ashlagi2023}, customer search \citep{HuXiao2025}, procurement restrictions \citep{Saban2021}, and joint advertising allocation and information provision \citep{ChenGallegoGaoLi2025}.

Our reporting analysis builds on iterated deletion and implementation under belief restrictions \citep{BergemannMorris2009,OllarPenta2017}. We separate the revenue costs of a sufficient certificate class and a positive reporting margin; neither must equal the minimum cost of unique implementation. Our margins strengthen reporting incentives while preserving exact truthfulness, complementing \citet{Balseiro2024}. For band reporting, we connect finite-message auctions \citep{Kos2012} to revenue equivalence with discontinuous allocations \citep{OllarPenta2025}. \citet{lopomo2021uncertaintymechanismdesign} show that small belief misspecification can discontinuously shrink the set of feasible mechanisms and select ex post incentive-compatible designs.
Our distinctive contribution is a fixed-interface revenue comparison: exact boundary indifference along a common affine marginal-belief path yields a polynomial identity that also holds under the objective law, even outside the protected interval, implying a revenue ceiling attained by a reserve with a positive interior certificate.

The broader robust mechanism design literature studies uncertainty about beliefs and information \citep{BergemannMorris2005,ChungEly2007,BrooksDu2021}. Distributionally robust screening develops dual formulations, randomized pricing, and finite-menu guarantees under valuation-distribution uncertainty \citep{ChenHuWang2024,WangLiuZhang2024,Wang2025,KocyigitKuhnRujeerapaiboon2024}. \citet{PhamYamashita2024} provide a worst-case foundation for dominant-strategy auctions under heterogeneous-prior uncertainty.
\citet{Anunrojwong2026Scale} shows that serving a lower bidder can improve prior-independent revenue guarantees with two i.i.d. regular bidders. Here lower-value service can supply inexpensive rent capacity or make misreporting costly. The nominal problem takes subjective laws as given; the interval analysis fixes the objective distribution and requires one contract and one certificate to remain valid across the prescribed subjective laws.

Projection and false-consensus evidence
\citep{Ross1977,Krueger1994,Madar2012,breitmoser2019knowing,bushong2024failures},
alongside representativeness
\citep{Kahneman1972,Bordalo2016,Bordalo2018},
motivate our belief families. Related studies explore mistaken beliefs about one's own preferences \citep{eliaz2006contracting,grubb2009selling}, analogy-based expectations \citep{Jehiel2011}, and endogenously misspecified learning \citep{FudenbergLanzaniStrack2021}. We model distorted beliefs about rivals' values while retaining correct conjectures about their strategy maps. Auction records discipline value magnitudes under this observation model, while identifying belief distortions requires additional evidence, such as belief elicitation \citep{Delavande2026}.

\noindent\textbf{Notation.} Boldface denotes vectors and matrices, with componentwise inequalities. Write $\mathbb Z_+$ for the nonnegative integers, $\mathcal M=\{1,\ldots,m\}$ and $\mathcal N=\{1,\ldots,n\}$ for buyer and type indices, $\mathbf e_j$ for the $j$th unit vector, $\ind\{\cdot\}$ for indicators, $\pos{z}=\max\{z,0\}$, and $\co(X)$ for the convex hull of $X$. For a profile $\mathbf s$ indexed by buyers, $\mathbf s_{-i}$ excludes buyer $i$.

\section{Environment and Comparable Design Problems}\label{sec:model}

A zero-value seller offers one object to $m\ge2$ risk-neutral buyers $i\in\mathcal M$, whose private values, or \emph{types}, are drawn independently from $\mathcal T=\{T_1>\cdots>T_n>0\}$, $n\ge2$, with probabilities $\mathbf q=(q_1,\ldots,q_n)$, where $q_j>0$ and $\sum_{j\in\mathcal{N}}q_j=1$. A rival configuration, or state, is $\mathbf k\in\cK:=\{\mathbf k\in\mathbb Z_+^n:\sum_jk_j=m-1\}$, where $k_j$ counts rivals of type $T_j$; its objective probability is
\[
 \Phi(\mathbf{k})=\frac{(m-1)!}{\prod_j k_j!}\prod_jq_j^{k_j}.
\]
A buyer of type $T_j$ assigns probability $\ph_j(\mathbf k)>0$ to each state $\mathbf k$; $\ph_j$ is her \emph{subjective law}, exchangeable over rivals. \emph{Subjective independence}, under which $\ph_j$ is multinomial with $m-1$ trials and cell probabilities $\boldsymbol\pi_j=(\pi_{j1},\ldots,\pi_{jn})$, is assumed only where stated. The laws need not derive from a common prior and do not vary with the mechanism or strategies; the seller knows them and evaluates revenue under $\Phi$. Buyers, values, and beliefs stay fixed when the seller changes the mechanism.

\noindent\textbf{Mechanisms, participation, and revenue.}
An \emph{anonymous direct mechanism} $(\mathbf p,\mathbf x)$ specifies, for report $T_j$ against rival reports with configuration $\mathbf k$, an allocation probability $p_j(\mathbf k)$ and an expected payment $x_j(\mathbf k)\in\mathbb R$. Feasibility becomes the allocation constraints:
\begin{equation}\label{eq:FEA}
 p_j(\mathbf{k})\ge0, \ \sum_{h\in\mathcal{N}}(k_h+\ind\{h=j\})\,p_h(\mathbf{k}+\mathbf{e}_j-\mathbf{e}_h)\le 1, \quad \forall j\in\mathcal{N}, \mathbf{k}\in\mathcal{K}. \tag{AC}
\end{equation}
Terms for absent reports with $k_h+\ind\{h=j\}=0$ are omitted. At the full profile $\mathbf k+\mathbf e_j$, each buyer reporting $T_h$ faces rivals with counts $\mathbf k+\mathbf e_j-\mathbf e_h$. Let $\cP$ be the feasible allocations. An allocation is \emph{monotone} if $p_h(\mathbf{k})\ge p_{h'}(\mathbf{k})$ for every $\mathbf{k}$ whenever $h<h'$. Write
\[
 u_j(\mathbf{k})=T_jp_j(\mathbf{k})-x_j(\mathbf{k}),\qquad
 P_{jh}(\mathbf{p})=\sum_{\mathbf{k}}\ph_j(\mathbf{k})p_h(\mathbf{k}),\qquad
 \Shat_j=\sum_{\mathbf{k}}\ph_j(\mathbf{k})u_j(\mathbf{k})
\]
for truthful rent, type $T_j$'s perceived winning probability from report $T_h$, and her perceived rent. As $x_j=T_jp_j-u_j$, we identify the mechanism with $(\mathbf p,\mathbf u)$, and write $\widehat{\boldsymbol U}=(\Shat_1,\ldots,\Shat_n)$.

Buyers play a \emph{subjective equilibrium} (SE): each type best responds under her own law of rival values and correct conjectures of rivals' strategy maps, which may be asymmetric and mixed (Definition~\ref{def:SE}). Rivals' strategies induce a law of rival-report configurations, and truthful rivals induce $\ph_j$. Reporting $T_h$ gives type $T_j$ the payoff $T_jp_h-x_h=u_h+(T_j-T_h)p_h$, so her gain from truthful reporting over mimicking $T_h$ at configuration $\mathbf k$, and under a law $\ph$ on $\cK$, is
\[
 g_{jh}(\mathbf k)=u_j(\mathbf k)-u_h(\mathbf k)-(T_j-T_h)p_h(\mathbf k),\qquad
 \mathrm{sl}_{jh}(\ph)=\sum_{\mathbf k}\ph(\mathbf k)g_{jh}(\mathbf k).
\]
A mechanism is \emph{truthful} if truthful reporting is an SE. Truthfulness and statewise participation are
\begin{align}
 \mathrm{sl}_{jh}(\ph_j)&\ge 0,&&\forall j, h\in\mathcal{N} \quad \text{and} \quad j\neq h, \tag{IC} \label{eq:ic}\\
 u_j(\mathbf{k})&\ge0,&&\forall j\in\mathcal{N}, \mathbf{k}\in\mathcal{K}. \tag{IR} \label{eq:ir}
\end{align}
Statewise participation applies before internal lotteries resolve; it admits a limited-liability interpretation \citep{Roberto2012} but neither limits subsidies nor makes incentives robust to other beliefs. For $\varsigma\ge0$, the \emph{subsidy cap} is
\begin{equation}\label{eq:cap}
x_j(\mathbf k)\ge-\varsigma p_j(\mathbf k), \quad \mbox{or equivalently,} \quad u_j(\mathbf{k})\le(T_j+\varsigma)p_j(\mathbf{k}),\quad \forall j\in\mathcal{N}, \mathbf{k}\in\mathcal{K}, \tag{SC}
\end{equation}
where $\varsigma=0$ requires \emph{nonnegative payments}. Charge $T_j-u_j/p_j$ only on winning if $p_j>0$, and transfer $u_j$ if $p_j=0$. This implements expected payments with nonnegative realized truthful utility. Under \eqref{eq:cap}, losers receive nothing and winner prices lie in $[-\varsigma,T_j]$. Every finite cap therefore excludes loser subsidies: relaxing it enlarges winning bidders' rent capacity, whereas \emph{unrestricted} transfers also permit rent delivery in losing states. 

An allocation $\mathbf p$ is \emph{implementable} if some $\mathbf u\ge0$ satisfies \eqref{eq:ic}, together with \eqref{eq:cap} when a subsidy cap is imposed. For $\mathbf p\in\cP$, allocated surplus and optimized revenue are
\[
G(\mathbf{p})=m\sum_jq_jT_j\sum_{\mathbf{k}}\Phi(\mathbf{k})p_j(\mathbf{k}), \quad \Rp(\mathbf{p})=\max_{\mathbf{u}\ge0:\,\eqref{eq:ic}}  \operatorname{Rev}(\mathbf p,\mathbf u),
\]
where $\operatorname{Rev}(\mathbf p,\mathbf u)= G(\mathbf{p})-m\sum_jq_j\sum_{\mathbf{k}}\Phi(\mathbf{k})u_j(\mathbf{k}) =m\sum_jq_j\sum_{\mathbf k}\Phi(\mathbf k)x_j(\mathbf k) $ denotes the seller’s revenue, that is, allocated surplus minus objective rent. Maximum revenue is characterized by the \emph{minimum-rent problem}, $\min_{\mathbf{u}\ge0:\,\eqref{eq:ic}} m\sum_jq_j\sum_{\mathbf{k}}\Phi(\mathbf{k})u_j(\mathbf{k})$, whose value is $+\infty$ if infeasible, and otherwise its solutions yield \emph{optimal payments} for a given allocation $\mathbf p$. The joint optimum $\Rstar=\max_{\mathbf p\in\cP}\Rp(\mathbf p)$ is obtained from the following LP:
\begin{equation}\label{eq:primal}
 \max_{\mathbf{p}\ge0,\,\mathbf{u}\ge0}\ \operatorname{Rev}(\mathbf{p},\mathbf{u}) \quad \text{s.t. \eqref{eq:ic} and \eqref{eq:FEA}.}
\end{equation}
An allocation $\mathbf p$ is \emph{globally optimal} if $\Rp(\mathbf p)=\Rstar$. A superscript `$\cdot^\varsigma$' corresponds to \eqref{eq:cap}, yielding $\Rp^\varsigma(\mathbf p)$ and $\mathcal R^{*,\varsigma}$, and the global optimality under a subsidy cap is defined likewise. An SE $\boldsymbol\xi$ of any mechanism is \emph{statewise participating} if every type's expected utility is nonnegative at every rival profile. Its \emph{anonymized outcome} $(\mathbf p^{\boldsymbol\xi},\mathbf u^{\boldsymbol\xi})$ averages its outcome over buyers and rival orderings and preserves expected revenue: $\operatorname{Rev}(\boldsymbol\xi)=\operatorname{Rev}(\mathbf{p}^{\boldsymbol\xi},\mathbf{u}^{\boldsymbol\xi})$ (see Appendix~\ref{app:model}).

\begin{lem}[Truthful design and revenue ranking]\label{lem:model}
(i) Every statewise-participating SE $\boldsymbol\xi$ of every mechanism has $\operatorname{Rev}(\boldsymbol\xi)\le\Rstar$, and a truthful anonymous direct mechanism satisfying \eqref{eq:ir} attains $\Rstar$;
(ii) Let $(\mathbf p,\mathbf u)$ satisfy \eqref{eq:ic}, \eqref{eq:ir} and $\operatorname{Rev}(\mathbf p,\mathbf u)=\Rstar$. Then, every statewise-participating SE $\boldsymbol\xi$ of $(\mathbf p,\mathbf u)$ has $\operatorname{Rev}(\boldsymbol\xi)\le\Rstar$, with equality if and only if $\mathbf p^{\boldsymbol\xi}$ is globally optimal and $\mathbf u^{\boldsymbol\xi}$ solves the minimum-rent problem at $\mathbf p^{\boldsymbol\xi}$.
Both parts hold with \eqref{eq:cap} and $\mathcal R^{*,\varsigma}$ in place of $\Rstar$, where in (i) the expected payments of $\boldsymbol\xi$ must satisfy the cap at every value profile.
\end{lem}

This is a nominal benchmark: simulating an equilibrium under direct reports establishes the existence of a truthful equilibrium, but not dominance or uniqueness. Other equilibria may generate lower revenue and be preferred by buyers. Section~\ref{sec:implementation} quantifies the cost of explicit reporting guarantees, while Appendix~\ref{app:model} discusses welfare rankings and their limitations.

\noindent\textbf{Benchmarks and belief families.}
Highest-value allocation with reserve $T_r$, $r\in\mathcal N$, is
\begin{equation}\label{eq:reserve}
 p_j^r(\mathbf{k})=\frac{\ind\{j\le r,\ \sum_{h<j}k_h=0\}}{k_j+1}. \tag{EF}
\end{equation}
With $T_{n+1}=0$, every reserve price in $(T_{r+1},T_r]$ induces $\mathbf p^r$. The \emph{efficient allocation} $\mathbf p^n$ attains the full surplus $S=G(\mathbf p^n)$, the objective expected highest value, while $G(\mathbf p^r)<S$ for $r<n$. Each reserve rule receives its own optimal payments, so it need not be a second-price auction. Write $\Rp_r=\Rp(\mathbf p^r)$, $\RE=\Rp_n$, and $\RR=\max_{r\in\mathcal{N}}\Rp_r$, with superscript $\varsigma$ under \eqref{eq:cap}. The \emph{allocation gain} $\Rstar-\RR$, or $\mathcal R^{*,\varsigma}-\RR^\varsigma$ under \eqref{eq:cap}, isolates the allocation value by comparing the joint allocation-payment design with the best reserve rule under nominal truthfulness. Superscript `$\cdot^\mathrm{rat}$' marks the \emph{rational benchmark} ($\ph_j=\Phi$ for all $j$), with \emph{rational optimum} $\max_{\mathbf p\in\cP}\Rp^{\mathrm{rat}}(\mathbf p)$. For a truthful mechanism,
\begin{equation}\label{eq:decomp}
 S-\operatorname{Rev}(\mathbf p,\mathbf u)
 =\underbrace{S-G(\mathbf p)}_{\text{allocative welfare loss}}
 +\underbrace{m\sum_jq_j\sum_{\mathbf k}\Phi(\mathbf k)u_j(\mathbf k)}_{\text{objective buyer rent}}.
\end{equation}
Under efficient allocation, the shortfall is buyer rent; a lower-value lottery raises revenue only if it saves more objective rent than the surplus it destroys. Buyers' perceived utilities can rank the same mechanisms differently.

Our general framework accommodates two belief families.  Under \emph{taste projection},
\begin{equation}\label{eq:projection}
 \pi_{jh}=(1-\alpha)q_h+\alpha\ind\{h=j\},\qquad 0\le\alpha<1, j,h\in\mathcal N,
\end{equation}
each buyer overweights rivals who share her value, with \emph{intensity} $\alpha$. The own-value shift follows taste-projection models \citep{Madar2012,breitmoser2019knowing,gagnon-bartsch2021projection} and is applied rival by rival to preserve independence. Under a \emph{shared category} $c\in\mathcal N$,
\begin{equation}\label{eq:sharedcat}
 \pi_{jh}=\beta^{\ind\{h=c\}}q_h/Z_c,\quad \mbox{where } Z_c=1+(\beta-1)q_c,\quad \beta>1, j,h\in\mathcal N,
\end{equation}
all buyers instead overweight rivals with the salient value $T_c$ and \emph{strength} $\beta$, as under representativeness \citep{Kahneman1972,Bordalo2016}. Thus, projection yields type-specific beliefs, whereas a shared category assigns all types the same belief law. We treat both kernels as scenario inputs rather than estimates.

%Relative to correct-belief auction design, the model separates the law evaluating revenue from the laws governing incentives. Relative to fixed-allocation payment design, it makes service endogenous under statewise participation and payment restrictions, so an allocation change affects both required rents and their delivery capacity. We analyze this interaction through common-law rent floors, fractional-knapsack delivery costs, and dual allocation scores that establish global optimality under the stated conditions. For reporting guarantees, finite deletion certificates yield matched joint-design and reserve programs, while a transport test identifies when admissible report laws restore objective-law incentive inequalities. The band extension uses boundary indifference across an affine belief interval and an explicit attaining reserve. Throughout, entry, belief families, reporting requirements, and correct strategy conjectures are maintained inputs.

\section{Allocation as Rent Capacity}\label{sec:allocation}
Allocation determines both the incentive rents a seller must provide and the states in which she can provide them. Under a subsidy cap, serving a report creates capacity to deliver utility, and that capacity is cheaper when buyers overweight the state relative to the seller. A common subjective law separates the required rent from its least-cost delivery and yields the global optimality conditions below for efficiency and lower-value lotteries. This section studies nominal truthfulness. Section~\ref{sec:implementation} then prices stronger reporting requirements.

Write $\mathbf D_h=(m-1)\mathbf e_h$ for the rival configuration in which every rival reports $T_h$, and define $
 \Lambda_j(\mathbf{k})=\ph_j(\mathbf{k})/\Phi(\mathbf{k})$ and $
 \Lambda_j=\max_{\mathbf{k}\in\mathcal{K}}\Lambda_j(\mathbf{k}).$
Delivering one unit of perceived rent at $\mathbf k$ costs $1/\Lambda_j(\mathbf k)$ objectively, so delivery is cheapest at type $T_j$'s \emph{cost peaks}, the states in $\argmax_{\mathbf k}\Lambda_j(\mathbf k)$. Under efficient allocation, report $T_h$ wins with probability $\bar p_h=\sum_{\mathbf{k}}\Phi(\mathbf{k})p_h^n(\mathbf{k})$, and the highest value is $T_h$ with probability $w_h=mq_h\bar p_h$.

\subsection{Required Rents and Delivery Capacity}\label{sec:common}
Suppose all types share a common subjective law, $\ph_j=\ph$, as under correct beliefs or shared-category beliefs \eqref{eq:sharedcat}. The likelihood ratios and their peaks are then independent of type, and
each report has a common perceived winning probability, $P_{jh}(\mathbf p)=P_h(\mathbf p):=\sum_{\mathbf k}\ph(\mathbf k)p_h(\mathbf k)$. Define the perceived \emph{rent floor} $\underline{\Shat}_j(\mathbf p)=\sum_{h>j}(T_{h-1}-T_h)P_h(\mathbf p)$ with $\underline{\Shat}_n(\mathbf p)=0$, and its least delivery cost under the cap:
\[
 U_j^\varsigma(\mathbf p)=\min\left\{\sum_{\mathbf k}\Phi(\mathbf k)u_j(\mathbf k):\ \sum_{\mathbf k}\ph(\mathbf k)u_j(\mathbf k)\ge\underline{\Shat}_j(\mathbf p),\ 0\le u_j(\mathbf k)\le(T_j+\varsigma)p_j(\mathbf k)\ \forall\mathbf k\in\mathcal{K}\right\},
\]
where $(T_j+\varsigma)p_j(\mathbf k)$ is the \emph{rent capacity} created by the allocation at $\mathbf k$.

\begin{thm}[Common laws]\label{thm:common}
Let $\ph_j=\ph$ for every $j$ and $\mathbf p\in\cP$.
(i) With or without \eqref{eq:cap}, $\mathbf p$ is implementable if and only if $P_1(\mathbf p)\ge\cdots\ge P_n(\mathbf p)$;
(ii) For an implementable $\mathbf p$, every optimal rent schedule, with or without \eqref{eq:cap}, satisfies $\Shat_j=\underline{\Shat}_j(\mathbf p)$ for all $j$, and
\[
 \Rp(\mathbf p)=G(\mathbf p)-m\sum_jq_j\frac{\underline{\Shat}_j(\mathbf p)}{\Lambda_j},\qquad
 \Rp^\varsigma(\mathbf p)=G(\mathbf p)-m\sum_jq_jU_j^\varsigma(\mathbf p).
\]
(iii) For an implementable $\mathbf p$, an uncapped rent schedule $\mathbf u\ge0$ is optimal if and only if, for every $j$,  $u_j(\mathbf k) =0$ for every $\mathbf k \not \in \argmax_{\mathbf k^\prime}\Lambda_j(\mathbf k^\prime)$ and $\Shat_j=\underline{\Shat}_j(\mathbf p)$.
\end{thm}

The cap leaves common-law implementability unchanged but can raise rent-delivery cost. A losing report has no capacity at its cost peak under \eqref{eq:cap}; the seller must use a costlier state or change allocation to create capacity. With correct beliefs, all states have the same delivery cost and both revenues equal $G(\mathbf p)-m\sum_jq_j\underline{\Shat}_j(\mathbf p)$.

\begin{prop}[Rent capacity]\label{prop:knapsack}
Let $\ph_j=\ph$ for every $j$ and let $\mathbf p$ be implementable under \eqref{eq:cap}. 
\begin{equation}\label{eq:knapdual}
 U_j^\varsigma(\mathbf p)=\max_{\vartheta\ge0}\Bigl\{\vartheta\,\underline{\Shat}_j(\mathbf p)-(T_j+\varsigma)\sum_{\mathbf k}\Phi(\mathbf k)p_j(\mathbf k)\pos{\vartheta\Lambda_j(\mathbf k)-1}\Bigr\}, \quad \mbox{for every $j$},
\end{equation}
with a maximizer in $\{0\}\cup\{1/\Lambda_j(\mathbf k):\mathbf k\in\cK\}$. A rent schedule $\mathbf u$ is optimal if and only if, for every $j$, $\Shat_j=\underline{\Shat}_j(\mathbf p)$, $0\le u_j\le(T_j+\varsigma)p_j$, and $u_j(\mathbf k)=(T_j+\varsigma)p_j(\mathbf k)$ whenever $u_j(\mathbf k')>0$ and $\Lambda_j(\mathbf k)>\Lambda_j(\mathbf k')$.
\end{prop}

The fractional-knapsack rule exhausts cheaper rent capacity before utilizing costlier states. The multiplier $\vartheta$ is  the marginal cost of perceived rent: service has incentive value only when it unlocks capacity cheaper than this cost. This exact reduction allows joint optimization of allocation and payment without tracking a separate rent variable for each state.

\begin{cor}[A compact program]\label{cor:compact}
Let $\ph_j=\ph$ for every $j$. Then $\Rstar= \max_{\mathbf p\in\cP} \left\{ \Rp(\mathbf p): \  P_1(\mathbf p)\ge\cdots\ge P_n(\mathbf p)\right\}$, and 
\[
\mathcal R^{*,\varsigma} =\left\{\begin{array}{ll}
\max_{\mathbf p\in\cP,\mathbf U\ge0} &G(\mathbf p)-m\sum_{j<n}q_jU_j \\
 \text{s.t.} & P_1(\mathbf p)\ge\cdots\ge P_n(\mathbf p),\\
 &U_j\ge\frac{\underline{\Shat}_j(\mathbf p)}{\Lambda_j(\mathbf k^\prime)}-(T_j+\varsigma)\sum_{\mathbf k}\Phi(\mathbf k)p_j(\mathbf k)\pos{\frac{\Lambda_j(\mathbf k)}{\Lambda_j(\mathbf k^\prime)}-1}, \ \forall j<n, \mathbf k^\prime\in\cK
\end{array}\right..
\]
\end{cor}

The program replaces the $n|\mathcal K|$ rents, $n(n-1)$ incentive constraints, and $n|\mathcal K|$ caps by $n-1$ variables and at most $(n-1)(L+1)$ inequalities, where $L$ is the number of distinct likelihood ratios. Allocation still ranges over $|\cK|=\binom{m+n-2}{n-1}$ rival configurations. This count grows polynomially in either $m$ or $n$ when the other is fixed.

\subsection{When a Lower-Value Lottery Is Optimal}\label{sec:sharedtop}
Let $n=3$ and let all types share the top category, $c=1$ in \eqref{eq:sharedcat}. Section~\ref{sec:common} applies with $\Lambda_j(\mathbf{k})=\beta^{k_1}Z_1^{1-m}$ and $\varphi(\mathbf{D}_1)=(\beta q_1/Z_1)^{m-1}$. The common cost peak $\mathbf{D}_1$ is a state where middle and low reports lose under every reserve rule, so \eqref{eq:cap} forbids delivering their rent there. Define 
\[
 \beta_c^{m-1}=1+\frac{(q_1+q_2)(T_1-T_2)}{q_2(T_2+\varsigma)},\qquad
 \lambda^\varsigma=\frac{T_2-T_3}{m(T_2+\varsigma)}\Bigl(\frac{q_3}{\beta q_1}\Bigr)^{m-1}.
\]
The rent-adjusted scores when no top report is present are
\[
 \bar\psi_2=T_2-\frac{q_1}{q_2}(T_1-T_2)\beta^{1-m},\qquad
 \bar\psi_3=T_3-\frac{T_2-T_3}{q_3}\Bigl[q_1\beta^{1-m}+q_2\min\Bigl\{1,\Bigl(\frac{\beta_c}{\beta}\Bigr)^{m-1}\Bigr\}\Bigr].
\]
Let $\mathbf{p}^{\mathrm{st}}$ equal $\mathbf{p}^n$ except that $p_2(\mathbf{D}_1)=\lambda^{\varsigma}$ and $p_1(\mathbf{D}_1-\mathbf{e}_1+\mathbf{e}_2)=(1-\lambda^{\varsigma})/(m-1)$: when $m-1$ top reports meet one middle report, the middle buyer wins with probability $\lambda^{\varsigma}$. Assume
\begin{equation}\tag{ST}\label{eq:ST}
 \bar\psi_2>\bar\psi_3>0,\qquad
 \underline{\Shat}_1(\mathbf p^n)+(T_1-T_2)\ph(\mathbf D_1)\lambda^\varsigma\ind\{\beta\ge\beta_c\}\le\frac{(T_1+\varsigma)\ph(\mathbf D_1)}m.
\end{equation}

The score inequalities preserve middle- and low-type service whenever no top report is present. The capacity inequality ensures that serving the middle report does not force the top type's rent into a costlier state. These conditions are checked at the specified strength and cap: $\beta_c$ classifies allocations within this region, rather than identifying the transition over all belief strengths.

\begin{thm}[Shared-top category under a subsidy cap]\label{thm:sharedtop}
Impose \eqref{eq:cap} and assume \eqref{eq:ST}.
(i) If $\beta<\beta_c$, $\mathbf p^n$ is the unique globally optimal allocation;
(ii) If $\beta>\beta_c$, $\mathbf p^{\rm st}$ is the unique globally optimal allocation, with unique optimal rents: $u_1(\mathbf k)=\underline{\Shat}_1(\mathbf p^{\rm st})\ind\{\mathbf k=\mathbf D_1\}/\ph(\mathbf D_1)$, $u_2(\mathbf k)=(T_2+\varsigma)\lambda^\varsigma\ind\{\mathbf k=\mathbf D_1\}$, and $u_3(\mathbf k)=0$, and allocative welfare loss $S-G(\mathbf p^{\rm st})=mq_1^{m-1}q_2(T_1-T_2)\lambda^\varsigma$;
(iii) If $\beta=\beta_c$, the globally optimal allocations are the convex combinations of $\mathbf p^n$ and $\mathbf p^{\rm st}$.
In every case, $\Rp^\varsigma(\mathbf p^1)=w_1T_1$, $\Rp^\varsigma(\mathbf p^2)=w_1T_1+w_2\bar\psi_2$, $\mathcal R^{*,\varsigma}=w_1T_1+w_2\bar\psi_2+w_3\bar\psi_3$, and
\[
 \mathcal R^{*,\varsigma}-\RE^\varsigma=q_2(T_2-T_3)q_3^{m-1}\pos{1-(\beta_c/\beta)^{m-1}},\qquad
 \mathcal R^{*,\varsigma}-\RR^\varsigma=\min\bigl\{\mathcal R^{*,\varsigma}-\RE^\varsigma,\ q_3^m\bar\psi_3\bigr\}.
\]
\end{thm}
%\noindent\textit{Proof sketch.} The common-law reduction expresses required rents as functions of allocation. The proposed lottery leaves the middle type’s floor unchanged, expands low-cost delivery capacity, and raises the top type’s floor. Dual scores upper-bound revenue for every feasible mechanism, and under \eqref{eq:ST}, the proposed allocation attains this bound. Complementary slackness establishes uniqueness and the threshold equality case; see Appendix~\ref{app:proofs}.

The lottery of allocating to a middle report against $m-1$ top reports creates rent capacity where every reserve rule provides none. A winning probability $\lambda$ supplies $(T_2+\varsigma)\lambda$ units of capacity, and the corresponding full profile occurs with objective probability $mq_1^{m-1}q_2$. Under \eqref{eq:ST}, for $\beta\ge\beta_c$ and $\lambda\in[0,\lambda^\varsigma]$, marginal revenue is $mq_1^{m-1}q_2(T_2+\varsigma)(\beta^{m-1}-\beta_c^{m-1})$. Thus, $\beta_c$ balances cheaper rent delivery against foregone surplus and additional rents owed to top types. At $\lambda^\varsigma$, the middle type's entire perceived rent floor is delivered at the peak. Gains over efficiency and the best reserve coincide unless excluding the low type raises reserve revenue.

With beliefs fixed, relaxing the cap within the lottery regime lowers both $\lambda^\varsigma$ and $\beta_c$, reducing the allocation distortion needed to deliver rent. If \eqref{eq:ST} remains valid as $\varsigma\to\infty$, the lottery vanishes while the gain over efficiency converges to $q_2(T_2-T_3)q_3^{m-1}(1-\beta^{1-m})>0$. Every finite cap still excludes transfers to sure losers. Under stricter reporting guarantees, service can also deter misreports without giving the intended type rent, so certified designs may use different states and payments. 

\noindent\textbf{Top-rent overflow.}
If top rent no longer fits at the cost peak, a lottery can displace rent previously allocated in states where top reports forgo service. Partition states by their top-rival count $t=k_1$. Write $d_{12}=T_1-T_2$, $d_{23}=T_2-T_3$, $B_j=T_j+\varsigma$, and $\Lambda^{[t]}=Z_1^{1-m}\beta^t$. For $0\le t\le m-1$, define the top type's perceived capacity $C_t(\mathbf p)=B_1\sum_{\mathbf k:k_1=t}\ph(\mathbf k)p_1(\mathbf k)$. Extend $\mathbf p^\lambda$ to $0\le\lambda\le1$ by adjusting the profile so that the middle report wins with probability $\lambda$, and the top reports share the residual. For $\ell=0,\ldots,m-2$, set
\[
 \zeta_\ell^{\rm ov}=\frac{\beta^{-\ell}}{B_2}
 \left(\frac{B_1}{\beta}+\frac{q_1d_{12}}{q_2}\right),\quad
 a_\ell=\min\{1,\zeta_\ell^{\rm ov}\},\quad
 \widetilde\psi_2=T_2-\frac{q_1d_{12}}{q_2}\beta^{-\ell},\quad
 \widetilde\psi_3=T_3-\frac{d_{23}}{q_3}(q_1\beta^{-\ell}+q_2a_\ell).
\]
The conditions at $(\ell,\lambda)$ are
\begin{equation}\tag{TC}\label{eq:overflow-tc}
 0\le\lambda\le1,\quad P_1(\mathbf p^\lambda)\ge P_2(\mathbf p^\lambda),\quad
 \widetilde\psi_2>\widetilde\psi_3>0,\quad
 \sum_{t>\ell} C_t(\mathbf p^\lambda)\le\underline{\Shat}_1(\mathbf p^\lambda)
       \le\sum_{t\ge\ell}C_t(\mathbf p^\lambda).
\end{equation}
\begin{thm}[Allocation with top-rent overflow]\label{thm:overflow}
Let $n=3$, assume shared-top beliefs with $\beta>1$, and impose \eqref{eq:cap}. Fix $\ell\in\{0,\ldots,m-2\}$.
(i) If $\zeta_\ell^{\rm ov}>1$ and \eqref{eq:overflow-tc} holds at $\lambda=0$, $\mathbf p^n$ is uniquely optimal;
(ii) If $\zeta_\ell^{\rm ov}<1$ and \eqref{eq:overflow-tc} holds at $\lambda=\lambda^\varsigma$, the lottery $\mathbf p^{\rm st}$ is optimal and uniquely maximizes $G$ among revenue-optimal allocations. It is uniquely optimal if $\ell=m-2$. If $\ell<m-2$ and the lower capacity inequality is strict at $\lambda=\lambda^\varsigma$, optimal allocations are nonunique;
(iii) If $\zeta_\ell^{\rm ov}=1$ and \eqref{eq:overflow-tc} holds at both $0$ and $\lambda^\varsigma$, every $\mathbf p^\lambda$ with $0\le\lambda\le\lambda^\varsigma$ is optimal.
For any $0\le\lambda\le\lambda^\varsigma$ satisfying \eqref{eq:overflow-tc}, optimized revenue at $\mathbf{p}^\lambda$ is given by \eqref{ec:overflow-rev}. Optimal top-rent delivery exhausts capacity in each tier $t>\ell$ and places the remainder in tier $\ell$. 
\end{thm}

Along a segment satisfying \eqref{eq:overflow-tc}, marginal revenue with respect to $\lambda$ is $mq_1^{m-1}q_2B_2\beta^{m-1}(1-\zeta_\ell^{\mathrm{ov}})$. The threshold prices both additional top rent and displaced delivery capacity. When several profiles supply the same perceived middle rent, concentrating service at the profile with the most top rivals sacrifices the least allocated surplus. Appendix~\ref{app:overflow} characterizes the optimal set in parts (ii)--(iii).

Theorems~\ref{thm:sharedtop} and~\ref{thm:overflow} establish global optima on the score and capacity regions specified by \eqref{eq:ST} and \eqref{eq:overflow-tc}, and together classify the calibrated belief path. Other three-type economies can reach a capacity boundary before the stated lottery exhausts the middle-type rent requirement; an intermediate lottery can then be optimal. The compact common-law program remains the exact characterization outside the theorem regions.

\section{Reporting Guarantees and Their Revenue Cost}\label{sec:implementation}
A nominal optimum ensures truth-telling is an equilibrium but can also permit other reporting equilibria. The operator first defines a sufficient reporting certificate and a utility margin, then compares joint and reserve designs under these criteria. Allocation reduces protection costs either by awarding truthful rent or by making false reports costly to mimic. We optimize within the specified certificate classes, maintaining fixed entry, given beliefs, and correct strategy conjectures.

\subsection{Pooling and Certificates}\label{sec:unique}
As in other multi-agent mechanisms \citep{DemskiSappington1984,MaMooreTurnbull1988}, buyers may prefer an untruthful SE. The next result characterizes all-top pooling equilibria.
\begin{prop}[Pooling at the top]\label{prop:pool}
Suppose $p_1(\mathbf{D}_1)=1/m$ and $p_h(\mathbf{D}_1)\le1/m$ for $h\ge2$. The profile in which every buyer always reports $T_1$ is an SE if and only if
\[
 mx_1(\mathbf{D}_1)\le T_n-m\max_{h\ge2}\{T_np_h(\mathbf{D}_1)-x_h(\mathbf{D}_1)\}.
\]
Its revenue equals the winner's price,
$mx_1(\mathbf{D}_1)=T_1-mu_1(\mathbf{D}_1)$.
Every reserve rule has $p_h(\mathbf{D}_1)=0$ for $h\ge2$, and under \eqref{eq:ir} and \eqref{eq:cap} also $x_h(\mathbf{D}_1)=0$, reducing the criterion to $mx_1(\mathbf{D}_1)\le T_n$.
\end{prop}

Pooling makes the discounted all-top profile certain, exposing the revenue risk of rent concentration. The lowest type's deviation constraints suffice because the top report's payoff advantage is nondecreasing in value. In the Introduction's two-type example, pooling is strict and preferred to truth by every type; it yields zero revenue at the optimum with nonnegative payments and $-\tfrac{16}{11}$ at the uncapped optimum (Section~\ref{sec:otherdist}). At the optimum of Theorem~\ref{thm:sharedtop}(ii), pooling is an SE exactly when $T_1-m\underline{\Shat}_1(\mathbf p^{\rm st})/\ph(\mathbf D_1)\le T_3-m(T_3+\varsigma)\lambda^\varsigma$. Under \eqref{eq:cap}, pooling revenue is at least $-\varsigma$.

Certificates with a positive margin rule out such equilibria by eliminating false reports in an order chosen by the designer. A strategy $\xi_i$ (Appendix~\ref{app:model}) is \emph{truthful on} $\mathcal N'\subseteq\mathcal N$ if $\xi_i(\{T_t\}\mid T_t)=1$ for every $t\in\mathcal N'$. For $\mathcal S=(\mathcal S_t)_{t\in\mathcal N}$ with $t\in\mathcal S_t\subseteq\mathcal N$, let $\mathcal F_j(\mathcal S)$ be the finite set of rival-report configuration laws induced from $\ph_j$ when rivals use possibly different pure strategies mapping each $T_t$ into $\{T_h:h\in\mathcal S_t\}$. Truthful reporting gives $\ph_j\in\mathcal F_j(\mathcal S)$. For a finite set $\mathcal X$ of laws on $\cK$, write $\co(\mathcal X)$ for its convex hull. Abbreviate $\mathcal F_j(\mathcal S)$ as $\mathcal F_j(\mathcal N')$ when $\mathcal S_t=\{t\}$ for $t\in\mathcal N'$ and $\mathcal S_t=\mathcal N$ otherwise. An \emph{ordering} is a strict total order $\prec$ on $\mathcal N$, with $\mathcal N_{\prec j}=\{h:h\prec j\}$. A \emph{deletion schedule} orders all $n(n-1)$ pairs $(j,h)$ with $j\ne h$ as $(j_\tau,h_\tau)_{\tau=1}^{n(n-1)}$. Starting from $\mathcal S_t^0=\mathcal N$, step $\tau$ sets $\mathcal S_{j_\tau}^{\tau}=\mathcal S_{j_\tau}^{\tau-1}\setminus\{h_\tau\}$ and leaves all other report sets unchanged.

\begin{defi}[Certificates]\label{def:certificate}
A mechanism has a \emph{whole-type certificate} with margin $\delta\ge0$ in the ordering $\prec$, or a \emph{report-by-report certificate} with margin $\delta$ in a deletion schedule, if, respectively,
\begin{equation}\label{eq:certificate}
 \mathrm{sl}_{jh}(\ph)\ge\delta\quad\bigl(j\ne h,\ \ph\in\mathcal F_j(\mathcal N_{\prec j})\bigr),\qquad
 \mathrm{sl}_{j_\tau h_\tau}(\ph)\ge\delta\quad\bigl(\ph\in\mathcal F_{j_\tau}(\mathcal S^{\tau-1}),\ \tau=1,\ldots,n(n-1)\bigr).
\end{equation}
\end{defi}

Whole-type certification tests every false report of a type against rivals truthful only at earlier types. Report-by-report certification also uses earlier partial deletions at any type. Processing each type's false reports consecutively turns a whole-type certificate into a report-by-report certificate with the same margin. The schedule is a proof of the contract's incentives, not a sequence of instructions that buyers must follow.

Let $\mathcal R_{\rm seq}(\delta)$ and $\mathcal R_{\rm del}(\delta)$ be the largest revenue $\operatorname{Rev}(\mathbf p,\mathbf u)$ over $\mathbf p,\mathbf u\ge0$ satisfying \eqref{eq:FEA} and a whole-type, respectively report-by-report, certificate with margin $\delta$, maximized over the $n!$ orderings, respectively the $\{n(n-1)\}!$ schedules, with $\max\varnothing=-\infty$. For $C\in\{{\rm seq},{\rm del}\}$, let $\mathcal E$ denote an \emph{elimination order} of class $C$: an ordering $\prec$ if $C={\rm seq}$ and a deletion schedule if $C={\rm del}$. Let $\mathcal R_C^{\mathcal E}(\delta)$ be the inner maximum for $\mathcal E$, and $\mathcal R_C(\delta;\mathbf p^r)$ and $\mathcal R_C^{\mathcal E}(\delta;\mathbf p^r)$ these values with $\mathbf p=\mathbf p^r$ fixed. A superscript $\varsigma$ adds \eqref{eq:cap}. Finally, a rival \emph{ranks} by a strict total order on $\mathcal N$ if she reports, at every value $T_t$, the first element of $\mathcal S_t$ under that order; $\mathcal V_j(\mathcal S)\subseteq\mathcal F_j(\mathcal S)$ denotes the laws induced when every rival ranks, by orders that may differ across rivals. Definition~\ref{def:certprograms} in Appendix~\ref{app:proofs} states these strategy sets, certificate constraints, and revenue programs formally.

\begin{thm}[Certificates and unique implementation]\label{thm:unique}
The following hold with and without \eqref{eq:cap}.
(i) If a mechanism has a whole-type or report-by-report certificate with $\delta>0$, truthful reporting is its unique SE; 
(ii) For each ordering or schedule $\mathcal E$, \eqref{eq:certificate} is a finite system of linear inequalities in $(\mathbf p,\mathbf u)$, so $\mathcal R_C^{\mathcal E}(\delta)$ and $\mathcal R_C^{\mathcal E}(\delta;\mathbf p^r)$ are LP values, attained whenever finite;
(iii) Under subjective independence, \eqref{eq:certificate} holds if and only if it holds with $\mathcal V_j$ replacing $\mathcal F_j$, and $|\mathcal V_j(\mathcal N_{\prec j})|\le|\cK|$.
\end{thm}
%\noindent\textit{Proof sketch.} Mixed rival strategies induce convex combinations of the tested laws. The first strict comparison eliminates a false report against every rival strategy; each subsequent comparison uses the restrictions already established. Induction eliminates every false report, including from mixed and asymmetric equilibria \citep[cf.][]{BergemannMorris2009}. Full proofs are in Appendix~\ref{app:proofs}.

For each fixed order or schedule, the LP compares allocations and payments subject to the specified reporting requirements. A positive margin requires a utility advantage at every deletion step, although participation may still bind. Maximizing over schedules compares the best joint and reserve designs within the same certificate class, even when their optimal schedules differ. These are sufficient certificates, so their revenue cost need not equal the minimum cost of unique implementation across all mechanisms. Ranking reduces the relevant laws within each schedule but leaves the outer optimization unchanged. Enumerating the $6!=720$ deletion schedules for three reports permits exact comparisons within the report-by-report class. The five-report whole-type and ten-report nominal calculations below impose different reporting requirements.

\begin{prop}[The cost of a positive certificate margin]\label{prop:margincost}
With and without \eqref{eq:cap}, for every $\delta\ge0$, $\mathcal R_{\rm seq}(\delta)\le\mathcal R_{\rm del}(\delta)\le\Rstar$, and $\mathcal R_C(0)-\mathcal R_C(\delta)=O(\delta)$ as $\delta\downarrow0$ for $C\in\{{\rm seq},{\rm del}\}$.
\end{prop}

For a fixed grid, a small margin costs arbitrarily little within either class. The witness has advantage $(T_j-T_h)^2/(2mT_1)$ at every configuration, so the conclusion is not uniform as value gaps vanish. Restricting nominal design to the certificate class has a separate cost:
\begin{equation}\label{eq:accounting}
 \Rstar-\mathcal R_C(\delta)=\underbrace{\Rstar-\mathcal R_C(0)}_{\text{certificate-class cost}}+\underbrace{\mathcal R_C(0)-\mathcal R_C(\delta)}_{\text{positive-margin cost}}.
\end{equation}
The identity also holds under \eqref{eq:cap}. A zero-margin certificate benchmarks the selected class without guaranteeing uniqueness. The first term is the cost of that class restriction, not the minimum cost of unique implementation across all mechanisms. The second prices a positive margin within the class. Section~\ref{sec:data} reports both, because choosing the class can cost much more than tightening its margin.

The \emph{matched allocation gain} $\mathcal R_C(\delta)-\max_r\mathcal R_C(\delta;\mathbf p^r)$ is the allocation gain within the same class and margin. Subtracting the larger zero-margin reserve value $\max_r\mathcal R_C(0;\mathbf p^r)$ instead gives the \emph{conservative difference}, a lower bound on the matched gain; a negative difference is inconclusive. If every reserve problem is infeasible, we report that fact separately from revenue.

\begin{prop}[The shadow price of a certificate margin]\label{prop:shadow}
With and without \eqref{eq:cap}, for each $C$, $\mathcal E$, and $r$, $\mathcal R_C^{\mathcal E}(\delta)\ge\mathcal R_C^{\mathcal E}(\delta;\mathbf p^r)$, and both are concave, piecewise linear, and nonincreasing in $\delta$ on the closed intervals where they are finite, with right derivative $-\bar y$ below the right end, where $\bar y$ is the largest sum of optimal dual variables of the inequalities \eqref{eq:certificate}. Hence the matched gain is nonnegative where finite; at $\delta$ below the right ends of the programs attaining the two maxima, its right derivative is the smallest $\bar y$ among the maximizing reserve programs minus the smallest $\bar y$ among the maximizing joint programs.
\end{prop}

The shadow-price comparison explains whether allocation makes a chosen margin cheaper; it does not give primitive conditions for a strict gain. Both revenue frontiers fall, and maximizing programs may change, so the allocation gain need not be monotone and a larger gap need not mean higher revenue. Section~\ref{sec:data} gives matched cases. Whole-type certification also restricts feasibility:

\begin{lem}[Reserve rules and whole-type margins]\label{lem:reservecert}
Let $n\ge3$ and impose \eqref{eq:cap}. For every collection of subjective laws, reserve $T_r$, and $\delta>0$, $\mathcal R^\varsigma_{\rm seq}(\delta;\mathbf p^r)=-\infty$.
\end{lem}

Under \eqref{eq:cap}, all reports below the top receive zero when rivals always report the top. The first lower type processed as a block cannot strictly prefer truth to every other losing report. Report-by-report certificates use partial deletions to certify reserves at small margins. Fixing a deletion schedule isolates allocation’s effect on the cost of a positive reporting margin. The construction below compares the joint design with every reserve design under that schedule.

\begin{prop}[A strict gain under a fixed certificate]\label{prop:fixedgain}
Let $m=2$, $n=3$, $\varsigma=0$, and $q_j=1/3$, with shared-top beliefs at $\beta=4$. Fix $\mathcal E=((2,1),(3,1),(3,2),(2,3),(1,3),(1,2))$. If $2T_1<5T_2$ and $0<\delta\le\min\{d_{12}/20,d_{23}/16,(T_1+9T_2-15T_3)/36\}$, then $\mathbf p^2$ uniquely maximizes reserve revenue under $\mathcal E$, and
\[
 \max_r\mathcal R_{\rm del}^{\mathcal E,0}(\delta;\mathbf p^r)
 =\frac{16T_1+24T_2}{45}-\frac{12}{5}\delta,\quad
 \mathcal R_{\rm del}^{\mathcal E,0}(\delta)
 -\mathcal R_{\rm del}^{\mathcal E,0}(\delta;\mathbf p^2)
 \ge \frac{2(5T_2-2T_1)}{9T_2}\delta>0.
\]
A witnessing allocation $\mathbf p^{\rm lot}$ agrees with $\mathbf p^2$ except at the top--middle profile, where the middle report wins with probability $\lambda=\delta/T_2$ and the top report wins with probability $1-\lambda$.
\end{prop}

The free middle-report win supplies rent against a top report. Relative to the best reserve, the witness saves $2\delta/3$ in expected middle-type rent, while foregone surplus and additional top-type rent each cost $2d_{12}\delta/(9T_2)$. The primitive inequality ensures a positive net gain, with all designs preserving the displayed schedule.

\subsection{When a Reporting Certificate Removes the Belief Premium}\label{sec:restore}
\begin{prop}[Objective-law restoration]\label{prop:transport}
Assume subjective independence and $t\in\mathcal S_t\subseteq\mathcal N$ for every $t$.
(i) $\Phi\in\co(\mathcal F_j(\mathcal S))$ if and only if
\begin{equation}\label{eq:hall}
 \sum_{t\in\mathcal N'}\pi_{jt}\le\sum_{h\in\cup_{t\in\mathcal N'}\mathcal S_t}q_h,\qquad\mathcal N'\subseteq\mathcal N;
\end{equation}
for $\mathcal F_j(\mathcal N_{\prec j})$, this reduces to $\pi_{jt}\le q_t$ for every $t\prec j$.
(ii) If \eqref{eq:hall} holds for $(j_\tau,\mathcal S^{\tau-1})$ at every step $\tau$ of a schedule, or for $(j,\mathcal N_{\prec j})$ for every $j$ in an ordering, every certificate with margin $\delta\ge0$ in it has $\mathrm{sl}_{jh}(\Phi)\ge\delta$ for all $j\ne h$, and its revenue is at most $\max_{\mathbf p\in\cP}\Rp^{\mathrm{rat}}(\mathbf p)$.
\end{prop}

View each rival type as a source of subjective probability mass and its remaining reports as destinations. Restoration is possible exactly when every subset's mass fits within the objective probability of its admissible destinations. This is the weighted Hall condition, checkable by maximum flow. When restoration holds at every step, the certificate must also protect incentives under the objective law and cannot earn more than the rational optimum. Failure removes this upper-bound argument; it does not itself establish a premium.
The rational optimum uses a monotone allocation with pointwise rational rents, $u_j(\mathbf{k})=\sum_{h>j}(T_{h-1}-T_h)p_h(\mathbf{k})$, and nonnegative payments (Lemma~\ref{lem:rational}). Those rents satisfy $g_{jh}\ge0$ statewise and hence every zero-margin certificate under every law. Therefore:

\begin{cor}[Certificates without overweighting]\label{cor:restore}
Assume subjective independence and $\pi_{jt}\le q_t$ for all $j$ and $t\ne j$. Then $\mathcal R^\varsigma_{\rm seq}(\delta)\le\mathcal R^\varsigma_{\rm seq}(0)=\max_{\mathbf p\in\cP}\Rp^{\mathrm{rat}}(\mathbf p)$ for all $\delta,\varsigma\ge0$, and likewise without \eqref{eq:cap}.
\end{cor}

Projection satisfies $\pi_{jt}=(1-\alpha)q_t$ for $t\ne j$, so whole-type certification removes the \emph{projection premium} above rational-optimal revenue. Report-by-report deletions can instead confine several types to too few reports for restoration. Under shared-category beliefs, only the salient category is overweighted, so every ordering placing that category last also faces the rational bound. The following example separates the two certificate classes.

\begin{exap}[A projection premium under unique implementation]\label{ex:premium}
Let $m=2$, $\mathcal T=\{30,20,10\}$, $\mathbf{q}=(1/3,1/3,1/3)$, and $\alpha=1/2$ in \eqref{eq:projection}. Let $0<\delta\le1/2$ and give $\mathbf{p}^2$ the rents:
\[
 \bigl(u_1(\mathbf D_h)\bigr)_{h=1}^3=\Bigl(2+\tfrac{36}{25}\delta,\ 7+\tfrac{66}{25}\delta,\ 0\Bigr),\qquad
 \bigl(u_2(\mathbf D_h)\bigr)_{h=1}^3=\Bigl(0,\ \tfrac65\delta,\ \tfrac65\delta\Bigr),\qquad u_3\equiv0.
\]
This mechanism has nonnegative payments and a report-by-report certificate with margin $\delta$ in the schedule $(2,1),(3,1),(2,3),(1,3),(1,2),(3,2)$, so for every $\varsigma\ge0$, and without a cap,
\[
 \mathcal R^\varsigma_{\rm seq}(\delta)\le20<\frac{64}3-\frac{36}{25}\delta\le\mathcal R^\varsigma_{\rm del}(\delta).
\]
\end{exap}

At $\delta=0.01$, the certificate earns 21.32 against the rational optimum 20 and nominal optimum 23.26. It retains reserve allocation $\mathbf{p}^2$: this example separates certificate classes, rather than establishing an allocation gain. Both binding steps violate restoration. At $(2,3)$, middle and low rivals are confined to $T_2,T_3$, with subjective mass $5/6$ exceeding objective mass $2/3$; rent $6\delta/5$ there deters understatement. A whole-type step cannot use this partial restriction. Section~\ref{sec:data} shows the calibrated obstruction and separately measures allocation value.

\subsection{One Contract under Belief Uncertainty}\label{sec:robust}
\noindent\textbf{Candidate laws.} Suppose the seller knows $\Phi$ and that each type's law lies in $\co(\mathcal L_j)$, the convex hull of a finite set $\mathcal L_j$ of laws on $\cK$. A mechanism is \emph{robustly truthful} if
\begin{equation}\label{eq:robustic}
 \mathrm{sl}_{jh}(\ph)\ge0,\qquad j\ne h,\ \ph\in\mathcal L_j. \tag{IC$^{\mathcal L}$}
\end{equation}
A superscript $\mathcal L$ denotes the same problem with \eqref{eq:ic} replaced by \eqref{eq:robustic}. Attach the multiplier $m\mu^{\ph}_{jh}\ge0$ to the constraint for $(j,h,\ph)$, write $\boldsymbol\mu=(\mu^{\ph}_{jh})_{j\ne h,\,\ph\in\mathcal L_j}$, and set
\[
 \ell^{\mathcal L}_h(\mathbf{k};\boldsymbol\mu)=-q_h+\sum_{\ph\in\mathcal L_h}\frac{\ph(\mathbf{k})}{\Phi(\mathbf{k})}\sum_{t\ne h}\mu^{\ph}_{ht}-\sum_{j\ne h}\sum_{\ph\in\mathcal L_j}\mu^{\ph}_{jh}\frac{\ph(\mathbf{k})}{\Phi(\mathbf{k})},~
 \psi^{\mathcal L}_h(\mathbf{k};\boldsymbol\mu)=T_h-\frac1{q_h}\sum_{j\ne h}\sum_{\ph\in\mathcal L_j}\mu^{\ph}_{jh}\frac{\ph(\mathbf{k})}{\Phi(\mathbf{k})}(T_j-T_h).
\]
These extend the single-law coefficients $\ell_h$ and $\psi_h$ developed in Section~\ref{sec:certificates}; they coincide when $\mathcal L_j=\{\ph_j\}$ for every $j$.

\begin{prop}[Robust truthfulness]\label{thm:robust}
(i) A mechanism satisfies \eqref{eq:robustic} if and only if $\mathrm{sl}_{jh}(\ph)\ge0$ for all $j\ne h$ and $\ph\in\co(\mathcal L_j)$. An optimal robustly truthful mechanism exists, and every monotone $\mathbf p\in\cP$ is implementable under \eqref{eq:robustic} with $0\le u_j\le T_jp_j$;
(ii) Theorem~\ref{thm:dual} and Proposition~\ref{prop:universal} hold with \eqref{eq:robustic}, $\ell^{\mathcal L}_h$, $\psi^{\mathcal L}_h$, and $\boldsymbol\mu=(\mu^{\ph}_{jh})$ in place of \eqref{eq:ic}, $\ell_h$, $\psi_h$, and $(\mu_{jh})$.
\end{prop}

This keeps one contract truthful across a specified belief set; unique implementation also needs a positive certificate across that set. Robust revenue cannot exceed nominal revenue at any included belief profile. Its shortfall prices protection against the stated set, whose size is an input rather than an estimate.

\begin{prop}[A total-variation check]\label{prop:tv}
For every mechanism, $j\ne h$, and law $\ph$ on $\cK$,
\[
 \mathrm{sl}_{jh}(\ph_j)\ge\mathrm{sl}_{jh}(\ph)-\bigl(\max_{\mathbf{k}}g_{jh}(\mathbf{k})-\min_{\mathbf{k}}g_{jh}(\mathbf{k})\bigr)\,\frac12\sum_{\mathbf{k}}|\ph_j(\mathbf{k})-\ph(\mathbf{k})|.
\]
Under \eqref{eq:ir} and \eqref{eq:cap}, $\max_{\mathbf k}g_{jh}(\mathbf k)-\min_{\mathbf k}g_{jh}(\mathbf k)\le2(T_1+\varsigma)$.
\end{prop}

The margin is a required utility advantage; its implication for belief error depends on the range of deviation payoffs. A tested slack of at least $\delta$ remains nonnegative under total-variation error at most $\delta/\{2(T_1+\varsigma)\}$; strict inequality preserves positive slack. Every deletion step requires its own check on induced rival-report laws. With $\delta=\$0.01$, $T_1\simeq\$27.45$, and nonnegative payments, this conservative radius is $1.82\times10^{-4}$. It is a distance between probability laws, not a percentage error in a belief parameter or a reporting-error probability. Direct interval checks can protect larger specified sets. For intervals of intensities, let $\pi_j(\alpha)$ be affine in $\alpha\in[\underline\alpha,\overline\alpha]$, as under projection or a shared category mixed with $q$, let $\varphi_j(\cdot;\alpha)$ be the multinomial law with $m-1$ trials and cell probabilities $\pi_j(\alpha)$, and write $f_{jh}(\alpha)=\operatorname{sl}_{jh}(\varphi_j(\cdot;\alpha))$. Throughout the interval comparison, a finite grid $A$ is a subset of $[\underline\alpha,\overline\alpha]$. The grid program therefore relaxes the full interval requirement.

\begin{prop}[Verification over an interval]\label{prop:verify}
(i) For every mechanism and $j\ne h$, $f_{jh}$ is a polynomial of degree at most $m-1$, minimized over $[\underline\alpha,\overline\alpha]$ on $\mathcal A_{jh}=\{\underline\alpha,\overline\alpha\}\cup\{\alpha\in(\underline\alpha,\overline\alpha):f_{jh}'(\alpha)=0\}$;
(ii) If a mechanism is optimal under \eqref{eq:robustic} with $\mathcal L_j=\{\ph_j(\cdot;\alpha):\alpha\in\mathcal A\}$ for a finite $\mathcal A$, possibly with further restrictions on $(\mathbf p,\mathbf u)$, and $f_{jh}\ge0$ on $\mathcal A_{jh}$ for all $j\ne h$, it remains optimal with $\mathcal L_j=\{\ph_j(\cdot;\alpha):\alpha\in[\underline\alpha,\overline\alpha]\}$;
(iii) Parts (i)--(ii) hold for the slacks of a whole-type certificate with $\mathcal F_j(\mathcal N_{\prec j})$ induced under $\boldsymbol\pi_j(\alpha)$, with $\delta$ in place of $0$, and a certificate imposed under finitely many cell-probability vectors holds at every mixture of the induced laws.
\end{prop}

Finite verification of a fixed contract checks at most $m$ points per slack (only endpoints for a constant slack). A failing intensity can be added before re-solving. A grid optimum that passes certifies interval optimality. For a fixed report-by-report schedule, the deletion slacks likewise have degree at most $m-1$ because pure report probabilities are affine in intensity. Section~\ref{sec:data} verifies one contract and schedule across an interval. The next example shows that a positive conservative difference can survive category uncertainty.

\begin{exap}[Category uncertainty]\label{ex:category}
Let $m=2$, $(T_1,T_2,T_3)=(28,18,12)$, $\mathbf q=(0.30,0.30,0.40)$, and $\varsigma=0$, and let every type's law lie in the convex hull of the shared-category laws \eqref{eq:sharedcat} with $\beta=16$ and $c\in\{1,3\}$, whose cell probabilities are $\boldsymbol\pi^{\rm top}=(48,3,4)/55$ and $\boldsymbol\pi^{\rm bot}=(3,3,64)/70$. The online appendix gives a mechanism that satisfies \eqref{eq:FEA} and $0\le u_j\le T_jp_j$, earns $2450067/139750\approx17.53$, and has a whole-type certificate with margin $\frac1{100}$ in the ordering $3\prec1\prec2$ under both candidate laws. Every reserve rule's best zero-margin whole-type certificate under both laws earns at most $\frac{87}5=17.40$, the rational optimum, so the conservative difference is at least $0.13$.
\end{exap}

The appendix verifies the contract exactly. Under each candidate, the last type overweights an earlier value, so restoration fails at her step. The comparison concerns this stated set; it gives no guarantee for arbitrary changes in salience.

\subsection{Finite Reporting Bands}\label{sec:continuous}
Let values be independent draws from an atomless distribution $F$ with full support $[\underline v,\overline v]\subset(0,\infty)$, partitioned into bands $I_j=[T_j,T_{j-1})$, with $\overline v=T_0>T_1>\cdots>T_n=\underline v$ and $\overline v$ included in $I_1$. Here $T_j$ is a band floor, buyers report a band, $(p_j(\mathbf k),x_j(\mathbf k))$ is as in Section~\ref{sec:model}, and $q_j=F(T_{j-1})-F(T_j)$. A value in $I_j$ uses law $\ph_j$ over rivals' bands; under subjective independence, she regards rivals' values as independent draws from $\sum_t\pi_{jt}F(\cdot\mid I_t)$, where $F(\cdot\mid I_t)$ is the conditional distribution of a value given that it lies in $I_t$, so truthful rivals induce $\ph_j$. The rent $u_j=T_jp_j-x_j$ gives value $v\in I_j$ the utility $u_j(\mathbf k)+(v-T_j)p_j(\mathbf k)$ at configuration $\mathbf k$, and band revenue is $\operatorname{Rev}(\mathbf p,\mathbf u)$. For $v\in[T_j,T_{j-1}]$, the closure of $I_j$, the advantage of truth over band $h$ at configuration $\mathbf k$, and under a law $\ph$ on $\cK$, is
\[
 g_{jh}(\mathbf k,v)=u_j(\mathbf k)+(v-T_j)p_j(\mathbf k)-u_h(\mathbf k)-(v-T_h)p_h(\mathbf k),\qquad
 \mathrm{sl}_{jh}(\ph,v)=\sum_{\mathbf k}\ph(\mathbf k)g_{jh}(\mathbf k,v),
\]
which are affine in $v$, with $g_{jh}(\mathbf k,T_j)=g_{jh}(\mathbf k)$ and $\mathrm{sl}_{jh}(\ph,T_j)=\mathrm{sl}_{jh}(\ph)$.

\begin{thm}[Band incentives]\label{thm:cvband}
(i) Participation at every value is equivalent to \eqref{eq:ir}; given \eqref{eq:ir}, \eqref{eq:cap} is equivalent, under the transfer rule of Section~\ref{sec:model}, to zero transfers to losers and winner prices of at least $-\varsigma$.
(ii) Truthful band reporting at every value is equivalent to $\mathrm{sl}_{jh}(\ph_j,T_j)\ge0$ and $\mathrm{sl}_{jh}(\ph_j,T_{j-1})\ge0$ for $j\ne h$, that is, to
\begin{equation}\label{ec:cvbandic}
 \sum_{\mathbf{k}}\ph_j(\mathbf{k})\{u_j(\mathbf{k})-u_h(\mathbf{k})\}\ge(T_j-T_h)P_{jh}(\mathbf p)+(T_{j-1}-T_j)\pos{P_{jh}(\mathbf p)-P_{jj}(\mathbf p)},\qquad j\ne h,
\end{equation}
so the band optimum is the value of \eqref{eq:primal} with \eqref{eq:ic} replaced by \eqref{ec:cvbandic}.
\end{thm}

Endpoint checks protect every value rather than a conditional-mean representative. Revenue uses floor rents; welfare uses actual values. The interior coefficient $\eta$ scales the reporting advantage by distance from band boundaries. It is a different requirement from the exact-type dollar margin $\delta$, so their revenue costs must be compared separately. A band mechanism has an \emph{interior certificate} with coefficient $\eta\ge0$ in $\prec$ if, for all $j\ne h$ and $\ph\in\mathcal F_j(\mathcal N_{\prec j})$,
\begin{equation}\label{eq:interior}
 \mathrm{sl}_{jh}(\ph,T_j)\ge0,\qquad\mathrm{sl}_{jh}(\ph,T_{j-1})\ge0,\qquad\mathrm{sl}_{jh}(\ph,T_j)+\mathrm{sl}_{jh}(\ph,T_{j-1})\ge\eta\,(T_{j-1}-T_j).
\end{equation}

\begin{prop}[Interior certificates]\label{prop:interior}
Assume subjective independence.
(i) If a band mechanism has an interior certificate with coefficient $\eta$ in $\prec$, then $\mathrm{sl}_{jh}(\ph,v)\ge\eta\min\{v-T_j,T_{j-1}-v\}$ for all $j\ne h$, $v\in I_j$, and $\ph\in\mathcal F_j(\mathcal N_{\prec j})$; truthful reporting is an SE at every value, and if $\eta>0$, every SE, including asymmetric and mixed ones, is truthful at $F$-almost every value;
(ii) For each $\prec$ and $\eta$, \eqref{eq:interior} consists of at most $3n(n-1)|\cK|$ linear inequalities, so the largest band revenue under \eqref{eq:FEA}, \eqref{eq:ir}, and \eqref{eq:interior} is an LP value, and positive coefficients cost $O(\eta)$ revenue as $\eta\downarrow0$.
\end{prop}

The truthfulness, almost-everywhere uniqueness, finite-LP, and $O(\eta)$ cost conclusions extend to report-by-report schedules, with $\mathcal F_{j_\tau}(\mathcal S^{\tau-1})$ in place of $\mathcal F_j(\mathcal N_{\prec j})$ and a schedule-dependent constraint count. The cost bound mixes in the mechanism $p_j(\mathbf k)=(n-j)\bar\eta$, $x_j(\mathbf k)=\bar\eta\sum_{t=j}^{n-1}T_t$, with $\bar\eta=1/\{m(n-1)\}$, which satisfies \eqref{eq:FEA}, \eqref{eq:ir}, and \eqref{eq:cap} for all $\varsigma\ge0$, and \eqref{eq:interior} at $\bar\eta$ for all orderings.

These cost conclusions hold for the fixed partition. Under \eqref{eq:reserve}, $\mathbf p^r$ is a \emph{band reserve rule}. If adjacent bands share a law, their incentives force indifference at the common boundary, so a positive uniform dollar margin is impossible near it. Interior protection may decline to zero there. With a common law, boundary indifference further restricts reserve allocations.
\begin{lem}[Band reserve rules]\label{lem:bandreserve}
Let $n\ge3$ and let all bands share a law $\ph$ with $\ph(\mathbf k)>0$ for every $\mathbf k$. Under \eqref{eq:ir}, with or without \eqref{eq:cap}, no band mechanism using a reserve rule admits an interior certificate for any ordering $\prec$ and coefficient $\eta>0$.
\end{lem}

The impossibility concerns whole-type certificates. Report-by-report certificates can use earlier partial deletions. Section~\ref{sec:dataresolution} gives a positive matched three-band gain at fixed beliefs. The next result identifies a limit when one band contract must protect an interval of common beliefs.

\begin{prop}[Common-law intervals and band revenue]\label{prop:bandinterval}
Let $n=3$. For every value, let rivals be subjectively independent with common band probabilities $\boldsymbol\pi(\kappa)=(1-\kappa)\boldsymbol\pi^0+\kappa\mathbf q$, where $\boldsymbol\pi^0$ has full support and $\kappa\in[0,\bar\kappa]$, $0<\bar\kappa\le1$. At the floors, set $b_j^{\mathrm{rat}}=T_j-(T_{j-1}-T_j)\sum_{t<j}q_t/q_j$. If $b_1^{\mathrm{rat}}>b_2^{\mathrm{rat}}>0>b_3^{\mathrm{rat}}$, every statewise-participating band mechanism truthful at every value throughout the interval satisfies
$\operatorname{Rev}(\mathbf p,\mathbf u)\le\mathcal R_B
 :=m\sum_{j,\mathbf k}q_j\Phi(\mathbf k)b_j^{\mathrm{rat}}p_j^2(\mathbf k)$.
The reserve $\mathbf p^2$ with $u_1=(T_1-T_2)p_2^2$ and $u_2=u_3=0$ attains $\mathcal R_B$ with nonnegative payments. It has a report-by-report interior certificate throughout the interval in $\mathcal E=((1,3),(1,2),(2,1),(2,3),(3,1),(3,2))$ for every $0<\eta\le\eta^*$, where
 $\eta^*=\frac1m\min_{0\le\kappa\le\bar\kappa}
 \bigl\{\pi_1(\kappa)^{m-1},\ \pi_2(\kappa)^{m-1},\ [1-\pi_1(\kappa)]^{m-1}\bigr\}$.
The matched joint and best-reserve revenues under those report-by-report requirements therefore equal $\mathcal R_B$.
\end{prop}

Exact incentives on both sides of a band boundary require the expected adjacent-report gain to be zero under each common belief law. The proof uses a polynomial of degree at most $m-1$ in $\kappa$: zero on any nondegenerate interval implies zero at $\kappa=1$, even if the objective law lies outside that interval. The same degree argument explains why $m$ distinct scenarios on this affine path suffice for the ceiling: each boundary polynomial has more roots than its degree. Four buyers require only four distinct scenarios, retaining the common-law and exact-incentive assumptions. The ceiling therefore applies for every $\bar\kappa>0$ under the proposition's conditions. A small interval need not imply a small revenue cost when feasibility requires these equalities exactly. The payment cap is not needed for the upper bound; the nonnegative-payment reserve and its positive certificate establish attainment.

The ceiling relies on exact value-by-value incentives and a common affine belief path. Proposition~\ref{prop:transport} transfers incentive inequalities across a convex hull of admissible report laws. Boundary equalities extend to their affine span, allowing a law outside the tested interval to impose the ceiling. Section~\ref{sec:dataresolution} contrasts this exact requirement with deviation incentives under perturbed, fixed contracts.

\section{Supporting Reductions and Allocation Bounds}\label{sec:structure}
This section provides the payment computations and revenue bounds used to evaluate allocation choices. Alignment reduces the fixed-allocation rent problem under type-specific beliefs, and LP duality supplies bounds for arbitrary beliefs. These tools verify the allocation comparisons and identify when an inexpensive bound supports retaining a reserve rule. 

\subsection{Aligned Beliefs and Rent Levels}\label{sec:aligned}
Define leakage factors $\rho_{jh}=\min_k\varphi_j(k)/\varphi_h(k)$ for $j\ne h$. Rent assigned to report $T_h$ benefits its imitators. Alignment identifies a state minimizing both its objective cost and this leakage.

\begin{defi}[Aligned beliefs]\label{def:alignment}
Beliefs are \emph{aligned} if the laws $\ph_1,\ldots,\ph_n$ are pairwise distinct and, for each $h$, there is a state $\mathbf{d}_h$ satisfying
\[
 \mathbf{d}_h\in\argmax_{\mathbf{k}}\Lambda_h(\mathbf{k})\ \cap\ \bigcap_{j\ne h}\argmin_{\mathbf{k}}\frac{\ph_j(\mathbf{k})}{\ph_h(\mathbf{k})}.
\]
The alignment is \emph{strict in cost} if $\mathbf{d}_h$ is the unique maximizer of $\Lambda_h(\cdot)$ for every $h$.
\end{defi}

Distinct full-support laws imply $0<\rho_{jh}<1$ and $\bar\rho=\max_{j\ne h}\rho_{jh}<1$. Under subjective independence, ratio extrema occur at diagonal configurations; positive projection is aligned. The $\mathbf{d}_h$ must be distinct, excluding the common laws of Section~\ref{sec:common}. Concentrating perceived rent $\widehat U_h$ at $\mathbf{d}_h$ costs $\widehat U_h/\Lambda_h$ objectively and gives imitator $T_j$ only $\rho_{jh}\widehat U_h$, yielding:
\begin{equation}\label{eq:reduced}
 \min_{\widehat{\boldsymbol U}\ge0}\ m\sum_j\frac{q_j}{\Lambda_j}\Shat_j
 \quad\text{subject to}\quad
 \Shat_j\ge\rh_{jh}\Shat_h+(T_j-T_h)P_{jh}(\mathbf{p}),\quad j\ne h,
\end{equation}
and the monotone contraction 
\begin{equation}\label{eq:fixedpoint}
 [\boldsymbol\Gamma_{\mathbf{p}}(\mathbf{v})]_j=\max\{0,\max_{h\ne j}[\rh_{jh}v_h+(T_j-T_h)P_{jh}(\mathbf{p})]\},
\end{equation}
with supremum-norm modulus $\bar \rho$ and fixed point $\widehat{\boldsymbol U}^*(\mathbf p)$.

\begin{thm}[Aligned joint design]\label{thm:compression}
Assume aligned beliefs and let $\mathbf p\in\cP$.
(i) The minimum-rent problem has the value of \eqref{eq:reduced}, whose unique solution is $\widehat{\boldsymbol U}^*(\mathbf p)$, and $x_j(\mathbf{k})=T_jp_j(\mathbf{k})-\Shat^*_j(\mathbf p)\ind\{\mathbf{k}=\mathbf{d}_j\}/\ph_j(\mathbf{d}_j)$ are optimal payments, uniquely if the alignment is strict in cost.
(ii) $\Rstar=\max\{G(\mathbf{p})-m\sum_jq_j\Shat_j/\Lambda_j:\ \mathbf p\in\cP,\ \widehat{\boldsymbol U}\ge\boldsymbol\Gamma_{\mathbf p}(\widehat{\boldsymbol U})\}$, a linear program (LP) in $(\mathbf p,\widehat{\boldsymbol U})$.
(iii) Under \eqref{eq:cap}, the payments in (i) remain optimal if $\Shat^*_j(\mathbf p)\le(T_j+\varsigma)\ph_j(\mathbf d_j)p_j(\mathbf d_j)$ for every $j$; under strict-in-cost alignment, this condition is also necessary for $\Rp^\varsigma(\mathbf p)=\Rp(\mathbf p)$.
\end{thm}

Part (i) applies the fixed-allocation argument of \citet{LuZhang2026}; part (ii) retains allocation while reducing rents from $n|\mathcal K|$ variables to $n$. Part (iii) tests whether the resulting schedule fits the allocation's capacity. Losing at a cost peak precludes rent delivery there under the cap. Monotone allocations admit a downward recursion.

\begin{cor}[Monotone allocations]\label{cor:recursion}
Assume aligned beliefs. If $\mathbf p$ is monotone, then $\Shat^*_n(\mathbf p)=0$ and
\begin{equation}\label{eq:downward}
 \Shat^*_j(\mathbf p)=\max_{h>j}\{\rh_{jh}\Shat^*_h(\mathbf p)+(T_j-T_h)P_{jh}(\mathbf{p})\},\qquad j<n.
\end{equation}
At reserve $T_r$, $\Shat_j^r:=\Shat^*_j(\mathbf p^r)$ vanishes for $j\ge r$, and, with $U_j^r=\Shat_j^r/\Lambda_j$,
\begin{equation}\label{eq:truncated}
 \Shat_j^r=\max_{j<h\le r}\{\rh_{jh}\Shat_h^r+(T_j-T_h)P_{jh}(\mathbf{p}^n)\}\ \ (j<r),\qquad
 \Rp_r=\sum_{h\le r}w_hT_h-m\sum_{j<r}q_jU_j^r.
\end{equation}
\end{cor}

As $\rho_{jh}\rho_{ht}\le\rho_{jt}$, upward deviations require no additional rents. Given $P_{jh}(\mathbf p^n)$, all reserve rules take $O(n^3)$ operations to evaluate. For projection, put $\omega=1-\alpha$, $\rho_h=(\omega q_h/\pi_{hh})^{m-1}$, and $\bar\rho_P=\max_{j<n}\rho_j$. Correct-belief rents are $U_j^{\mathrm{rat},r}=\sum_{h=j+1}^r(T_{h-1}-T_h)\bar p_h$ for $j<r$ and zero otherwise.

\begin{cor}[Projection payments]\label{cor:projection}
Under \eqref{eq:projection} with $\alpha>0$, beliefs are aligned, strictly in cost, with $\mathbf d_j=\mathbf D_j$, $\Lambda_j=(\pi_{jj}/q_j)^{m-1}$, and $\rh_{jh}=\rh_h$. At $\mathbf p^r$, the optimal objective rents and the unique optimal payments are
\begin{equation}\label{eq:projrec}
 U_j^r=\rh_j\max_{j<h\le r}\{U_h^r+(T_j-T_h)\bar p_h\}\ (j<r),\ U_j^r=0\ (j\ge r),\ 
 x_j^r(\mathbf k)=T_jp_j^r(\mathbf k)-U_j^rq_j^{1-m}\ind\{\mathbf k=\mathbf D_j\}.
\end{equation}
For $j<r$, $U_j^r$ is continuous and strictly decreasing in $\alpha\in[0,1)$, equals $U_j^{\mathrm{rat},r}$ at $\alpha=0$, and satisfies $U_j^r\le\rh_jU_j^{\mathrm{rat},r}$. Hence $0\le\Rstar-\RR\le S-\RE\le\bar\rh_P(S-\RE^{\mathrm{rat}})$, and $\mathbf x^n\ge0$ for all $\alpha<1$ sufficiently close to one.
\end{cor}

Projection places rent on own-type matches while understatement wins against lower rivals. Stronger projection reduces rent cost and the remaining value of allocation redesign; at moderate intensities, the rare rent contingencies can require large subsidies.

\subsection{Finite Settlement at a Fixed Allocation}\label{sec:settlement}
For nonmonotone allocations, rent dependence can contain cycles. The sweep
\begin{equation}\label{eq:sweep}
 \widehat{\boldsymbol U}^{(t)}=\boldsymbol\Gamma_{\mathbf{p}}(\widehat{\boldsymbol U}^{(t-1)}),\quad t\ge1, \quad\mbox{where } \widehat{\boldsymbol U}^{(0)}=\mathbf{0},
\end{equation}
is nondecreasing and converges to $\widehat{\boldsymbol U}^*(\mathbf p)$. Equation~\eqref{eq:fixedpoint} is a deterministic discounted Markov decision problem: type $j$ stops for zero or moves to $h$ with reward $(T_j-T_h)P_{jh}(\mathbf p)$ and discount $\rh_{jh}$. Established results imply strongly polynomial solvability \citep{PostYe2015,Karczmarz2022}; finite termination additionally requires: 
$\widehat{\boldsymbol U}^{(n)}=\widehat{\boldsymbol U}^{(n-1)}$.
In a directed graph on $\mathcal N$, an arc $j\to h$ is \emph{tight} when $\Shat^*_j(\mathbf p)=\rh_{jh}\Shat^*_h(\mathbf p)+(T_j-T_h)P_{jh}(\mathbf p)$. Call an acyclic graph of tight arcs a \emph{supporting graph} if an arc leaves $j$ exactly when $\Shat^*_j(\mathbf p)>0$.  When $\widehat{\boldsymbol U}^{(n)}=\widehat{\boldsymbol U}^{(n-1)}$ holds,  $\widehat{\boldsymbol U}^{(n-1)}=\widehat{\boldsymbol U}^*(\mathbf p)$; let $t_j=\min\{t:\Shat_j^{(t)}=\Shat^*_j(\mathbf p)\}$ be the \emph{settling round} of $T_j$, and $\mathcal G=\{j\to h:\ t_h<t_j\}$. 

%\medskip
%%%%\noindent\begin{minipage}{\textwidth}
%%%\noindent\refstepcounter{algorithm}\label{alg:settling}\textbf{Algorithm~\thealgorithm\ (Settling sweep).} %%\emph{Input:} $\rh_{jh}$ and $(T_j-T_h)P_{jh}(\mathbf p)$ for $j\ne h$.
%\begin{enumerate}[label=\emph{Step \arabic*.},leftmargin=*,topsep=2pt,itemsep=0pt]
%\item Compute \eqref{eq:sweep} for $t=1,\ldots,n$.
%\item If \eqref{eq:settlement} fails, continue \eqref{eq:sweep} until $\norm{\boldsymbol\Gamma_{\mathbf p}(\widehat{\boldsymbol U}^{(t)})-\widehat{\boldsymbol U}^{(t)}}_\infty/(1-\bar\rh)$ falls below a tolerance. Otherwise set $\widehat{\boldsymbol U}^*(\mathbf p)=\widehat{\boldsymbol U}^{(n-1)}$.
%\item In the latter case, set $t_j=\min\{t:\Shat_j^{(t)}=\Shat^*_j(\mathbf p)\}$, the \emph{settling round} of %$T_j$, and return $\widehat{\boldsymbol U}^*(\mathbf p)$ and $\mathcal G=\{j\to h:\ t_h<t_j\}$.
%\end{enumerate}
%\end{minipage}

\begin{prop}[Finite settlement]\label{prop:settlement}
Assume aligned beliefs and let $\mathbf p\in\cP$. 
$\widehat{\boldsymbol U}^{(n)}=\widehat{\boldsymbol U}^{(n-1)}$ if and only if a supporting graph exists, which is guaranteed by monotone allocations. 
When $\widehat{\boldsymbol U}^{(n)}=\widehat{\boldsymbol U}^{(n-1)}$, graph $\mathcal G$ is acyclic, its sinks are exactly the zero-rent types, and $\widehat{\boldsymbol U}^*(\mathbf p)$ is the unique solution of
\begin{equation}\label{eq:settlingrecursion}
 \Shat_j=0\ \ \text{at sinks of }\mathcal G,\qquad
 \Shat_j=\max_{h:\,j\to h\in\mathcal G}\{\rh_{jh}\Shat_h+(T_j-T_h)P_{jh}(\mathbf p)\}\ \ \text{elsewhere}.
\end{equation}

%The following are equivalent: (i) \eqref{eq:settlement} holds; (ii) $\widehat{\boldsymbol U}^{(t)}=\widehat{\boldsymbol U}^{(t-1)}$ for some $t\ge1$; (iii) some acyclic graph of tight arcs, a \emph{supporting graph}, has an arc leaving $j$ if and only if $\Shat^*_j(\mathbf p)>0$. Monotone allocations satisfy (i)--(iii). If they hold, Algorithm~\ref{alg:settling} returns $\widehat{\boldsymbol U}^*(\mathbf p)$ and an acyclic graph $\mathcal G$ whose sinks are the zero-rent types, and $\widehat{\boldsymbol U}^*(\mathbf p)$ is the unique solution of
%\begin{equation}\label{eq:settlingrecursion}
% \Shat_j=0\ \ \text{at sinks of }\mathcal G,\qquad
% \Shat_j=\max_{h:\,j\to h\in\mathcal G}\{\rh_{jh}\Shat_h+(T_j-T_h)P_{jh}(\mathbf p)\}\ \ \text{elsewhere}.
%\end{equation}
\end{prop}

Finite settlement requires an acyclic subgraph supporting the rents; the full graph of tight arcs may contain cycles. Given the coefficients $\rh_{jh}$ and $P_{jh}(\mathbf{p})$, the criterion can be checked in $O(n^3)$ arithmetic operations. It can fail even with two types: for $n=2$, $p_1\equiv0$, and $p_2\equiv1/m$, the uncapped map $\boldsymbol\Gamma_{\mathbf p}(\mathbf v)=(\rh_{12}v_2+(T_1-T_2)/m,\rh_{21}v_1)$ is injective and does not fix zero, so its iterates from zero converge without reaching the fixed point in finitely many steps. The reduced LP remains exact.

\begin{exap}[Finite settlement without monotonicity]\label{ex:settlement}
Let $m=2$, $(T_1,T_2,T_3)=(4,3,1)$, $\mathbf{q}=(\tfrac13,\tfrac13,\tfrac13)$, and $\alpha=\tfrac14$ in \eqref{eq:projection}. Each type assigns probability $\tfrac12$ to a rival of her value and $\tfrac14$ to each other value, so beliefs are aligned, strictly in cost, with $\rh_{jh}=\tfrac12$. At the feasible, nonmonotone allocation $\mathbf{p}=(\tfrac38,\tfrac12,\tfrac18)$, deviation gains $(T_j-T_h)P_{jh}(\mathbf p)$ equal $\tfrac12$, $\tfrac38$, $\tfrac14$ for $(j,h)=(1,2)$, $(1,3)$, $(2,3)$, respectively, and are negative for $j>h$. The sweep gives $\widehat{\boldsymbol U}^{(1)}=(\tfrac12,\tfrac14,0)$ and $\widehat{\boldsymbol U}^{(2)}=\widehat{\boldsymbol U}^{(3)}=(\tfrac58,\tfrac14,0)$. Thus, $(t_1,t_2,t_3)=(2,1,0)$ and $\mathcal G=\{1\to2,1\to3,2\to3\}$. The unique supporting graph is $\{1\to2,2\to3\}$, with $1\to3$ slack as $\rh_{13}\Shat^*_3(\mathbf p)+\tfrac38=\tfrac38<\tfrac58=\Shat^*_1(\mathbf p)$. Optimal payments are nonnegative. Hence, a nonmonotone allocation can admit exact finite rent computation without subsidies.
\end{exap}

\subsection{Dual Scores and Allocation Certificates}\label{sec:certificates}
Without alignment, duality in \eqref{eq:primal} still bounds allocation value. Attach $m\mu_{jh}\ge0$ to \eqref{eq:ic}, put $M_h=\sum_{t\ne h}\mu_{ht}$, and define
\begin{align*}
 \ell_h(\mathbf{k};\boldsymbol\mu)&=-q_h+\Lambda_h(\mathbf{k})M_h-\sum_{j\ne h}\mu_{jh}\Lambda_j(\mathbf{k}),\qquad \psi_h(\mathbf{k};\boldsymbol\mu)=T_h-\frac{1}{q_h}\sum_{j\ne h}\mu_{jh}\Lambda_j(\mathbf{k})(T_j-T_h),\\
 \psi_h^\varsigma(\mathbf{k};\boldsymbol\mu)&=\psi_h(\mathbf{k};\boldsymbol\mu)+\frac{T_h+\varsigma}{q_h}\pos{\ell_h(\mathbf{k};\boldsymbol\mu)},\qquad \chi_j(\mathbf{k};\boldsymbol\mu)=\max\Bigl\{0,\max_{h:\,k_h+\ind\{h=j\}>0}\psi_h(\mathbf{k}+\mathbf{e}_j-\mathbf{e}_h;\boldsymbol\mu)\Bigr\}.
\end{align*}
The \emph{score} $\psi_h$ deducts incentive cost of serving $T_h$. A positive net rent return, $\Phi(\mathbf k)\ell_h(\mathbf k;\boldsymbol\mu)$, can be earned only within capacity $(T_h+\varsigma)p_h(\mathbf k)$ under the cap, so $\psi^\varsigma_h$ credits that capacity. Let $\cD=\{\boldsymbol\mu\ge0:\ell_h(\mathbf k;\boldsymbol\mu)\le0\text{ for all }h,\mathbf k\}$ and
$H(\boldsymbol\mu)=\sum_jq_j\sum_{\mathbf{k}}\Phi(\mathbf{k})\chi_j(\mathbf{k};\boldsymbol\mu)$,
the expected largest nonnegative score at a full profile. Define $H^\varsigma$ and $\chi^\varsigma_j$ using $\psi^\varsigma$. For implementable $\mathbf p$, let $\Delta^\varsigma(\mathbf p;\boldsymbol\mu)=H^\varsigma(\boldsymbol\mu)-\Rp^\varsigma(\mathbf p)$ for $\boldsymbol\mu\ge0$ and $\Delta(\mathbf p;\boldsymbol\mu)=H(\boldsymbol\mu)-\Rp(\mathbf p)$ for $\boldsymbol\mu\in\cD$.

\begin{thm}[Allocation certificates]\label{thm:dual}
Let beliefs be arbitrary and $\varsigma\ge0$.
(i) $\Rstar=\min_{\boldsymbol\mu\in\cD}H(\boldsymbol\mu)$ and $\mathcal R^{*,\varsigma}=\min_{\boldsymbol\mu\ge0}H^\varsigma(\boldsymbol\mu)$, with the minima attained;
(ii) For implementable $\mathbf p$,
\begin{equation}\label{eq:capdual}
 \Rp^\varsigma(\mathbf{p})=\min_{\boldsymbol\mu\ge0}\ m\sum_jq_j\sum_{\mathbf{k}}\Phi(\mathbf{k})p_j(\mathbf{k})\psi_j^\varsigma(\mathbf{k};\boldsymbol\mu),\ 
 \Rp(\mathbf{p})=\min_{\boldsymbol\mu\in\cD}\ m\sum_jq_j\sum_{\mathbf{k}}\Phi(\mathbf{k})p_j(\mathbf{k})\psi_j(\mathbf{k};\boldsymbol\mu),
\end{equation}
with the minima attained on the sets $\cD^\varsigma(\mathbf p)$ and $\cD(\mathbf p)$ of multipliers \emph{attached} to $\mathbf p$;
(iii) For $\boldsymbol\mu\ge0$, $0\le\mathcal R^{*,\varsigma}-\Rp^\varsigma(\mathbf p)\le\Delta^\varsigma(\mathbf p;\boldsymbol\mu)$, with equality if and only if $\boldsymbol\mu$ minimizes $H^\varsigma$; and $\mathbf p$ is globally optimal under \eqref{eq:cap} if and only if $\Delta^\varsigma(\mathbf p;\boldsymbol\mu)=0$ for some $\boldsymbol\mu\in\cD^\varsigma(\mathbf p)$. Likewise without the cap, with $\boldsymbol\mu\in\cD$.
\end{thm}

At attached multipliers, the certificate measures the expected shortfall from each profile's largest nonnegative score: selling at a negative score, withholding at a positive one, or serving a lower score. Under correct beliefs, $\mu_{h-1,h}=\sum_{j<h}q_j$ is attached to every reserve and yields $\psi_h = b^{\mathrm{rat}}_h$, with $b^{\mathrm{rat}}_1=T_1$ and $b_h^{\mathrm{rat}}=T_h-(T_{h-1}-T_h)\sum_{j<h}q_j/q_h$ being \emph{discrete virtual values} \citep{Vohra2012,Saban2021}. Two further constructions follow.

\begin{cor}[Multipliers for common laws]\label{cor:commondual}
Let $\ph_j=\ph$ for every $j$, and let $\mathbf p$ be implementable. For $j<n$, let $\vartheta_j$ attain the maximum in \eqref{eq:knapdual}, or let $\vartheta_j=1/\Lambda_j$ without \eqref{eq:cap}, and let $\vartheta_n=0$. Then $\mu_{j,j+1}=\sum_{t\le j}q_t\vartheta_t$ for $j<n$, with all other $\mu_{jh}=0$, belongs to $\cD^\varsigma(\mathbf p)$, or to $\cD(\mathbf p)$ without \eqref{eq:cap}, and
\[
 \psi_h^\varsigma(\mathbf k;\boldsymbol\mu)=T_h+(T_h+\varsigma)\pos{\vartheta_h\Lambda_h(\mathbf k)-1}-\ind\{h\ge2\}\frac{\mu_{h-1,h}}{q_h}(T_{h-1}-T_h)\Lambda_h(\mathbf k);
\]
without \eqref{eq:cap}, the middle term vanishes and $\psi_h^\varsigma=\psi_h$.
\end{cor}

Knapsack shadow costs determine adjacent IC multipliers. Capacity adds $(T_h+\varsigma)(\vartheta_h\Lambda_h(\mathbf k)-1)$ to service's score when positive. Correct beliefs give $\vartheta_j=1$ and the rational virtual values.

\begin{cor}[Multipliers from finite settlement]\label{cor:settlingdual}
Assume aligned beliefs. Let $\mathbf p\in\cP$ satisfy $\widehat{\boldsymbol U}^{(n)}=\widehat{\boldsymbol U}^{(n-1)}$. Let $\boldsymbol\mu\ge0$ satisfy $\mu_{jh}=0$ unless $j\to h$ is a tight arc of $\mathcal G$, and
\[
 M_j=\ind\{\Shat^*_j(\mathbf p)>0\}\Bigl(\frac{q_j}{\Lambda_j}+\sum_{h\ne j}\mu_{hj}\rh_{hj}\Bigr),\qquad j\in\mathcal N.
\]
Such $\boldsymbol\mu$ exist and can be computed in decreasing order of $t_j$, because $\mu_{hj}>0$ requires $t_h>t_j$. Every such $\boldsymbol\mu$ belongs to $\cD(\mathbf p)$.
\end{cor}

A positive-rent type passes $q_j/\Lambda_j$ plus leakage-discounted incoming multipliers along tight arcs; zero-rent types pass nothing. For Example~\ref{ex:settlement}'s beliefs at efficient allocation, $\mu_{12}=2/9$ and $\mu_{23}=1/3$ give the highest report the largest score, but the all-low score is $-1/2$. Reserve $T_2$ withholds there and earns $55/18=H(\boldsymbol\mu)$, certifying global optimality and gain $1/18$ over efficient allocation.

\subsection{When a Bound Supports Retaining a Reserve Rule}\label{sec:bound}
Alignment also gives a bound directly in primitives. For $h=2,\ldots,n$, define
\[
 \theta_h=\max_{j<h}\max_{\mathbf k:p_h^n(\mathbf k)>0}\frac{\Lambda_j(\mathbf k)}{\Lambda_j}, \
 K_h=\frac{q_h(T_h-T_{h+1})}{\sum_{j<h}q_j(T_j-T_h)},\
 B_{\rm loc}=\sum_{h=2}^n w_h\pos{\theta_h\sum_{j<h}\frac{q_j}{q_h}(T_j-T_h)-(T_h-T_{h+1})}.
\]
Here $\theta_h$ compares higher types' likelihood weights where $T_h$ wins with their cost peaks. The ratio $K_h$ compares the gap to the next lower value with the objectively weighted understatement gains; at $h=n$, the alternative is withholding.

\begin{thm}[Performance of reserve rules]\label{thm:primitive}
Under aligned beliefs and unrestricted transfers, (i)
$0\le\Rstar-\RR\le\Rstar-\RE\le B_{\rm loc}$; (ii) Efficient allocation is globally optimal if $\theta_h\le K_h$ for every $h=2,\ldots,n$; 
(iii) If $\theta_h< K_h$ for every $h=2,\ldots,n$, the efficient allocation is the unique globally optimal allocation up to tie-breaking among tied highest reports.
\end{thm}

Every $\boldsymbol\mu\in\cD(\mathbf p^n)$ with $\mu_{jh}=0$ for $j>h$ satisfies
\begin{equation}\label{eq:scorelower}
 \psi_h(\mathbf k;\boldsymbol\mu)\ge T_h-\theta_h\sum_{j<h}\frac{q_j}{q_h}(T_j-T_h),
 \qquad h\ge2,\quad p_h^n(\mathbf k)>0.
\end{equation}
Every lower report scores at most $T_{h+1}$, giving $B_{\rm loc}$ by comparison with the highest report's bound. Failure of this sufficient test is inconclusive. The capped conclusions also hold when Theorem~\ref{thm:compression}(iii)'s capacity test holds at $\mathbf p^n$. Under projection, $\theta_h=\max_{j<h}\rho_j\to0$: sufficiently strong projection makes efficient allocation uniquely optimal, with eventually nonnegative payments.

At any monotone allocation, \emph{pointwise rational rents} $u_j(\mathbf k)=\sum_{h>j}(T_{h-1}-T_h)p_h(\mathbf k)$ satisfy IC under every law. Applied to efficient allocation, they give a belief-independent fallback.

\begin{prop}[A belief-free bound]\label{prop:universal}
For arbitrary beliefs and $\varsigma\ge0$, $\mathcal R^{*,\varsigma}\le\Rstar$, with equality if some solution of \eqref{eq:primal} satisfies \eqref{eq:cap}, and
\begin{equation}\label{eq:universal}
 \mathcal R^{*,\varsigma}-\RR^\varsigma\le S-\RE^\varsigma\le S-\RE^{\mathrm{rat}}\le m(T_1-T_n)(1-q_1)^{m-1},
\end{equation}
which also holds without \eqref{eq:cap}.
\end{prop}

Even valid multipliers can give loose bounds. Define the smallest attached certificate by $\Delta^\varsigma_*(\mathbf p)=\min_{\boldsymbol\mu\in\cD^\varsigma(\mathbf p)}\Delta^\varsigma(\mathbf p;\boldsymbol\mu)$, or $\Delta_*(\mathbf p)$ without a cap. This LP value is exact only if an attached multiplier also minimizes the global bound.

\begin{cor}[Certificates at efficient allocation]\label{cor:stcert}
Impose \eqref{eq:cap}, assume \eqref{eq:ST}, and let $\beta>\beta_c$. Then $\cD^\varsigma(\mathbf p^n)=\{\boldsymbol\mu^\circ\}$, where $\boldsymbol\mu^\circ$ is given by Corollary~\ref{cor:commondual} with $(\vartheta_1,\vartheta_2,\vartheta_3)=(1/\Lambda_1,Z_1^{m-1},0)$, and
\[
 \Delta^\varsigma_*(\mathbf p^n)\ge mq_1^{m-1}q_2(T_2+\varsigma)(\beta^{m-1}-\beta_c^{m-1})=\left(\mathcal R^{*,\varsigma}-\RE^\varsigma\right)/\lambda^\varsigma.
\]
Moreover, $S-\RE^\varsigma\le T_1+\varsigma$, and, for fixed values, probabilities, $m$, and $\varsigma$, \eqref{eq:ST} and $\beta>\beta_c$ hold for all sufficiently large $\beta$.
\end{cor}

At efficiency, complementary slackness fixes $\vartheta_2=Z_1^{m-1}$, crediting capacity at a state where the middle report never wins. The bound extrapolates the lottery's marginal return to probability one. The joint optimum instead uses $\vartheta_2=Z_1^{m-1}(\beta_c/\beta)^{m-1}$. As $\beta\to\infty$, $\lambda^{\varsigma}\to0$ and the bound/gain ratio diverges. An operator should screen with $\min\{\Delta^\varsigma(\mathbf p;\boldsymbol\mu),S-\Rp^\varsigma(\mathbf p)\}$. A bound below redesign cost supports retaining $\mathbf p$; a large bound  is inconclusive. In the calibration, the attached bound is \$49.19, the full-surplus bound is 29.08 cents, and the actual gain is 24.72 cents. Evaluating given multipliers takes one profile pass; optimizing the global bound is as large as joint design.

\begin{exap}[Projection payments and a certificate]\label{ex:proj}
Revisiting Example~\ref{ex:premium} with $\pi_{jj}=\tfrac23$, $\Lambda_j=2$, and $\rh_j=\tfrac14$ for every $j$, and setting $\bar p_2=\tfrac12$ and $\bar p_3=\tfrac16$, \eqref{eq:projrec} at $r=3$ gives $\mathbf{U}^3=(\tfrac{65}{48},\tfrac5{12},0)$. Winners pay their values except at ties between two buyers with values $30$ or $20$, where payments are $21.88$ or $17.50$, respectively. Revenue is $23.26$, exceeding $18.89$ under correct beliefs for the same allocation. Full surplus is $S=24.44$. The nonzero multipliers are $\mu_{12}=\tfrac16$ and $\mu_{23}=\tfrac5{24}$. Their associated scores, $\psi_1=30$, $\psi_2(\mathbf k)=20-5\Lambda_1(\mathbf k)$, and $\psi_3(\mathbf k)=10-\tfrac{25}4\Lambda_2(\mathbf k)$ ensure the highest report's score is the largest nonnegative score at every profile. thus, $\Delta(\mathbf{p}^3;\mu)=0$, confirming the efficient allocation is globally optimal and $\mathcal{R}^*=23.26$. With correct beliefs, $\mathbf{b}^{\mathrm{rat}}=(30,10,-10)$, and the reserve price of $20$ attains the rational optimum revenue of $20$.
\end{exap}

\subsection{Two Further Distortions under Projection}\label{sec:otherdist}
Aligned beliefs do not make reserve rules optimal. For $m=n=3$, let $\mathbf p(\lambda)$ be efficient except at $(T_2,T_3,T_3)$, where the middle buyer wins with probability $1-\lambda$ and each lowest buyer with $\lambda/2$. It is monotone for $0\le\lambda\le2/3$, with downward-report winning probabilities
\[
 P_{23}=\frac{\omega^2q_3^2}{3}+\pi_{22}\omega q_3\lambda,\qquad
 P_{12}=\omega^2\Bigl(\frac{q_2^2}{3}+q_2q_3+q_3^2(1-\lambda)\Bigr),\qquad
 P_{13}=\omega^2\Bigl(\frac{q_3^2}{3}+q_2q_3\lambda\Bigr).
\]
Write $P_{jh}=P_{jh}(\mathbf p(\lambda))$ and $\Shat_j^{(h)}(\lambda)=\rh_h\Shat_h(\lambda)+(T_j-T_h)P_{jh}$ for $j<h$. Corollary~\ref{cor:recursion} gives optimal perceived rents $\Shat_3(\lambda)=0$, $\Shat_2(\lambda)=\Shat_2^{(3)}(\lambda)$, and $\Shat_1(\lambda)=\max\{\Shat_1^{(2)}(\lambda),\Shat_1^{(3)}(\lambda)\}$. Both branches of $\Shat_1$ are affine in $\lambda$, and $\Shat_1^{(3)}$ has the larger slope. Let $\lambda^*$ solve $\Shat_1^{(2)}(\lambda)=\Shat_1^{(3)}(\lambda)$, and let
\[
 \bar\mu_{12}=\frac{(T_2-T_3)+q_1\omega^2(T_1-T_3)/(\Lambda_1q_3)+\pi_{22}\omega(T_2-T_3)/(\Lambda_2q_3)}
 {\omega^2(T_1-T_2)/q_2+\omega^2(T_1-T_3)/q_3-\rh_2\pi_{22}\omega(T_2-T_3)/(q_2q_3)},
\]
the multiplier $\mu_{12}$ that equalizes the middle and lowest scores at $(T_2,T_3,T_3)$. Assume
\begin{equation}\tag{KR}\label{eq:KR}
 \Shat_1^{(2)}(0)>\Shat_1^{(3)}(0),\quad \lambda^*<2/3,\quad \bar\mu_{12}<q_1/\Lambda_1,\quad T_2-\bar\mu_{12}\omega^2(T_1-T_2)/q_2>0.
\end{equation}

\begin{prop}[A lower-value lottery under projection]\label{thm:kink}
Let $m=n=3$, $0<\alpha<1$ in \eqref{eq:projection}, and assume \eqref{eq:KR}. The allocation $\mathbf{p}(\lambda^*)$, with the rents above concentrated at $\mathbf{D}_1$ and $\mathbf{D}_2$, is globally optimal, strictly outperforms every reserve rule, and has
\[
 \Rstar-\RE=3\Bigl(\frac{q_1}{\Lambda_1}-\bar\mu_{12}\Bigr)\bigl(\Shat_1^{(2)}(0)-\Shat_1^{(3)}(0)\bigr)>0,\qquad
 S-G(\mathbf{p}(\lambda^*))=3q_2q_3^2(T_2-T_3)\lambda^*.
\]
It is also globally optimal with nonnegative payments if $3\Shat_1(\lambda^*)/\pi_{11}^2\le T_1$ and $3\Shat_2(\lambda^*)/\pi_{22}^2\le T_2$.
\end{prop}

This lottery makes reporting the middle value less attractive to the top type while making the low report more attractive. Both constraints bind at the optimum. For $(T_1,T_2,T_3)=(10,9,7)$, $q=(0.60,0.10,0.30)$, and $\alpha=0.10$, $\lambda^*=0.31$ raises revenue from $9.70$ to $9.71$ (exactly $151098763/15564800$), reduces surplus from $9.88$ to $9.87$, and keeps payments nonnegative. The active rent constraints determine the allocation scores.

A cap can instead make partial sale optimal. With two types, $\mathbf p^1$ and $\mathbf p^2$ differ only at the all-low profile $\mathbf D_2+\mathbf e_2$; $K_2$ compares its surplus with the high type's understatement gain.

\begin{prop}[Partial sale under a subsidy cap]\label{prop:twocap}
Let $n=2$ and $0<\alpha<1$ in \eqref{eq:projection}, and impose \eqref{eq:cap}. Every optimal mechanism has $p_1(\mathbf k)=1/(k_1+1)$ for all $\mathbf k$, $p_2(\mathbf k)=0$ for $\mathbf k\ne\mathbf D_2$, and $u_2\equiv0$. If $K_2\Lambda_1(\mathbf{k})\ne\Lambda_1(\mathbf{D}_2)$ for every $\mathbf{k}$, all optimal mechanisms sell the all-low profile with the same probability
\begin{equation}\label{eq:capsale}
 mp_2(\mathbf{D}_2)=\min\biggl\{1,\ \frac{m(T_1+\varsigma)}{(T_1-T_2)\ph_1(\mathbf{D}_2)}\sum_{\mathbf{k}:\,K_2\Lambda_1(\mathbf{k})>\Lambda_1(\mathbf{D}_2)}\frac{\ph_1(\mathbf{k})}{k_1+1}\biggr\},
\end{equation}
which is nondecreasing in $\varsigma$. If $0<mp_2(\mathbf D_2)<1$, every optimal allocation is a \emph{partial sale}, a strict convex combination of $\mathbf p^1$ and $\mathbf p^2$, and $\Rp^\varsigma(\mathbf p^r)<\mathcal R^{*,\varsigma}$ for $r=1,2$.
\end{prop}

The seller fills high-type capacity in decreasing likelihood ratio, selling at the all-low profile while surplus exceeds marginal rent cost. In the second opening example, $m=2$, $T=(10,1)$, $q=(0.20,0.80)$, and $\alpha=0.30$. Unrestricted transfers support efficiency and revenue $208/55\approx3.78$, with a winner subsidy $16/11$ at all-high reports. Nonnegative payments give sale probability $55/63$ and revenue $1184/315\approx3.76$, versus $128/35\approx3.66$ under efficiency and $3.60$ under the high reserve. Full sale returns at $\varsigma\ge16/11$. These comparisons are nominal; Section~\ref{sec:implementation} adds reporting guarantees.

\section{A Matched Numerical Case}\label{sec:data}
The calibrated case measures the value of allocation flexibility after an operator specifies reporting resolution and the required incentive protection. Every reserve receives optimized payments under the same beliefs, payment restrictions, and incentives as joint design. Table~\ref{tab:main} separates nominal allocation value, certificate-class cost, and positive-margin cost. Revenues are source-period dollars per auction; gains and welfare losses are cents, computed before rounding.

\subsection{Data, Monetary Scale, and Scenario Inputs}\label{sec:datacal}
We use 749 screened listings from the 1,000 public eBay U.S.\ Mint proof-set auctions of \citet{Wegmann2011}, available from \citet{SUdata,VillaniData}. Unlike that common-value study, our counterfactual maintains a private-value observation process: conditional on the reserve and released bidder count, price including shipping is the second-highest entrant value above the reserve. The value-to-book ratio is lognormal, with log mean depending on six item and seller indicators and log book value. Appendix~\ref{app:caldetail} gives the screening-window-normalized likelihood, estimates, and specification checks. These checks assess the maintained price model, not identification of beliefs or counterfactual entry.

At the modal profile and book value \$22.50, conditional means of the bottom 50\%, next 30\%, and top 20\% of the retained central 99\% give computational values $\mathcal{T}=(27.45,18.16,11.37)$ and $\mathbf q=(0.20,0.30,0.50)$. Four buyers give surplus \$23.22. Shared-top strength four implies perceived probabilities $(0.50,0.1875,0.3125)$: a top rival is perceived as 50\% likely rather than 20\%. Projection at $\alpha=0.15$ instead gives own-type probabilities 32\%, 40.5\%, and 57.5\%. Refitting the 493 listings unaffected by the source's bid-removal screen gives matched gains of 1.49 and 3.34 cents.

\subsection{Nominal Allocation and Rent Capacity}\label{sec:datapay}
Correct-belief optimal payments earn \$20.64, against \$18.70 for the best second-price auction with a reserve in $\mathcal T$. Under projection and nonnegative payments, efficient allocation earns \$22.57: the \$3.87 advantage comprises \$1.94 from payment optimization and \$1.93 from belief disagreement, with zero allocation gain. Unrestricted transfers earn \$22.83 but require a \$177.87 winner subsidy at the all-top profile, whose probability is 0.16\%. Banning subsidies costs 25.11 cents.

Shared-top beliefs activate rent capacity. At $\beta=4$, the scores are ordered and positive and top rent \$0.73 fits peak capacity \$0.86, satisfying Theorem~\ref{thm:sharedtop}. Without a cap, efficient allocation delivers \$0.41 to a losing middle type against three top rivals. Nonnegative payments instead require a 2.28\% chance of a free object at that profile, which occurs with probability 0.96\%. Revenue rises by 24.72 cents over the best reserve and allocated value falls by 0.20 cents. A \$50 cap lowers the lottery to 0.61\% and raises the gain to 24.97 cents, below its 25.06-cent limit; every finite cap still excludes transfers to sure losers.

\begin{figure}[!t]
\vspace{-0.1in}
\centering
\includegraphics[width=0.96\textwidth]{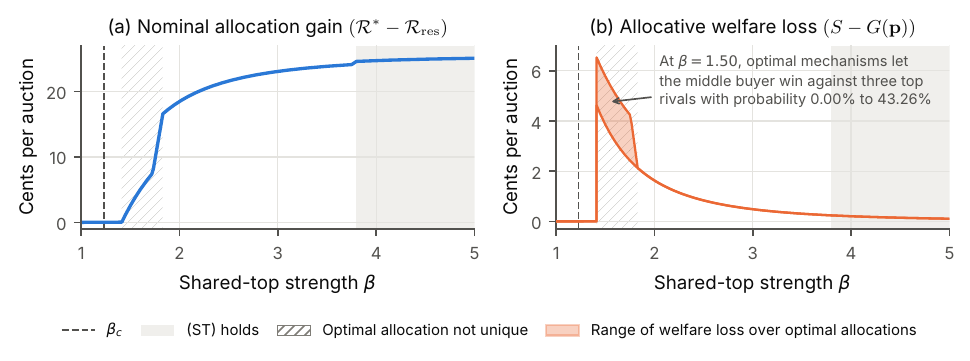}
\caption{Nominal allocation across shared-top belief strengths with four buyers and nonnegative payments. Hatching marks nonuniqueness. Panel (b) reports welfare-loss ranges over optimal allocations. Theorems~\ref{thm:sharedtop} and~\ref{thm:overflow} classify this path; $\beta_c$ applies only within the shaded \eqref{eq:ST} region.
}\label{fig:regimes}
\vspace{-0.2in}
\end{figure}

Theorems~\ref{thm:sharedtop} and~\ref{thm:overflow} classify this calibrated path (Figure~\ref{fig:regimes}). Efficiency is uniquely optimal below the cutoff near 1.41. At the cutoff, the lottery segment is optimal. The stated lottery then uniquely maximizes welfare among revenue optima and is uniquely revenue optimal from the cutoff near 1.83. Top rents overflow until \eqref{eq:ST} first holds near 3.80, where $\beta_c\simeq1.23$ applies. At each $\beta$, the orange band in Figure~\ref{fig:regimes}(b) spans the smallest and largest welfare losses $S-G(\mathbf p)$ among allocations that attain the same maximum revenue. Its maximum width is 1.93 cents per auction.
The gain is 54.67, 24.72, and 1.59 cents with two, four, and eight buyers. The optimal dual multipliers certify the measured gain. The certificate attached to efficiency is instead \$49.19, exceeding the 29.08-cent full-surplus fallback; a large attached bound is inconclusive (Corollary~\ref{cor:stcert}). This distinguishes a bound that screens out costly redesign from a verified gain that can support it.

\subsection{Certificate-Class Cost and Positive-Margin Cost}\label{sec:dataunique}
Nominal revenue does not include a guarantee of truthful reporting: the projection and shared-top optima admit strict all-top pooling equilibria earning \$0.00 and \$3.96, preferred by every type to truth. Restriction to zero-margin report-by-report certificates makes joint and best-reserve revenues coincide at \$21.56 and \$22.13. Joint certificate-class costs are therefore \$1.02 and \$1.04; zero-margin certificates can still permit multiplicity. Adding a one-cent margin costs joint design 3.73 and 2.37 cents, versus 5.21 and 5.69 cents for reserves. Allocation saves the differences, 1.48 and 3.32 cents. Under correct beliefs both classes earn \$20.62, with no allocation gain.

The service changes implement different incentives (Table~\ref{tab:policy}). With shared-top beliefs, a free middle-report lottery supplies truthful rent $\delta$ against top rivals and removes the middle type's low report. With projection, charging a middle-report winner $T_2$ makes that deviation lose $(T_2-T_3)p_2=\delta$ for a low type. Remaining payments and deletions complete the certificates. The calibrated projection schedule violates the transport condition at two steps and retains a \$0.87 premium over correct-belief optimum. Failure of restoration alone would not establish that premium; the verified contract does.

\begin{figure}[!t]
\vspace{-0.1in}
\centering
\includegraphics[width=0.96\textwidth]{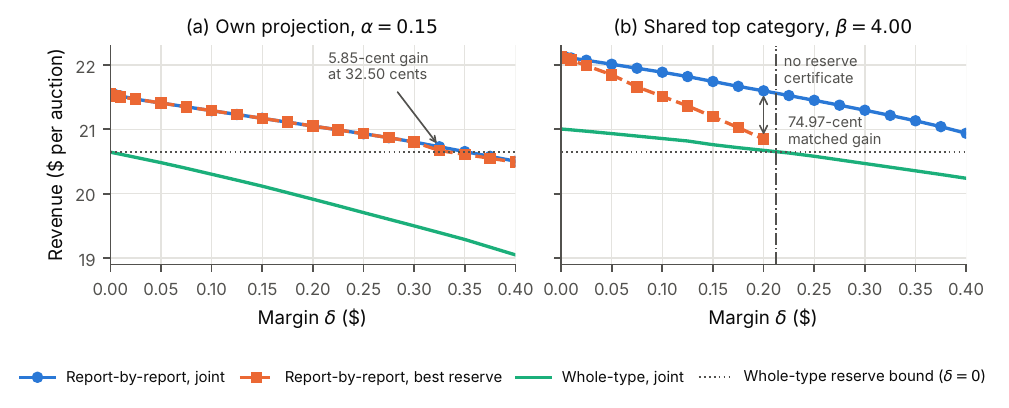}
\caption{Certified revenue across utility margins. Report-by-report joint and best-reserve designs face identical margins. The whole-type reserve line is a zero-margin bound. A wider gap between declining revenue frontiers measures cost savings at stronger reporting protection.}\label{fig:frontier}
\vspace{-0.2in}
\end{figure}

Whole-type certification is a separate comparison: no reserve has a positive dollar-margin certificate in that class (Lemma~\ref{lem:reservecert}). Its \$20.64 zero-margin bound gives conservative differences. Figure~\ref{fig:frontier} separates them from matched gains. The shared-top matched gap rises to 37.53 and 74.97 cents at margins \$0.10 and \$0.20 as reserve revenue falls faster; projection gains are nonmonotone. A larger gap measures savings at a stronger requirement, not an increase in revenue.

\subsection{Reporting Bands and Belief Protection}\label{sec:dataresolution}
With the retained population and salient top quintile fixed, 3, 5, and 10 reports yield nominal gains of 24.72, 52.69, and 81.36 cents for conditional-mean exact types, and 31.18, 47.35, and 75.84 cents for every-value bands (Table~\ref{tab:ECresolution}). Coarsening itself loses 92.99, 67.20, and 19.83 cents of continuum surplus \$24.15; each nominal joint design adds less than one cent. Reporting resolution can therefore matter more for welfare than the lottery.

Under three-band report-by-report certification, joint design and the best reserve both earn \$17.83 at $\eta=0$ (Table~\ref{tab:main}). Requiring $\eta=0.10\%$ lowers revenues to \$17.82 and \$17.80, costing 0.48 and 2.99 cents and leaving a 2.50-cent matched gain. This dimensionless requirement guarantees an advantage of at least $\eta\min\{v-T_j,T_{j-1}-v\}$: the minimum guarantee is \$0.001 one dollar from the nearest boundary and vanishes at the boundary. It therefore differs from the exact-type dollar margin. The joint contract gives a middle-band report a 0.10\% chance of winning against three top reports, charging its \$15.33 floor on winning. The resulting value-dependent rent supports its six-step certificate. Joint design adds 0.51 cents of allocation loss relative to the reserve, beyond their common 92.99-cent coarsening loss and the reserve's 71.07-cent allocation loss.

Now retain the bands and $\eta=0.10\%$, but require one contract and schedule to remain valid for every common belief law
$\boldsymbol\pi(\kappa)=(1-\kappa)(1/2,3/16,5/16)+\kappa\mathbf{q}$
over $[0,\bar\kappa]$. Proposition~\ref{prop:bandinterval} gives both designs revenue \$16.67 at $\bar\kappa=0.10$ and $0.20$. Protection costs approximately \$1.15 for joint design and \$1.13 for the reserve, eliminating the allocation gain (Table~\ref{tab:main}). Under the proposition's conditions, this ceiling holds for every positive interval width. At $\kappa=0.001$, the fixed point-belief joint and reserve contracts violate endpoint IC by only 0.088 and 0.086 cents, respectively (Appendix~\ref{app:calresolution}).

\subsection{Reoptimization, Fixed Contracts, and Belief Uncertainty}\label{sec:databelief}
The 300 paired refits use the documented common-seed protocol, reconstruct values, and reoptimize both allocation classes over all 720 schedules. The one-cent matched gains are all positive, with central 95\% refit ranges of $[1.31,1.66]$ and $[3.28,3.37]$ cents under projection and shared-top beliefs, respectively. These ranges measure sampling sensitivity under the maintained calibration. In six shared-top refits, the joint design nearly withholds at the all-low profile, with welfare losses of 62.57--64.89 cents. In five, the matched reserve excludes the low type and loses more; in the sixth, the reserve is efficient, so the joint design loses an additional 63.76 cents. Welfare-loss variation across revenue-optimal joint schedule faces is below 0.01 cent in each of the six refits (Appendix~\ref{app:calrefit}).

Belief protection addresses another source of uncertainty while holding exact values fixed. Contracts retaining the one-cent margin throughout $[0,0.10]$ and $[0,0.20]$ have matched gains of 3.38 and 3.45 cents, with joint revenue costs of 6.49 and 14.04 cents. Every optimal point-certified joint contract fails endpoint truthful IC by at least 10.82 and 23.97 cents across all optimal schedule faces. The operating decision begins with the reporting interface and guarantee, followed by comparison with the matched reserve. At point beliefs, allocation flexibility covers at most 1.48, 3.32, or 2.50 cents of additional cost per auction---approximately 0.07\%, 0.15\%, or 0.14\% of the corresponding reserve revenue. These gains must cover incremental service, communication, and verification costs. Joint band and belief protection leaves no incremental revenue to fund allocation redesign. The reporting requirement needs its own justification, and adoption must also account for welfare and fixed-contract robustness.

\section{Conclusion}\label{sec:conclusion}
When payments cannot deliver rents freely across states, allocation determines both the incentive rents a seller must provide and the capacity available to deliver them. Our common-law formulation separates these two roles. Conditional three-type characterizations identify optimal lower-value lotteries and price the high-type rent capacity they displace. Under a fixed strict reporting certificate, a constructive two-buyer result also establishes a positive allocation gain. Transport and boundary arguments then identify reporting and belief restrictions that restore the objective-law revenue ceiling and, under the stated conditions, allow a reserve to attain it.

The collectible-auction case measures the recurring costs these choices can cover. Nominal shared-top allocation flexibility earns 24.72 cents per auction; matched one-cent reporting requirements yield gains of 1.48 and 3.32 cents. The point-belief band gain is 2.50 cents and becomes zero under exact common-interval protection. Reoptimized gains coexist with service changes and frozen-contract failures. The evidence therefore supports modest savings at chosen reporting requirements; measuring beliefs, entry responses, and deployment costs requires separate evidence.

\noindent\textbf{Code and data availability.} The accompanying reproduction package contains the attributed public data and license, calibration and optimization code, exact inputs, full paired-refit records, seeds, verification, and execution instructions. All sampling summaries use the same paired records. Appendix~\ref{app:caldetail}  distinguishes reoptimized comparisons from fixed-contract diagnostics.

%============================ BACK MATTER ============================

\bibliographystyle{informs2014}
\bibliography{reference_bd}

\newpage

\begin{table}[t]
\caption{Distinctive contributions relative to the closest literature}
\label{tab:contribution_comparison}
\centering
\begingroup
\fontsize{10}{12}\selectfont
\setlength{\tabcolsep}{5pt}
\renewcommand{\arraystretch}{1.05}
\begin{tabularx}{\linewidth}{@{}
  >{\raggedright\arraybackslash}p{0.17\linewidth}
  >{\raggedright\arraybackslash}p{0.34\linewidth}
  >{\raggedright\arraybackslash}X@{}}
\toprule
Design question & Closest antecedents & Distinct contribution and scope\\
\midrule
Allocation and rent capacity & Payment design at fixed allocation \citep{LuZhang2026}; virtual values and mechanism-design LPs \citep{Myerson1981,Vohra2012}. & Allocation changes required rents and delivery capacity. Under stated three-value shared-top conditions, global optima account for overflow and welfare differences among revenue optima (Theorems~\ref{thm:common}--~\ref{thm:overflow}).\\[4pt]
Cost of a reporting guarantee & Implementation under belief restrictions \citep{OllarPenta2017}; gains from approximate IC \citep{Balseiro2024}. & Separate certificate-class and positive-margin costs. A lottery strictly beats every reserve under the same fixed strict schedule (Propositions~\ref{prop:margincost}--~\ref{prop:fixedgain}).\\[4pt]
Reporting certificates and revenue ceilings & Iterated-deletion \citep{BergemannMorris2009}; belief-dependent implementation \citep{OllarPenta2017}. & Under subjective independence, a transport test restores objective-law incentive inequalities. Whole-type certificates eliminate the projection premium (Proposition~\ref{prop:transport} and Corollary~\ref{cor:restore}).\\[4pt]
Band reporting across common beliefs & Finite messages \citep{Kos2012}; robust IC and revenue restrictions \citep{lopomo2021uncertaintymechanismdesign,OllarPenta2025}. & Exact band incentives on a common belief interval imply a revenue ceiling that a positively certified reserve attains (Proposition~\ref{prop:bandinterval}).\\
\bottomrule
\end{tabularx}
\endgroup
\vspace{-0.2in}
\end{table}

\begin{table}[!t]
\vspace{-0.1in}
\centering
\caption{Allocation value under matched information and reporting requirements}\label{tab:main}
\begin{small}\setlength{\tabcolsep}{4pt}
\begin{tabular*}{\textwidth}{@{\extracolsep{\fill}}lrrrrr}\toprule
 & \multicolumn{2}{c}{Revenue (\$)} & \multicolumn{2}{c}{Allocation gain (cents)} & Welfare loss (cents)\\\cmidrule(lr){2-3}\cmidrule(lr){4-5}\cmidrule(lr){6-6}
Beliefs or requirement & Joint & Best reserve & Estimate & Refit range & Joint\\\midrule
\multicolumn{6}{@{}l}{\textit{A. Exact types: nominal truthfulness at point beliefs}}\\
Correct & 20.64 & 20.64 & 0.00 & --- & 0.00 \\
Projection & 22.57 & 22.57 & 0.00 & $[0.00,\ 0.00]$ & 0.00 \\
Shared-top & 23.17 & 22.93 & 24.72 & $[23.49,\ 25.80]$ & 0.20 \\[2pt]
\multicolumn{6}{@{}l}{\textit{B.  Exact types: report-by-report certificate, zero margin}}\\
Correct    &20.64   &20.64    &0.00&---&0.00\\
Projection & 21.56 & 21.56 & 0.00 & --- & 0.00 \\
Shared-top & 22.13 & 22.13 & 0.00 & --- & 0.00 \\[2pt]
\multicolumn{6}{@{}l}{\textit{C. Exact types: report-by-report certificate, $\delta=\$0.01$}}\\
Correct    &20.62   &20.62 &0.00    &---&0.00\\
Projection & 21.52 & 21.50 & 1.48 & $[1.31,\ 1.66]$ & 0.80 \\
Shared-top & 22.11 & 22.08 & 3.32 & $[3.28,\ 3.37]$ & 0.30 \\[2pt]
\multicolumn{6}{@{}l}{\textit{D. Exact types: one-cent certificate over a shared-top belief interval}}\\
$\kappa\in[0.00,0.10]$ & 22.04 & 22.01 & 3.38 & --- & 0.30 \\
$\kappa\in[0.00,0.20]$ & 21.97 & 21.93 & 3.45 & --- & 0.30 \\
\multicolumn{6}{@{}l}{\textit{E. Three every-value bands: shared-top point beliefs}}\\
Nominal truthfulness         &18.43 &18.12 &31.18&---&0.36\\
Report certificate, $\eta=0$ &17.83&17.83&0.00&---&71.07\\
Report certificate, $\eta=0.10\%$ &17.82 &17.80 &2.50&---&71.58\\
\multicolumn{6}{@{}l}{\textit{F. Same three bands and $\eta=0.001$: one contract over common beliefs}}\\
$\bar\kappa=0.10$ or $0.20$ & 16.67 & 16.67 & 0.00 & --- & 71.07\\
\bottomrule\end{tabular*}\end{small}
\begin{minipage}{\textwidth}\footnotesize\medskip
\textit{Notes.} Four buyers; nonnegative payments; $\alpha=0.15$, $\beta=4.00$. Payments optimized under matched requirements. Zero-margin certificates permit multiplicity. Panel~D: one contract and schedule throughout each interval. Panel~E: band floors for participation and payments; incentives at every value; losses exclude common coarsening loss. Ranges: central 95\% of 300 paired refits. Exact types: joint loss $S-G$; best baseline reserves are efficient. Differences use unrounded values.
\end{minipage}
\vspace{-0.2in}
\end{table}

\clearpage

\begin{APPENDICES}
\newpage
\setcounter{page}{1}
\setcounter{section}{0}

\begin{center}
{\large Online Appendix \\
\vspace{0.1in}
\PaperTitle
}
\end{center}

\renewcommand{\thesection}{O\Alph{section}}
\renewcommand{\theequation}{OA-\arabic{equation}}
\renewcommand{\thefigure}{OA-\arabic{figure}}
\renewcommand{\thetable}{OA-\arabic{table}}
\renewcommand{\thethm}{OA-\arabic{thm}}
\renewcommand{\thelem}{OA-\arabic{lem}}
\renewcommand{\theprop}{OA-\arabic{prop}}
\renewcommand{\theexap}{OA-\arabic{exap}}
\renewcommand{\thecor}{OA-\arabic{cor}}
\renewcommand{\therem}{OA-\arabic{rem}}
\renewcommand{\thedefi}{OA-\arabic{defi}}
\renewcommand{\theas}{OA-\arabic{as}}
\setcounter{equation}{0}
\setcounter{thm}{0}
\setcounter{lem}{0}
\setcounter{prop}{0}
\setcounter{cor}{0}
\setcounter{figure}{0}
\setcounter{table}{0}
\setcounter{exap}{0}
\setcounter{rem}{0}
\setcounter{defi}{0}
\setcounter{as}{0}
Appendix~\ref{app:model} specifies the mechanism class and equilibrium convention, Appendix~\ref{app:proofs} contains the proofs, Appendix~\ref{app:caldetail} documents calibration and numerical checks, and Appendix~\ref{app:examples} verifies the motivating examples.

\section{Mechanism Class and Equilibrium Ranking}\label{app:model}
Rivals' values $\mathbf s_{-i}\in\mathcal T^{m-1}$ have configuration $\mathbf k(\mathbf s_{-i})$, with $k_h(\mathbf s_{-i})=\sum_{i'\ne i}\ind\{s_{i'}=T_h\}$; an ordered profile has objective probability $\Phi(\mathbf s_{-i}):=\Phi(\mathbf k(\mathbf s_{-i}))\prod_hk_h(\mathbf s_{-i})!/(m-1)!$ and subjective probability $\ph_j(\mathbf s_{-i})$, defined likewise. A mechanism $(A_i,\tilde p_i,\tilde x_i)_{i=1}^m$ gives each buyer $i$ a measurable action set $A_i$ and specifies, at each action profile $\mathbf a\in A:=\prod_iA_i$, her allocation probability $\tilde p_i(\mathbf a)\ge0$, with $\sum_i\tilde p_i(\mathbf a)\le1$, and expected payment $\tilde x_i(\mathbf a)\in\mathbb R$, with measurable $\tilde p_i$ and $\tilde x_i$, the latter integrable under the laws below. A strategy $\xi_i$ assigns to each value $s\in\mathcal T$ a distribution $\xi_i(\cdot\mid s)$ on $A_i$; it is \emph{pure} if every $\xi_i(\cdot\mid s)$ is a point mass. A profile $\boldsymbol\xi=(\xi_1,\ldots,\xi_m)$ and values $\mathbf s\in\mathcal T^m$ induce the law $\boldsymbol\xi(\cdot\mid\mathbf s)=\bigotimes_i\xi_i(\cdot\mid s_i)$ on $A$, its marginal $\boldsymbol\xi_{-i}(\cdot\mid\mathbf s_{-i})$ on $A_{-i}$, and expected outcomes $\tilde p_i^{\boldsymbol\xi}(\mathbf s)=\int_A\tilde p_i\,d\boldsymbol\xi(\cdot\mid\mathbf s)$ and $\tilde x_i^{\boldsymbol\xi}(\mathbf s)=\int_A\tilde x_i\,d\boldsymbol\xi(\cdot\mid\mathbf s)$. Type $T_j$ of buyer $i$ perceives the utility of action $a_i\in A_i$ against $\boldsymbol\xi_{-i}$ as
\[
 \Shat_{ij}(a_i;\boldsymbol\xi_{-i})=\sum_{\mathbf s_{-i}\in\mathcal T^{m-1}}\ph_j(\mathbf s_{-i})\int_{A_{-i}}\bigl\{T_j\tilde p_i(a_i,\mathbf a_{-i})-\tilde x_i(a_i,\mathbf a_{-i})\bigr\}\,\boldsymbol\xi_{-i}(d\mathbf a_{-i}\mid\mathbf s_{-i}).
\]
\begin{defi}[Subjective equilibrium]\label{def:SE}
A strategy profile $\boldsymbol\xi$ is a \emph{subjective equilibrium (SE) with correct strategy conjectures} if, for all $i\in\mathcal M$ and $j\in\mathcal N$,
\begin{equation}\label{eq:SE}
 \int_{A_i}\Shat_{ij}(a_i;\boldsymbol\xi_{-i})\,\xi_i(da_i\mid T_j)\ge\Shat_{ij}(a';\boldsymbol\xi_{-i}),\qquad\forall a'\in A_i. \tag{SE}
\end{equation}
It is \emph{strict} if each $\xi_i(\cdot\mid T_j)$ is a point mass at some $a_i$ with $\Shat_{ij}(a_i;\boldsymbol\xi_{-i})>\Shat_{ij}(a';\boldsymbol\xi_{-i})$ for all $a'\ne a_i$, and \emph{statewise participating} if $T_j\tilde p_i^{\boldsymbol\xi}(T_j,\mathbf s_{-i})-\tilde x_i^{\boldsymbol\xi}(T_j,\mathbf s_{-i})\ge0$ for all $i\in\mathcal M$, $j\in\mathcal N$, and $\mathbf s_{-i}\in\mathcal T^{m-1}$.
\end{defi}
Buyers conjecture rivals' strategies correctly but weight rivals' values by their own laws; by linearity in $\xi_i(\cdot\mid T_j)$, \eqref{eq:SE} also excludes profitable mixed deviations.

An \emph{anonymous direct mechanism} has $A_i=\mathcal T$ and, for $s_i=T_j$, $\tilde p_i(\mathbf s)=p_j(\mathbf k(\mathbf s_{-i}))$ and $\tilde x_i(\mathbf s)=x_j(\mathbf k(\mathbf s_{-i}))$, where $p_j(\mathbf k)\ge0$ and $x_j(\mathbf k)\in\mathbb R$ are the allocation and expected payment for report $T_j$ against rival reports with configuration $\mathbf k$. Feasibility is \eqref{eq:FEA}. Rivals' strategies $\boldsymbol\xi_{-i}$ induce type $T_j$'s law of rival-report configurations
\[
 \ph_j^{\boldsymbol\xi_{-i}}(\mathbf k)=\sum_{\mathbf s_{-i}\in\mathcal T^{m-1}}\ph_j(\mathbf s_{-i})\,\boldsymbol\xi_{-i}\bigl(\{\mathbf s_{-i}':\mathbf k(\mathbf s_{-i}')=\mathbf k\}\bigm|\mathbf s_{-i}\bigr),
\]
so that $\Shat_{ij}(T_h;\boldsymbol\xi_{-i})=\sum_{\mathbf k}\ph_j^{\boldsymbol\xi_{-i}}(\mathbf k)\{T_jp_h(\mathbf k)-x_h(\mathbf k)\}$. Truthful reporting is $\xi_i(\{T_h\}\mid T_h)=1$ for all $i\in\mathcal M$ and $h\in\mathcal N$; truthful rivals induce $\ph_j^{\boldsymbol\xi_{-i}}=\ph_j$ and $\Shat_{ij}(T_j;\boldsymbol\xi_{-i})=\Shat_j$.

A strategy profile $\boldsymbol\xi$ of a mechanism has the \emph{anonymized outcome} $(\mathbf p^{\boldsymbol\xi},\mathbf u^{\boldsymbol\xi})$, its outcome averaged over buyers and over the ordered rival profiles of each configuration:
\begin{equation}\label{eq:anon}
 (p_j^{\boldsymbol\xi},x_j^{\boldsymbol\xi})(\mathbf k)=\frac{\prod_hk_h!}{m\,(m-1)!}\sum_{i=1}^m\sum_{\mathbf s_{-i}:\,\mathbf k(\mathbf s_{-i})=\mathbf k}(\tilde p_i^{\boldsymbol\xi},\tilde x_i^{\boldsymbol\xi})(T_j,\mathbf s_{-i}),
\end{equation}
with $u_j^{\boldsymbol\xi}=T_jp_j^{\boldsymbol\xi}-x_j^{\boldsymbol\xi}$. As $\Phi(\mathbf s_{-i})=\Phi(\mathbf k)\prod_hk_h!/(m-1)!$ when $\mathbf k(\mathbf s_{-i})=\mathbf k$, the expected revenue and allocated surplus under $\boldsymbol\xi$ are $\operatorname{Rev}(\boldsymbol\xi)=\operatorname{Rev}(\mathbf p^{\boldsymbol\xi},\mathbf u^{\boldsymbol\xi})$ and $G(\boldsymbol\xi)=G(\mathbf p^{\boldsymbol\xi})$. Truthful reporting in an anonymous direct mechanism $(\mathbf p,\mathbf u)$ has anonymized outcome $(\mathbf p,\mathbf u)$.

\begin{prop}[Revenue and welfare ranking of equilibria]\label{prop:select}
(i) If equality holds in Lemma~\ref{lem:model}(ii), then $\mathbf p^{\boldsymbol\xi}=\mathbf p$ if $\mathbf p$ is the unique globally optimal allocation, $(\mathbf p^{\boldsymbol\xi},\mathbf u^{\boldsymbol\xi})=(\mathbf p,\mathbf u)$ if moreover the minimum-rent problem at $\mathbf p$ has a unique solution, and $G(\boldsymbol\xi)\le G(\mathbf p)$ if $G(\mathbf p)=\max\{G(\mathbf p'):\Rp(\mathbf p')=\Rstar\}$.
(ii) For every $\mathbf u$ and every strategy profile $\boldsymbol\xi$ of $(\mathbf p^n,\mathbf u)$, $G(\boldsymbol\xi)\le S$, with equality if and only if $\boldsymbol\xi$ is truthful reporting.
Part~(i) holds likewise under \eqref{eq:cap}, with \eqref{eq:cap} added to the minimum-rent problem and $\Rp^\varsigma$ and $\mathcal R^{*,\varsigma}$ in place of $\Rp$ and $\Rstar$.
\end{prop}
Part~(i) and Lemma~\ref{lem:model}(ii) concern anonymized outcomes, averages over buyers, not strategies: an untruthful SE with the truthful anonymized outcome earns the same revenue, and outside efficient allocation the welfare comparison cannot exclude it. Part~(ii) identifies strategies under efficient allocation but ranks by objective surplus, which buyers need not share. Neither part predicts that buyers coordinate on truthful reporting.

\section{Technical Proofs of Main Results}\label{app:proofs}

\subsubsection*{Proof of Lemma~\ref{lem:model}.}
(i) \emph{Direct mechanism.} The expected outcomes $(\tilde p_i^{\boldsymbol\xi},\tilde x_i^{\boldsymbol\xi})$ define a feasible direct mechanism: the expectation of $\sum_i\tilde p_i$ is at most one. Truthful reports preserve the original revenue, statewise participation, and cap. Against truthful rivals, type $T_j$ reporting $T_h$ obtains $\int_{A_i}\Shat_{ij}(a_i;\boldsymbol\xi_{-i})\xi_i(da_i\mid T_h)$ by Fubini. This is bounded by $\sup_{a_i\in A_i}\Shat_{ij}(a_i;\boldsymbol\xi_{-i})$, attained at $h=j$ by \eqref{eq:SE}.

\emph{Anonymity.} Conditional on $\mathbf k$, each ordered rival profile has the same probability under both subjective and objective laws. Thus averaging as in \eqref{eq:anon} preserves revenue and gives \eqref{eq:ic}, \eqref{eq:ir}, and \eqref{eq:cap} from the corresponding conditions for each buyer. To obtain \eqref{eq:FEA}, average $\sum_i\tilde p_i(\mathbf s)\le1$ uniformly over full profiles with counts $k_h+\ind\{h=j\}$: buyer $i$ has value $T_h$ with frequency $(k_h+\ind\{h=j\})/m$, and her rivals then range uniformly over profiles of configuration $\mathbf k+\mathbf e_j-\mathbf e_h$.

\emph{Attainment and (ii).} The finite LP \eqref{eq:primal} contains the zero mechanism and is bounded by $G(\mathbf p)\le S$, so its value is attained, also with \eqref{eq:cap}. Any optimum is truthful and statewise participating by \eqref{eq:ic}--\eqref{eq:ir}. By direct reduction and averaging, $\operatorname{Rev}(\boldsymbol\xi)=\operatorname{Rev}(\mathbf p^{\boldsymbol\xi},\mathbf u^{\boldsymbol\xi})\le\Rp(\mathbf p^{\boldsymbol\xi})\le\Rstar$. The first inequality binds exactly when $\mathbf u^{\boldsymbol\xi}$ minimizes rent at $\mathbf p^{\boldsymbol\xi}$, and the second exactly when $\mathbf p^{\boldsymbol\xi}$ is globally optimal. The same reasoning uses the capped programs if \eqref{eq:cap} holds.

\emph{Transfers.} Truthful realized utility is $u_j/p_j\ge0$ on winning and zero on losing for $p_j>0$, and $u_j\ge0$ if $p_j=0$. Expected payment is $T_jp_j-u_j$. The cap forces $u_j=0$ when $p_j=0$ and bounds the winner's price below by $-\varsigma$. Statewise participation alone permits arbitrarily negative winner prices. \qed

\subsubsection*{Proof of Proposition~\ref{prop:select}.}
(i) Lemma~\ref{lem:model}(ii) gives $\Rp(\mathbf p^{\boldsymbol\xi})=\Rstar$ and rent minimization at $\mathbf p^{\boldsymbol\xi}$ for every revenue-optimal SE. This gives the surplus bound and pins down the allocation if the optimum is unique. The rents then agree with $\mathbf u$ whenever the minimum-rent solution is unique. Under the cap, $x_h\ge-\varsigma p_h$ is preserved by averaging over reports, giving the capped conclusions.

(ii) For $\bar s=\max_i s_i$, the realized surplus loss at $\mathbf s$ is
\[
 \bar s-\sum_i s_i\tilde p_i^{\boldsymbol\xi}(\mathbf s)
 =\bar s\Bigl(1-\sum_i\tilde p_i^{\boldsymbol\xi}(\mathbf s)\Bigr)
    +\sum_i(\bar s-s_i)\tilde p_i^{\boldsymbol\xi}(\mathbf s)\ge0.
\]
Its objective expectation is $S-G(\boldsymbol\xi)$. Full support and $T_n>0$ make equality equivalent to full allocation to highest-valued buyers in every profile, which truth attains under $\mathbf p^n$. Conversely, under $\mathbf p^n$, give buyers $i\ne i'$ values $T_j>T_h$ and all others $T_n$. Equality requires $i$ to win surely, so every report of $i$ in support must exceed every report of $i'$ in support. Alternate $i,i'$ as $j$ ranges from $1$ to $n$ and apply this separation to consecutive values. Both the support minima and maxima are strictly decreasing $n$-tuples in $\mathcal T$, hence each equals $(T_1,\ldots,T_n)$. Exchanging $i,i'$ covers both buyers at every type. As the pair was arbitrary, reporting is truthful. \qed

%\subsection{Allocation and Rent Capacity}\label{app:proofscap}
\subsubsection*{Proof of Theorem~\ref{thm:common}.}
\emph{Step 1 (implementability).} Under a common law, the constraint for $(j,h)$ is $\Shat_j-\Shat_h\ge(T_j-T_h)P_h(\mathbf p)$. Adding the two directions gives $(T_j-T_h)(P_j-P_h)\ge0$. Conversely, write each difference of floors as the signed sum of adjacent gaps weighted by $P_t$. For $j<h$, each $P_t$ in that sum is at least $P_h$; for $j>h$, each $P_t$ in the subtracted sum is at most $P_h$. Thus the floors satisfy all IC constraints. Monotonicity also gives $\underline{\Shat}_j\le(T_j-T_n)P_j$, so $u_j=p_j\underline{\Shat}_j/P_j$ (zero if $P_j=0$) delivers the floor within \eqref{eq:cap}.

\emph{Step 2 (exact floors).} The downward adjacent constraints and $\Shat_n\ge0$ imply $\Shat_j\ge\underline{\Shat}_j$ by induction. The type-wise minimum cost $c_j(t)$ of delivering at least $t$ is strictly increasing on its feasible domain: scaling a minimizer for $t'>t$ by $t/t'$ preserves the bounds and strictly reduces its positive cost. Therefore each type-wise minimizer delivers exactly its target. Since the floors jointly satisfy IC, type-wise cost minimization at the floors is necessary and sufficient for revenue optimality. Without a cap, $\sum_{\mathbf k}\Phi(\mathbf k)u_j(\mathbf k)\ge\Shat_j/\Lambda_j$, with equality precisely for rents supported on maximum-likelihood-ratio states, where any floor can be delivered. With the cap, $c_j(\underline{\Shat}_j)=U_j^\varsigma(\mathbf p)$ by definition. \qed

\subsubsection*{Proof of Proposition~\ref{prop:knapsack}.}
Fix $j$ and let $t=\underline{\Shat}_j(\mathbf p)$ and $c(\mathbf k)=(T_j+\varsigma)p_j(\mathbf k)$. The feasible, bounded rent LP has dual objective $\vartheta t-\sum_{\mathbf k}c(\mathbf k)\eta(\mathbf k)$, with $\vartheta,\eta\ge0$ and $\vartheta\ph-\eta\le\Phi$. Eliminating $\eta=\Phi\pos{\vartheta\Lambda_j-1}$ gives \eqref{eq:knapdual} by strong duality. The piecewise-linear objective attains a maximum at zero or at some kink $1/\Lambda_j(\mathbf k)$.

Moving a small perceived rent mass $\epsilon$ from a lower-ratio state $\mathbf k'$ with $u_j>0$ to a higher-ratio state $\mathbf k$ below cap reduces cost by $\epsilon\{1/\Lambda_j(\mathbf k')-1/\Lambda_j(\mathbf k)\}$. Conversely, if no such pair exists and $u_j$ delivers $t>0$, take $\vartheta$ as the inverse of the smallest ratio on its positive support. States with $\vartheta\Lambda_j>1$ are at capacity and those with $\vartheta\Lambda_j<1$ have zero rent, giving complementary slackness with the displayed dual choice. For $t=0$ only the zero schedule minimizes cost. Then apply Theorem~\ref{thm:common}(ii). \qed

\subsubsection*{Proof of Corollary~\ref{cor:compact}.}
Theorem~\ref{thm:common} gives the monotonicity constraints and objective, with $U_n^\varsigma=0$. Proposition~\ref{prop:knapsack} expresses $U_j^\varsigma(\mathbf p)$ as the maximum of zero and the affine expressions in the LP; the objective coefficient $-mq_j<0$ makes the epigraph variable bind. Without the cap, the rent costs in Theorem~\ref{thm:common} are affine in $\mathbf p$. \qed

\subsubsection*{Proof of Theorem~\ref{thm:sharedtop}.}
Let $\bar\lambda=\lambda^\varsigma\ind\{\beta\ge\beta_c\}$ and $L=\Lambda_1=(\beta/Z_1)^{m-1}$.

\emph{Step 1 (floors).} Along $\mathbf p^\lambda$, the middle floor is $(T_2-T_3)\ph(\mathbf D_3)/m=(T_2+\varsigma)\ph(\mathbf D_1)\lambda^\varsigma$, and the top floor rises by $(T_1-T_2)\ph(\mathbf D_1)\lambda$. Since the efficient top floor is at least the middle floor, \eqref{eq:ST} implies $\bar\lambda\le1/m$. Hence for $0\le\lambda\le\bar\lambda$, $p_2^\lambda(\mathbf D_1)\le1/m=p_1^\lambda(\mathbf D_1)$ and the rule is pointwise monotone. The top floor fits at $\mathbf D_1$ by \eqref{eq:ST} and affinity.

\emph{Step 2 (multipliers).} The knapsack supports $\vartheta_1=1/L$. Starting with $\mathbf D_1$, type~2 fills capacity $(T_2+\varsigma)\ph(\mathbf D_1)\lambda$, then uses states with $k_1=0$, of ratio $Z_1^{1-m}$. Their total capacity $(T_2+\varsigma)P_2(\mathbf p^n)$ exceeds the floor $(T_2-T_3)P_3(\mathbf p^n)$. Thus $\vartheta_2=Z_1^{m-1}$ is optimal, as is any $\vartheta_2\in[1/L,Z_1^{m-1}]$ at $\lambda=\lambda^\varsigma$. Set $a=\min\{1,(\beta_c/\beta)^{m-1}\}$. Corollary~\ref{cor:commondual} attaches $(\vartheta_1,\vartheta_2,\vartheta_3)=(1/L,aZ_1^{m-1},0)$ at $\bar\lambda$, and along $[0,\lambda^\varsigma]$ when $\beta=\beta_c$.

\emph{Step 3 (scores).} With $y=\beta^{k_1}$, the scores from these multipliers are $T_1$,
$g(y)=T_2+(T_2+\varsigma)\pos{ay-1}-(q_1/q_2)(T_1-T_2)\beta^{1-m}y$, and $T_3-(T_2-T_3)(q_1\beta^{1-m}+q_2a)y/q_3$. By construction, $g(1)=\bar\psi_2<T_1$, and $g(\beta^{m-1})\le T_1$, with equality just when $\beta\ge\beta_c$. Convexity gives $g(y)<T_1$ at intermediate $y$. The low score is at most $\bar\psi_3$. Thus the highest positive score follows the highest report, with a top--middle tie only at $\mathbf D_1+\mathbf e_2$ when $\beta\ge\beta_c$. With no top report, the scores are $\bar\psi_2>\bar\psi_3>0$.

\emph{Step 4 (optimality).} The candidate $\mathbf p^{\bar\lambda}$ fully assigns to highest scores, as does the specified segment at $\beta=\beta_c$. Attachment and Theorem~\ref{thm:dual}(iii) give optimality with value $w_1T_1+w_2\bar\psi_2+w_3\bar\psi_3$.

\emph{Step 5 (equality).} Attainment forces highest-score service at each profile and capacity filled where $\ell_h>0$ and zero rent where $\ell_h<0$, by \eqref{eq:lagidentity} and \eqref{eq:cap}. Hence with $\beta<\beta_c$ the allocation is $\mathbf p^n$; otherwise it is $\mathbf p^\lambda$ for some $\lambda\in[0,1]$. Since $\ell_2(\mathbf D_1)=q_2(\beta_c^{m-1}-1)>0$, type~2 delivers $(T_2+\varsigma)\ph(\mathbf D_1)\lambda$ there, so the exact floor from Step~1 forces $\lambda\le\lambda^\varsigma$. If $\beta>\beta_c$, $\ell_2<0$ at $k_1=0$, so no other capacity delivers rent and $\lambda=\lambda^\varsigma$. Also $\ell_1<0$ off $\mathbf D_1$, and the exact floors determine all rents. At $\beta=\beta_c$, the segment in Step~4 exhausts the possibilities.

\emph{Step 6 (reserves).} At $\mathbf p^1$ every floor vanishes, giving $w_1T_1$. At $\mathbf p^2$ and $\mathbf p^n$ the top floor fits at $\mathbf D_1$ and any middle service has $k_1=0$; these rules attach the scores of Step~3 with $a=1$. Their revenues are the assigned score expectations, $w_1T_1+w_2\bar\psi_2$ and that amount plus the all-low contribution $w_3\{T_3-(T_2-T_3)(q_1\beta^{1-m}+q_2)/q_3\}$. Subtracting from the optimum, using $w_3=q_3^m$, gives both reserve gaps since $\bar\psi_2>0$ rules out $\mathbf p^1$ as the best reserve.

\emph{Revenue path and surplus.} Under $\beta\ge\beta_c$, the changed profile has probability $mq_1^{m-1}q_2$ and each middle win replaces value $T_1$ by $T_2$. Along $0\le\lambda\le\lambda^\varsigma$, the top rent cost $U_1^\varsigma$ increases by $(T_1-T_2)q_1^{m-1}\lambda$, while $U_2^\varsigma$ falls by $(T_2+\varsigma)q_1^{m-1}(\beta^{m-1}-1)\lambda$, from the delivery tiers in Step~2. Combining these with surplus and the definition of $\beta_c$ gives the stated slope. \qed

\subsubsection*{Proof of Theorem~\ref{thm:overflow}.}\label{app:overflow}
Write $\boldsymbol\pi=(\beta q_1,q_2,q_3)/Z_1$. At $\mathbf p^n$, $m\pi_jP_j$ is the subjective probability of largest value $T_j$. The changed profile lowers $P_1$ by $\pi_1^{m-2}\pi_2\lambda$, increases $P_2$ by $\pi_1^{m-1}\lambda$, and leaves $P_3$ unchanged, so \eqref{eq:overflow-tc} gives monotonicity. Its top floor rises by $d_{12}\pi_1^{m-1}\lambda$ and its middle floor remains $B_2\pi_1^{m-1}\lambda^\varsigma$. Binomial tier probabilities give
\[
 C_t(\mathbf p^\lambda)=B_1\left\{
 \binom{m-1}{t}\frac{\pi_1^t(1-\pi_1)^{m-1-t}}{t+1}
 -\ind\{t=m-2\}\pi_1^{m-2}\pi_2\lambda\right\}.
\]
Under \eqref{eq:overflow-tc}, Proposition~\ref{prop:knapsack} supports $\vartheta_1=1/\Lambda^{[\ell]}$. For $\lambda\le\lambda^\varsigma$, type~2 fills the capacity at $\mathbf D_1$, then uses $k_1=0$ where $B_2P_2(\mathbf p^n)>d_{23}P_3(\mathbf p^n)$. The multiplier is $1/\Lambda^{[0]}$, or any value in $[1/\Lambda^{[m-1]},1/\Lambda^{[0]}]$ at $\lambda^\varsigma$. Thus, with $U_j=U_j^\varsigma(\mathbf p^\lambda)$ and $C_t=C_t(\mathbf p^\lambda)$,
\begin{equation}\label{ec:overflow-rev}
\begin{aligned}
 \Rp^\varsigma(\mathbf p^\lambda)=S-mq_1^{m-1}q_2d_{12}\lambda-m(q_1U_1+q_2U_2),\quad&U_1=\sum_{t>\ell}C_t/\Lambda^{[t]}
       +(\underline{\Shat}_1(\mathbf p^\lambda)-\sum_{t>\ell}C_t)/\Lambda^{[\ell]},\\
 U_2=B_2\pi_1^{m-1}\bigl\{\lambda/\Lambda^{[m-1]}+
                      (\lambda^\varsigma-\lambda)/\Lambda^{[0]}\bigr\},\quad &U_3=0.
\end{aligned}
\end{equation}
Differentiating and using $B_1-B_2=d_{12}$ gives the stated slope.

\emph{Global certificate.} Attach multipliers by Corollary~\ref{cor:commondual} with $\vartheta_2=a_\ell/\Lambda^{[0]}$, $\vartheta_3=0$. This $\vartheta_2$ is permitted at each candidate because $\zeta_\ell^{\rm ov}\beta^{\ell+1}=(B_1+\beta q_1d_{12}/q_2)/B_2>1$. At a profile with $h\ge1$ top reports, the top and middle scores are
\[
 T_1+B_1\pos{\beta^{h-1-\ell}-1},\qquad
 T_2+B_2\pos{a_\ell\beta^h-1}-(q_1/q_2)d_{12}\beta^{h-\ell}.
\]
Low scores are below $T_1$; at $h=0$ the scores are $\widetilde\psi_2>\widetilde\psi_3>0$. For $h\ge\ell+1$, $a_\ell\beta^h>1$ and the top score exceeds the middle score by $B_2\beta^h\pos{\zeta_\ell^{\rm ov}-1}$. For $1\le h\le\ell$, the middle score is at most $T_2$ if its capacity term vanishes; otherwise it equals $(B_2a_\ell\beta^\ell-q_1d_{12}/q_2)\beta^{h-\ell}-\varsigma<T_1$, since the coefficient is at most $B_1/\beta$. Every candidate thus sells exclusively to highest positive scores. The inequalities in \eqref{eq:overflow-tc} extend across the segment in (iii) by affinity. Theorem~\ref{thm:dual} gives optimality and strict rankings give uniqueness in (i).

\emph{Optimal face in (ii)--(iii).} Equality requires efficient service at profiles with at most $\ell$ top reports and full sale among highest-scoring top and middle reports otherwise. Let $z_h=\sum_{\mathbf k:k_1=h}\ph(\mathbf k)p_2(\mathbf k)$ on mixed profiles with $h\ge\ell+1$, and put $z_h=0$ otherwise and $z=\sum_h z_h$. The remaining necessary and sufficient conditions are
\begin{equation}\label{ec:overflow-face}
\begin{gathered}
 z=\underline{\Shat}_2(\mathbf p^n)/B_2\ \text{in (ii)},\qquad
 z\le\underline{\Shat}_2(\mathbf p^n)/B_2\ \text{in (iii)},\\[-1pt]
 \sum_{t>\ell}C_t(\mathbf p)\le\underline{\Shat}_1(\mathbf p^n)+d_{12}z
                            \le\sum_{t\ge\ell}C_t(\mathbf p).
\end{gathered}
\end{equation}
Indeed, the tier-$h$ increase $m\pi_2z_h$ in subjective middle service equals the top loss, giving $C_t(\mathbf p)=C_t(\mathbf p^n)-B_1(\pi_2/\pi_1)z_{t+1}$. The interim probabilities match $\mathbf p^\lambda$ at $\lambda=z/\pi_1^{m-1}$. Optimality requires the exact floors and, by \eqref{eq:lagidentity}, zero rent where $\vartheta_j\Lambda^{[t]}<1$ and filled capacity where $\vartheta_j\Lambda^{[t]}>1$. Thus the top floor gives the tier inequalities and the middle floor gives the condition on $z$: mixed-profile capacity is filled and no-top rent is permitted only in (iii). Conversely the condition on $z$ implies $\lambda\in[0,\lambda^\varsigma]$, so monotonicity follows from the endpoints. The tier inequalities permit exact top delivery; fill mixed middle capacity and, in (iii), use the unchanged no-top capacity for the remainder. These rents with $u_3=0$ satisfy IC and the equality conditions of the score bound.

In (ii) with $\ell=m-2$, the only mixed profile allowing middle service has $m-1$ top reports, so $z$ pins down $\mathbf p^{\rm st}$. For $\ell<m-2$, transfer a small $\epsilon>0$ of $z_{m-1}$ to the profile of $\ell+1$ top and $m-\ell-1$ middle reports; both donating reports have positive service. This increases capacity above $\ell$ by $B_1(\pi_2/\pi_1)\epsilon$ and leaves capacity at or above $\ell$ fixed, preserving \eqref{ec:overflow-face} when the lower bound is strict. Objective loss is $S-G(\mathbf p)=mq_2d_{12}\sum_hz_h/\Lambda^{[h]}$. At fixed $z$ this is uniquely minimized at the highest tier, attainable only by $\mathbf p^{\rm st}$. \qed

\subsubsection*{The calibrated shared-top path.}\label{app:overflowpath}
Use the exact calibrated values with $m=4$, $\varsigma=0$, $\mathbf q=(1/5,3/10,1/2)$ and $\boldsymbol\pi=(\beta,3/2,5/2)/(\beta+4)$. Scaled by $(\beta+4)^3$, the top floors at efficiency and the lottery are $E=(1157d_{12}+125d_{23})/32$ and $L=E+d_{12}z_*$, with $z_*=125d_{23}/(32T_2)$ and $\lambda^0=z_*/\beta^3$. Lottery tier capacities are $T_1(64,24\beta,4\beta^2-3z_*/(2\beta),\beta^3/4)$; omit the negative term at efficiency. The three cutoffs are the positive roots of
\[
 T_2\beta^2-\tfrac23d_{12}\beta-T_1=0,\qquad
 T_1\{\beta^3/4+4\beta^2-3z_*/(2\beta)\}=L,\qquad T_1\beta^3/4=L .
\]
The first polynomial has one positive root. Both capacity sums strictly increase on $\beta>0$ and cross $L$. Exact substitution brackets the roots in $(1.41,283/200)$, $(365/200,1.83)$, and $(759/200,3.80)$, respectively.

Monotonicity gives the tier tests. At efficiency, capacity above tier~1 is below $E$ at $1.42$, and capacity at or above tier~1 exceeds $E$ at $1$. For the lottery, the latter capacity exceeds $L$ at $1.41$. The last two equations give the later tier boundaries. $\zeta_1^{\rm ov}$ strictly decreases through one at the first root, and $\zeta_2^{\rm ov}(1.82)<9/25$. The no-top scores are ordered and positive since $\widetilde\psi_3\ge T_3-d_{23}>0$ and $\widetilde\psi_2-\widetilde\psi_3\ge d_{23}+\min\{0,2d_{23}/5-2d_{12}/3\}>0$; the same bounds hold for the scores in \eqref{eq:ST}. For the lottery at or above the first root, $\lambda^0\le z_*/1.41^3<1$, and $\beta(\beta+4)^3(P_1-P_2)=\beta^4/4+4\beta^3+24\beta^2+(891/32-z_*)\beta-3z_*/2$ is positive and increasing for $\beta\ge1$. Finally $3.79^3>1+5d_{12}/(3T_2)$ places the third root above $\beta_c$.

At $\beta=1$, strictly ordered positive correct-belief virtual values give unique efficiency, as does Theorem~\ref{thm:overflow}(i) for $\beta>1$ below the first root. Parts (iii) and (ii) give an optimal segment at the first root and nonuniqueness strictly between the first two, respectively, with full optimal sets in \eqref{ec:overflow-face}. Tier~2 gives a unique lottery from the second root to the third. From the third root onward, \eqref{eq:ST} and Theorem~\ref{thm:sharedtop} give the same lottery with peak delivery. \qed

\subsubsection*{Proof of Proposition~\ref{prop:pool}.}
Under all-$T_1$ reporting, each buyer faces $\mathbf D_1$ surely, whatever her law. A type $T_j$ earns $T_j/m-x_1(\mathbf D_1)$ from report $T_1$ and $T_jp_h(\mathbf D_1)-x_h(\mathbf D_1)$ from report $T_h$. The difference, $T_j\{1/m-p_h(\mathbf D_1)\}-x_1(\mathbf D_1)+x_h(\mathbf D_1)$, is nondecreasing in $T_j$ because $p_h(\mathbf D_1)\le1/m$. It is therefore nonnegative, or positive, for all $j$ and $h\ge2$ if and only if it is for $j=n$, which is the stated condition; by linearity, the comparison extends to lotteries over reports. The object is sold surely, and total payment, also the winner's price under the transfer rule, is $mx_1(\mathbf D_1)=T_1-mu_1(\mathbf D_1)$. Every reserve rule has $p_1^r(\mathbf D_1)=1/m$ and $p_h^r(\mathbf D_1)=0$ for $h\ge2$; \eqref{eq:ir} and \eqref{eq:cap} then force $u_h(\mathbf D_1)=0$, hence $x_h(\mathbf D_1)=0$, for $h\ge2$, and $mx_1(\mathbf D_1)\ge-\varsigma$. \qed

\begin{defi}[Certificate classes and revenue programs]\label{def:certprograms}
(i) Given $\mathcal S$ as in Section~\ref{sec:unique}, set $\Sigma(\mathcal S)=\{\sigma:\mathcal N\to\mathcal N:\sigma(t)\in\mathcal S_t\ \forall t\}$. A profile $\boldsymbol\sigma=(\sigma_{i'})_{i'\ne i}$ of such maps sends rival values $s_{i'}=T_{t_{i'}}$ to reports $T_{\sigma_{i'}(t_{i'})}$ and induces
\begin{equation}\label{ec:inducedlaw}
 \ph_j^{\boldsymbol\sigma}(\mathbf k)=\sum_{\mathbf s_{-i}}\ph_j(\mathbf s_{-i})
 \ind\{\mathbf k(\boldsymbol\sigma(\mathbf s_{-i}))=\mathbf k\}.
\end{equation}
Thus $\mathcal F_j(\mathcal S)$ uses all $\boldsymbol\sigma\in\Sigma(\mathcal S)^{m-1}$ and $\mathcal V_j(\mathcal S)$ uses only maps that rank as in Section~\ref{sec:unique}.

(ii) Let $\mathfrak E_{\rm seq}$ and $\mathfrak E_{\rm del}$ be the orderings and deletion schedules of Section~\ref{sec:unique}. With $X=\{(\mathbf p,\mathbf u)\ge\mathbf0:\eqref{eq:FEA}\}$, the certificate sets are
\begin{align}
 \Pi^{\prec}_{\rm seq}(\delta)&=\{(\mathbf p,\mathbf u)\in X:\ \mathrm{sl}_{jh}(\ph)\ge\delta\quad(j\ne h,\ \ph\in\mathcal F_j(\mathcal N_{\prec j}))\},\label{ec:wtset}\\
 \Pi^{\mathcal E}_{\rm del}(\delta)&=\{(\mathbf p,\mathbf u)\in X:\ \mathrm{sl}_{j_\tau h_\tau}(\ph)\ge\delta\quad(\tau,\ \ph\in\mathcal F_{j_\tau}(\mathcal S^{\tau-1}))\}.\label{ec:rrset}
\end{align}
The capped sets $\Pi^{\mathcal E,\varsigma}_C(\delta)$ additionally impose \eqref{eq:cap}.

(iii) With $\sup\varnothing=-\infty$,
\begin{equation}\label{ec:certprog}
 \mathcal R^{\mathcal E}_C(\delta)=\sup_{(\mathbf p,\mathbf u)\in\Pi^{\mathcal E}_C(\delta)}\operatorname{Rev}(\mathbf p,\mathbf u),
 \qquad \mathcal R_C(\delta)=\max_{\mathcal E\in\mathfrak E_C}\mathcal R^{\mathcal E}_C(\delta).
\end{equation}
The fixed-allocation and capped programs restrict these sets as specified in Section~\ref{sec:unique}.
\end{defi}

\subsubsection*{Proof of Theorem~\ref{thm:unique}.}
\emph{Step 1 (induced laws).} Conditional on the full rival-value profile, independently mixed strategies induce a product law over rival reports. For strategies supported on $\mathcal S$, averaging over values makes $\ph_j^{\boldsymbol\xi_{-i}}$ separately affine in each rival's report distribution at each value. Expanding each distribution into point masses gives a convex combination of admissible pure-map laws, in $\co(\mathcal F_j(\mathcal S))$.

\emph{Step 2 (elimination).} In a report-by-report certificate, if all equilibrium strategies are supported on $\mathcal S^{\tau-1}$, the slack $\mathrm{sl}_{j_\tau h_\tau}(\ph_j^{\boldsymbol\xi_{-i}})$ for $j=j_\tau$ is at least $\delta$ by Step~1 and linearity. Thus $\delta>0$ excludes $h_\tau$ from every type-$j_\tau$ strategy support by \eqref{eq:SE}. Induction from $\mathcal S^0$ leaves only truth. Truthful rivals are admissible throughout, so the certificate implies truth is an SE. Processing types consecutively in a whole-type ordering tests subsets of the original rival laws at each deletion, proving (i) for that class as well.

\emph{Step 3 (LPs and ranking).} Finiteness of $\mathcal F_j$ gives linear programs, bounded above by $S$ and attaining their values when feasible. At $\delta=0$ the zero mechanism, or \eqref{ec:dsicrent} at $\mathbf p^r$, is feasible. Under subjective independence, start with a minimizing pure-map profile for a given slack. Holding other rivals' maps fixed, rank the reports of one rival by their conditional expected slack, which depends on her report alone. Choosing the first admissible report at each value cannot increase the slack. Replacing rivals in turn preserves minimality and produces a profile of ranking maps, so testing $\mathcal V_j$ suffices. At a whole-type step, a ranking rival makes one common report at all unprocessed values. Under i.i.d. subjective values, the induced law thus depends only on the counts of rivals choosing each common report, yielding at most $|\cK|$ laws. \qed

\subsubsection*{Proof of Proposition~\ref{prop:margincost}.}
By Theorem~\ref{thm:unique}'s proof, the two certificate classes are nested and imply nominal IC. The witness $p_j=T_j/(mT_1)$, $x_j=T_j^2/(2mT_1)$ is feasible, with nonnegative rents and payments. Its statewise slack is $(T_j-T_h)^2/(2mT_1)$, so every schedule has margin at least $\bar\delta=\min_{t<n}(T_t-T_{t+1})^2/(2mT_1)$. Mix the witness with weight $\delta/\bar\delta$ into a best zero-margin mechanism using the same order. The feasible mixture achieves margin $\delta$ and loses at most $(\delta/\bar\delta)\{\mathcal R_C(0)-\sum_jq_jT_j^2/(2T_1)\}$ in revenue. The same argument applies with the cap. \qed

\subsubsection*{Proof of Proposition~\ref{prop:shadow}.}
Fix $\mathcal E$. The inner problem has form $\max\{\mathbf c^\top\mathbf z:\mathbf A\mathbf z\le\mathbf b-\delta\boldsymbol\iota,\ \mathbf z\ge0\}$, where $\boldsymbol\iota$ indicates the certificate rows. Fixing $\mathbf p$ leaves the same form up to an objective constant. The feasible margins are a closed interval containing zero (projection of a polyhedron and monotonicity in $\delta$). The value is bounded by $S$ and equals $\min_{\mathbf y\in Y}(\mathbf b-\delta\boldsymbol\iota)^\top\mathbf y$, where $Y=\{\mathbf y\ge0:\mathbf A^\top\mathbf y\ge\mathbf c\}$. Since $Y$ is pointed, the minimum is attained at a vertex. The resulting finite lower envelope is concave, piecewise linear and nonincreasing. Below the right endpoint, the right derivative is the smallest slope among active vertices. This slope equals $-\bar y$: for any optimal $\mathbf y$ and feasible $\delta'>\delta$, the value at $\delta'$ is at most the value at $\delta$ minus $(\delta'-\delta)\boldsymbol\iota^\top\mathbf y$, so $\boldsymbol\iota^\top\mathbf y$ is bounded on the optimal face and attains its maximum at a vertex there.

Fixing the allocation restricts the program, giving the nonnegative matched gain when finite. Each finite maximum over inner programs has right derivative equal to the largest right derivative among its active programs below their right endpoints, since inactive affine pieces remain strictly below locally. Subtract the derivatives of the joint and reserve maxima to obtain the formula. \qed

\subsubsection*{Proof of Lemma~\ref{lem:reservecert}.}
Let $j$ be the first type below the top in $\prec$ and $h\ne j$ any other type below the top. All-top rival reporting is truthful on $\mathcal N_{\prec j}\subseteq\{1\}$, so $\ind_{\mathbf D_1}\in\mathcal F_j(\mathcal N_{\prec j})$. At $\mathbf D_1$, reserve service and capped rent are zero for both reports, yielding $\mathrm{sl}_{jh}(\ind_{\mathbf D_1})=0<\delta$. \qed

\subsubsection*{Proof of Proposition~\ref{prop:fixedgain}.}
Here $\boldsymbol\pi=(2/3,1/6,1/6)$ and matrix entry $(j,k)$ refers to own report $T_j$ facing $\mathbf e_k$. Under \eqref{eq:ir} and \eqref{eq:cap}, losing reports have zero rent. For $\mathbf p^2$ and $\mathbf p^3$, write the rent rows as $(a,b,c)$, $(0,e,f)$, and $(0,0,z)$ with nonnegative entries and $z=0$ for $\mathbf p^2$. Expected rent is $2(a+b+c+e+f+z)/9$.

\emph{Reserve bounds.} At step~4, the truthful rival map gives $e+f\ge6\delta$ under $\mathbf p^2$. At step~6, let the rival's type~1 report either~1 or~2, with all other types truthful. The two inequalities in \eqref{eq:certificate} give
\begin{equation}\label{ec:fixedgainbounds}
 4a+b+c-e-f\ge\tfrac32d_{12}+6\delta,\qquad
 5b+c-5e-f\ge\tfrac72d_{12}+6\delta.
\end{equation}
Weight them by $1/4$ and $3/20$ and use $\tfrac35c+2e+\tfrac75f\ge\tfrac75(e+f)\ge42\delta/5$. This gives $a+b+c+e+f\ge9d_{12}/10+54\delta/5$. Subtracting the rent cost from $G(\mathbf p^2)=(5T_1+3T_2)/9$ gives the stated upper bound.

Under $\mathbf p^3$, truth at step~4 gives $e+f-z\ge d_{23}/2+6\delta$. The map that sends only type~1 to report~3 at step~5 gives $b+5c-5z\ge5(d_{12}+d_{23})/2+6\delta$. Add this to the second inequality of \eqref{ec:fixedgainbounds}, which also applies here, and divide by six. After adding $a+e+f+z$ and using nonnegativity,
\[
 a+b+c+e+f+z\ \ge\ d_{12}+\tfrac5{12}d_{23}+2\delta+\tfrac76(e+f-z)
 \ \ge\ d_{12}+d_{23}+9\delta.
\]
Since $G(\mathbf p^3)=(5T_1+3T_2+T_3)/9$, its revenue is at most $(T_1+T_2+T_3)/3-2\delta$. The $\mathbf p^2$ upper bound exceeds this by $(T_1+9T_2-15T_3)/45-2\delta/5\ge(T_1+9T_2-15T_3)/90>0$. Under $\mathbf p^1$, reports~2 and~3 always lose and pay zero, so step~3 has zero slack.

\emph{Attainment and the lottery.} Put $a_\delta=d_{12}/5+12\delta/5$ and $b_\delta=7d_{12}/10+12\delta/5$. Pair $\mathbf p^2$ and $\mathbf p^{\rm lot}$, respectively, with
\[
 \mathbf u^{\rm res}=\begin{pmatrix}a_\delta&b_\delta&0\\0&0&6\delta\\0&0&0\end{pmatrix},
 \qquad
 \mathbf u^{\rm lot}=\begin{pmatrix}a_\delta+d_{12}\lambda&b_\delta&0\\
 \delta&\delta&\delta\\0&0&0\end{pmatrix}.
\]
Feasibility follows from $\lambda=\delta/T_2<1/16$. The nonnegative rent entries satisfy their caps: $a_\delta+d_{12}\lambda\le79d_{12}/200<T_1/2$, $b_\delta\le41d_{12}/50<15T_1/16<T_1(1-\lambda)$, and $6\delta\le3d_{23}/8<T_2$. In the lottery, $\delta=T_2\lambda$ against report~1 and $\delta<T_2/16$ against reports~2 and~3.

At step $\tau$, the least slack over admissible rival maps is $\sum_t\pi_t\min_{k\in\mathcal S_t^{\tau-1}}g_{j_\tau h_\tau}(\mathbf e_k)$. Substitution gives the following bounds, with equality in rows 3--6.
\[
\begin{array}{c|cc}
(j_\tau,h_\tau)&(\mathbf p^2,\mathbf u^{\rm res})&(\mathbf p^{\rm lot},\mathbf u^{\rm lot})\\ \hline
(2,1)&\tfrac3{10}d_{12}-\tfrac{12}5\delta
 &\tfrac3{10}d_{12}-(\tfrac75+d_{12}/T_2)\delta\\
(3,1)&\delta&\delta\\
(3,2)&d_{23}/6&d_{23}/6-(\tfrac13+\tfrac{2T_3}{3T_2})\delta\\
(2,3)&\delta&\delta\\
(1,3)&b_\delta/6&b_\delta/6\\
(1,2)&\delta&\delta
\end{array}
\]
At step~2 the reserve slack against reports~1 and~2 is the pointwise step~1 slack plus $d_{23}p_1^2$, and against report~3 it is $d_{12}+d_{23}$. For the lottery the adjustment is $d_{23}p_1^{\rm lot}-\delta\ge0$ at every report. At step~3 only type~1 can report~1. The pointwise slacks are $(0,d_{23}/2,d_{23}-6\delta)$ and $(-T_3\lambda,d_{23}/2-\delta,d_{23}-\delta)$, respectively. At steps~5 and~6, types~2 and~3 report truthfully, while type~1 can report any type and then only~1 or~2, respectively. Thus the displayed bounds all exceed or equal $\delta$ by $d_{12}\ge20\delta$, $d_{23}\ge16\delta$, and $d_{12}/T_2<3/2$.

The entries of $\mathbf u^{\rm res}$ sum to $9d_{12}/10+54\delta/5$, attaining the best reserve bound. The lottery changes allocated surplus by $-2d_{12}\lambda/9$ and total expected rent by $2(d_{12}\lambda-3\delta)/9$. Its certified revenue therefore exceeds the reserve revenue by $2(5T_2-2T_1)\delta/(9T_2)$. \qed

\subsubsection*{Proof of Proposition~\ref{prop:transport}.}
(i) Let $\mathcal Q$ be the report marginals obtained by sending masses $\pi_{jt}$ from values $t$ to reports in $\mathcal S_t$. If $\Phi\in\co(\mathcal F_j(\mathcal S))$, its expected rival counts divided by $m-1$ give $\mathbf q$ as a convex average in $\mathcal Q$. The mass from any subset $\mathcal N'$ must land in $\cup_{t\in\mathcal N'}\mathcal S_t$, yielding \eqref{eq:hall}.

Conversely, in the transport network with source capacities $\pi_{jt}$ into the values, infinite capacities on admissible value-to-report arcs, and sink capacities $q_h$ from the reports, a finite cut with values $\mathcal N'$ on its source side has capacity at least $\sum_{t\notin\mathcal N'}\pi_{jt}+\sum_{h\in\cup_{t\in\mathcal N'}\mathcal S_t}q_h\ge1$. Max-flow min-cut \citep{FordFulkerson1956} supplies a unit flow. Dividing each arc flow by $\pi_{jt}>0$ gives an admissible randomized map with marginal $\mathbf q$; used independently by all rivals, it induces $\Phi\in\co(\mathcal F_j(\mathcal S))$ by Step~1 of Theorem~\ref{thm:unique}'s proof. At a whole-type step, any $\mathcal N'$ containing an unprocessed type has all reports as neighbors, while the processed types have singleton neighbors. Thus \eqref{eq:hall} reduces to the stated componentwise bounds.

(ii) By linearity every tested slack is at least $\delta$ at $\Phi$. All pairs are tested, so correct-belief IC gives the revenue bound. \qed

\subsubsection*{Proof of Corollary~\ref{cor:restore}.}
Lemma~\ref{lem:rational} supplies a rational-optimal mechanism at monotone $\mathbf p^*$ with rents \eqref{ec:dsicrent}. Proposition~\ref{prop:universal}'s proof gives statewise nonnegative slack and payments, hence every zero-margin certificate under any cap. The assumed componentwise bounds imply \eqref{eq:hall} at every whole-type step by Proposition~\ref{prop:transport}(i). Its part (ii) bounds the revenue above by the rational optimum, attained at zero margin by the stated mechanism. \qed

The proof uses the following lemma, which irons the virtual values $b_h^{\mathrm{rat}}$ \citep{Myerson1981}; write $q_{\ge h}=\sum_{j\ge h}q_j$, with $q_{\ge n+1}=0$.
\begin{lem}[Rational optimum]\label{lem:rational}
Let $Y_0=0$ and $Y_h=(1-q_{\ge h+1})T_h$ for $h\in\mathcal N$; let $\bar Y_h$ be the value at $1-q_{\ge h+1}$ of the least concave majorant on $[0,1]$ of the points $(1-q_{\ge h+1},Y_h)$, $h=0,\ldots,n$; and let $\bar b_h^{\mathrm{rat}}=(\bar Y_h-\bar Y_{h-1})/q_h$. Then $\bar b_1^{\mathrm{rat}}\ge\cdots\ge\bar b_n^{\mathrm{rat}}$ and $\max_{\mathbf{p}\in\cP}\Rp^{\mathrm{rat}}(\mathbf{p})=\sum_hw_h\pos{\bar b_h^{\mathrm{rat}}}$. The maximum is attained by the rents \eqref{ec:dsicrent} at the monotone allocation $\mathbf{p}^*$ that divides the object equally among the buyers whose reports have the highest $\bar b_h^{\mathrm{rat}}$ if this value is positive, and does not sell otherwise. If $b_1^{\mathrm{rat}}\ge\cdots\ge b_n^{\mathrm{rat}}$, then $\bar b_h^{\mathrm{rat}}=b_h^{\mathrm{rat}}$ for every $h$ and $\max_{\mathbf{p}\in\cP}\Rp^{\mathrm{rat}}(\mathbf{p})=\max_r\Rp_r^{\mathrm{rat}}$.
\end{lem}

\subsubsection*{Proof of Lemma~\ref{lem:rational}.}
Let $\mathbf p\in\cP$ and $\mathbf u\ge0$ satisfy \eqref{eq:ic} under correct beliefs, and write $P_h=\sum_{\mathbf k}\Phi(\mathbf k)p_h(\mathbf k)$ and $U_j=\sum_{\mathbf k}\Phi(\mathbf k)u_j(\mathbf k)$. Then \eqref{eq:ic} reads $U_j-U_h\ge(T_j-T_h)P_h$. Adding the constraints for $(j,h)$ and $(h,j)$ gives $P_1\ge\cdots\ge P_n$, and chaining the adjacent downward constraints from $U_n\ge0$ gives $U_j\ge\sum_{h>j}(T_{h-1}-T_h)P_h$. Since $Y_h-Y_{h-1}=q_hb_h^{\mathrm{rat}}$, summation by parts, with $P_{n+1}=0$ and $Y_0=\bar Y_0=0$, gives
\[
 \operatorname{Rev}\le m\sum_hq_hb_h^{\mathrm{rat}}P_h
 =m\sum_hq_h\bar b_h^{\mathrm{rat}}P_h
  +m\sum_h(Y_h-\bar Y_h)(P_h-P_{h+1})
 \le m\sum_hq_h\bar b_h^{\mathrm{rat}}P_h.
\]
The slopes $\bar b_h^{\mathrm{rat}}$ of the concave majorant are nonincreasing, so the largest score present at a profile is that of its highest value. By \eqref{ec:regroupsum} and \eqref{eq:FEA}, the expected allocated score is therefore at most the expected positive part of that score, namely $\sum_hw_h\pos{\bar b_h^{\mathrm{rat}}}$.

The stated $\mathbf p^*$ attains that score bound. It is monotone: equal own-report scores give equal probabilities against any fixed rivals, while a strictly higher score than a winning report wins outright. Thus \eqref{ec:dsicrent} implements it with nonnegative payments and attains the adjacent rent lower bound. The majorant is piecewise affine with kinks only at abscissae where it meets a point. If $Y_h<\bar Y_h$ for some $h<n$, it is therefore affine on $[1-q_{\ge h},1-q_{\ge h+2}]$, so $\bar b_h^{\mathrm{rat}}=\bar b_{h+1}^{\mathrm{rat}}$; then $\mathbf p^*$ treats reports $T_h$ and $T_{h+1}$ alike, and their winning probabilities satisfy $P_h^*=P_{h+1}^*$. Also $Y_n=\bar Y_n$. Hence the summation-by-parts correction vanishes, proving attainment of the revenue bound. If $b^{\mathrm{rat}}$ is nonincreasing, its polygon is already concave and $\bar b^{\mathrm{rat}}=b^{\mathrm{rat}}$; the reserve formula $\Rp_r^{\mathrm{rat}}=\sum_{h\le r}w_hb_h^{\mathrm{rat}}$ then attains $\sum_hw_h\pos{b_h^{\mathrm{rat}}}$, since $b_1^{\mathrm{rat}}=T_1>0$. \qed

\subsubsection*{Verification of Example~\ref{ex:premium}.}
The reserve winning rows are $(1/2,1,1)$, $(0,1/2,1)$, and $(0,0,0)$, so the stated rents obey $0\le u_j\le T_jp_j^2$. With one rival, the least deletion slack is $\sum_t\pi_{jt}\min_{k\in\mathcal S_t}g_{jh}(\mathbf D_k)$. Substitution in schedule order yields
\[
 3-36\delta/25,\quad 53/6-41\delta/25,\quad \delta,\quad
 7/6+11\delta/25,\quad \delta,\quad 25/6-\delta,
\]
all at least $\delta$ for $0<\delta\le1/2$. Surplus is $70/3$ and expected rent $2+36\delta/25$, giving the stated revenue. The correct-belief virtual values are $(30,10,-10)$, so Lemma~\ref{lem:rational} gives optimal revenue $30(5/9)+10(3/9)=20$ at $\mathbf p^2$. Corollary~\ref{cor:restore} gives the whole-type bound. \qed

\subsubsection*{Proof of Proposition~\ref{thm:robust}.}
(i) The slack $\mathrm{sl}_{jh}(\ph)$ is linear in $\ph$, so nonnegativity on $\mathcal L_j$ is equivalent to nonnegativity on its convex hull. The robust finite program is feasible at the zero mechanism and has revenue bounded by $S$, hence attains its optimum. For monotone $\mathbf p$, the rents \eqref{ec:dsicrent} satisfy $0\le u_j^{\mathrm{rat}}\le T_jp_j$ and $g_{jh}\ge0$ at every configuration (proof of Proposition~\ref{prop:universal}), hence \eqref{eq:robustic}.

(ii) Give each robust IC constraint multiplier $m\mu^{\ph}_{jh}\ge0$. Collecting terms in the normalized Lagrangian gives coefficients $\Phi(\mathbf k)\ell_h^{\mathcal L}(\mathbf k;\boldsymbol\mu)$ on $u_h$ and $q_h\Phi(\mathbf k)\psi_h^{\mathcal L}(\mathbf k;\boldsymbol\mu)$ on $p_h$. The fixed-allocation dual's constant term is $m\sum_{j\ne h}\sum_{\ph\in\mathcal L_j}\mu^{\ph}_{jh}(T_j-T_h)\sum_{\mathbf k}\ph(\mathbf k)p_h(\mathbf k)=G(\mathbf p)-m\sum_{h,\mathbf k}q_h\Phi(\mathbf k)p_h(\mathbf k)\psi_h^{\mathcal L}(\mathbf k;\boldsymbol\mu)$. These identities are the only changes to Steps~2--6 of the proof of Theorem~\ref{thm:dual}. The remaining arguments use feasibility, nonnegative multiplier--slack products, and finite LP duality, and therefore give the robust version of Theorem~\ref{thm:dual}, including its capped statements. The proof of Proposition~\ref{prop:universal} uses only that a cap restricts the program and that the rents \eqref{ec:dsicrent} at $\mathbf p^n$, which cost $S-\RE^{\mathrm{rat}}$, satisfy the incentive constraints; by part~(i) they satisfy \eqref{eq:robustic}, so it applies verbatim. \qed

\subsubsection*{Proof of Proposition~\ref{prop:tv}.}
With $g=g_{jh}$, $\mathrm{sl}_{jh}(\ph_j)-\mathrm{sl}_{jh}(\ph)=\sum_{\mathbf k}\{\ph_j(\mathbf k)-\ph(\mathbf k)\}g(\mathbf k)$. As both laws sum to one, the positive and negative parts of $\ph_j-\ph$ each have mass $\tfrac12\sum_{\mathbf k}|\ph_j(\mathbf k)-\ph(\mathbf k)|$, so this sum is at least $(\min g-\max g)$ times that mass.
Under \eqref{eq:ir} and \eqref{eq:cap}, $0\le u_j\le T_j+\varsigma$ and
$(T_j-T_h)p_h\le u_h+(T_j-T_h)p_h\le(T_j+\varsigma)p_h$.
Hence $-(T_j+\varsigma)\le g\le T_j+\varsigma+\pos{T_h-T_j}$, so $\max g-\min g\le2(T_1+\varsigma)$. \qed

\subsubsection*{Proof of Proposition~\ref{prop:verify}.}
(i) Each $\ph_j(\mathbf k;\alpha)=(m-1)!\prod_t\{\pi_{jt}(\alpha)^{k_t}/k_t!\}$ is a product of $m-1$ affine factors, so the slack is a polynomial of degree at most $m-1$. If the slack is constant, endpoint evaluation suffices; otherwise its minimum occurs at an endpoint or an interior derivative zero, giving at most $m$ evaluations. (ii) The finite-grid program relaxes interval IC; if its optimum passes these polynomial checks, it is interval-feasible by (i) and therefore interval-optimal.

(iii) Under $\boldsymbol\pi_j(\alpha)$, a rival's pure strategy $\xi$ reports $T_h$ with probability $\sum_t\pi_{jt}(\alpha)\,\xi(\{T_h\}\mid T_t)$, affine in $\alpha$, and independent rivals make each configuration probability a sum of products of $m-1$ such factors, so each certificate slack is a polynomial of degree at most $m-1$ and the arguments of (i)--(ii) apply with $\delta$ in place of $0$. Slack is linear in the law, giving the claim on mixtures. \qed

\subsubsection*{Verification of Example~\ref{ex:category}.}
With one rival write $p_{jk}=p_j(\mathbf e_k)$ and $u_{jk}=u_j(\mathbf e_k)$. Set $p_{jj}=1/2$ and $p_{kj}=1-p_{jk}$ for $j<k$, with $(p_{12},p_{13},p_{23})=(163951/167700,1,75833/111800)$. The nonzero rents $(u_{11},u_{12},u_{13},u_{21},u_{23})=(14996/41925,501/100,138192/13975,6939/55900,43261/13975)$ satisfy $0\le u_j\le T_jp_j$ and give the stated revenue. For each candidate, substituting into $\min_{h\ne j}\sum_t\pi_t\min_{k\in\mathcal S_t}g_{jh}(\mathbf e_k)$ yields $1/100$ at each step of $3\prec1\prec2$; linearity covers mixtures. The replication package gives 18 exact dual certificates, one per reserve and whole-type ordering, imposing both laws at zero margin. The resulting bound $87/5$ equals the rational optimum by Lemma~\ref{lem:rational} with scores $(28,8,3)$. \qed

\subsubsection*{Proof of Theorem~\ref{thm:cvband}.}
(i) A truthful value $v\in I_j$ receives $u_j+(v-T_j)p_j$, minimized at $v=T_j$, so participation is equivalent to \eqref{eq:ir}. Under the transfer rule of Section~\ref{sec:model}, when $p_j>0$ the winner pays $T_j-u_j/p_j$ and losers receive nothing; the winner's price is at least $-\varsigma$ exactly when $u_j\le(T_j+\varsigma)p_j$. When $p_j=0$, the buyer receives $u_j$, so zero transfers to losers require $u_j=0$, also exactly the cap under \eqref{eq:ir}. This proves the stated equivalence.

(ii) Against report $I_h$, a value $v\in I_j$ has perceived gain from truth $\mathrm{sl}_{jh}(\ph_j,v)=\sum_{\mathbf{k}}\ph_j(\mathbf{k})\{u_j(\mathbf{k})-u_h(\mathbf{k})\}+(v-T_j)P_{jj}-(v-T_h)P_{jh}$. This affine function is nonnegative throughout $I_j$ exactly when it is nonnegative at both endpoints, including the upper endpoint by continuity. These two linear inequalities give \eqref{ec:cvbandic} and the band LP. \qed

\subsubsection*{Proof of Proposition~\ref{prop:interior}.}
Fix $j\ne h$ and $\ph$, and write $a=\mathrm{sl}_{jh}(\ph,T_j)$ and $b=\mathrm{sl}_{jh}(\ph,T_{j-1})$, which are nonnegative by \eqref{eq:interior}. Since $\mathrm{sl}_{jh}(\ph,v)$ is affine in $v$, every $v\in I_j$ satisfies
\[
 \mathrm{sl}_{jh}(\ph,v)=\frac{(T_{j-1}-v)\,a+(v-T_j)\,b}{T_{j-1}-T_j}\ge\frac{\min\{v-T_j,T_{j-1}-v\}}{T_{j-1}-T_j}\,(a+b)\ge\eta\min\{v-T_j,T_{j-1}-v\},
\]
where the last step is the third inequality in \eqref{eq:interior}.

\emph{Induced laws.} Under subjective independence, a rival $i'$ with strategy $\xi_{i'}$ reports $T_h$, from the viewpoint of a value in $I_j$, with probability $\sum_t\pi_{jt}z_{th}$, where $z_{th}=\int\xi_{i'}(\{T_h\}\mid v')\,F(dv'\mid I_t)$, and rivals' reports are independent. If $\xi_{i'}$ is truthful at $F$-almost every value of the bands in $\mathcal N'$, then $z_{tt}=1$ for $t\in\mathcal N'$. For $t\notin\mathcal N'$, $(z_{th})_h$ is a convex combination of unit vectors, and the configuration law is affine in each rival's vectors; as in Step~1 of the proof of Theorem~\ref{thm:unique}, it lies in $\co(\mathcal F_j(\mathcal N'))$.

(i) The first claim is the display above. Truthful rivals induce $\ph_j\in\mathcal F_j(\mathcal N_{\prec j})$, so $\mathrm{sl}_{jh}(\ph_j,v)\ge0$ at every $v\in I_j$, and truth is an SE at every value. Let $\eta>0$ and take an SE. Suppose that all strategies are truthful at $F$-almost every value of the bands ordered before $j$, as is vacuous for the first band. By the preceding paragraph and linearity, truth beats every other report by at least $\eta\min\{v-T_j,T_{j-1}-v\}>0$ at every interior value $v$ of $I_j$, for every buyer. Best responses at $F$-almost every value of $I_j$ are therefore truthful, and induction along $\prec$ proves the claim.

(ii) Each inequality in \eqref{eq:interior} is linear in $\ph$. Under subjective independence, the ranking argument in Step~3 of the proof of Theorem~\ref{thm:unique}, applied to each of the three inequalities, shows that it suffices to impose them at the at most $|\cK|$ laws in $\mathcal V_j(\mathcal N_{\prec j})$ for each $(j,h)$, which gives at most $3n(n-1)|\cK|$ inequalities. Revenue is bounded by allocated value and the zero mechanism is feasible at $\eta=0$, so each program attains its maximum when feasible. For the mechanism $p_j(\mathbf k)=(n-j)\bar\eta$, $x_j(\mathbf k)=\bar\eta\sum_{t=j}^{n-1}T_t$, with $\bar\eta=1/\{m(n-1)\}$, $p_j\le(n-1)\bar\eta=1/m$ gives \eqref{eq:FEA}, $x_j\ge0$ gives \eqref{eq:cap}, and $u_j=T_jp_j-x_j=\bar\eta\sum_{t=j}^{n-1}(T_j-T_t)\ge0$ gives \eqref{eq:ir}. At every configuration, a value $v$ gains $\bar\eta\sum_{t=j}^{h-1}(v-T_t)$ from truth over a lower report $T_h$, $h>j$, and $\bar\eta\sum_{t=h}^{j-1}(T_t-v)$ over a higher report, $h<j$. At $v\in\{T_j,T_{j-1}\}$ both sums are nonnegative, and the first is at least $\bar\eta(T_{j-1}-T_j)$ at $v=T_{j-1}$, the second at $v=T_j$. Hence \eqref{eq:interior} holds with $\eta=\bar\eta$ against every law and in every ordering. Let $\mathcal R^{\mathcal I}(\eta)$ be the largest band revenue under \eqref{eq:FEA}, \eqref{eq:ir}, and \eqref{eq:interior}, maximized over orderings. Mixing this mechanism with weight $\eta/\bar\eta$ into an optimal mechanism at coefficient zero, as in the proof of Proposition~\ref{prop:margincost}, gives $0\le\mathcal R^{\mathcal I}(0)-\mathcal R^{\mathcal I}(\eta)\le(\eta/\bar\eta)\{\mathcal R^{\mathcal I}(0)-\operatorname{Rev}^\circ\}$ for $0<\eta\le\bar\eta$, where $\operatorname{Rev}^\circ$ is this mechanism's revenue, and likewise under \eqref{eq:cap}. For a schedule, replace $\mathcal F_j(\mathcal N_{\prec j})$ by $\mathcal F_{j_\tau}(\mathcal S^{\tau-1})$. The same affine-slack argument and the induction of Step~2 of the proof of Theorem~\ref{thm:unique}, applied at almost every value, give the truthfulness and uniqueness conclusions. The ranking argument represents the certificate by at most $3\sum_{\tau=1}^{n(n-1)}|\mathcal V_{j_\tau}(\mathcal S^{\tau-1})|$ linear inequalities, so the revenue problem is a finite LP. The bound $|\mathcal V_j(\mathcal N_{\prec j})|\le|\cK|$ used above is specific to whole-type steps. The witness above satisfies \eqref{eq:interior} against every law, so the same mixing argument yields an $O(\eta)$ revenue cost for each schedule and after maximization over schedules, also under \eqref{eq:cap}. \qed

\subsubsection*{Proof of Lemma~\ref{lem:bandreserve}.}
For a band mechanism $(\mathbf p^r,\mathbf u)$, the advantages $g_{jh}(\mathbf k,v)$ of Section~\ref{sec:continuous} satisfy the slope identity $g_{jh}(\mathbf k,T_{j-1})-g_{jh}(\mathbf k,T_j)=(T_{j-1}-T_j)\{p^r_j(\mathbf k)-p^r_h(\mathbf k)\}$. Suppose, to the contrary, that \eqref{eq:interior} holds in $\prec$ with $\eta>0$; truthful rivals induce $\ph\in\mathcal F_j(\mathcal N_{\prec j})$ for every $j$.

\emph{Step 1 (boundary indifference).} For $j<n$ and every $\mathbf k$, $g_{j+1,j}(\mathbf k,T_j)=u_{j+1}(\mathbf k)+(T_j-T_{j+1})p^r_{j+1}(\mathbf k)-u_j(\mathbf k)=-g_{j,j+1}(\mathbf k,T_j)$. The first inequality of \eqref{eq:interior} for the pair $(j,j+1)$ at the floor $T_j$ of $I_j$, and the second for $(j+1,j)$ at the top $T_j$ of $I_{j+1}$, both at $\ph$, give $\mathrm{sl}_{j,j+1}(\ph,T_j)\ge0\ge\mathrm{sl}_{j,j+1}(\ph,T_j)$, so $\mathrm{sl}_{j,j+1}(\ph,T_j)=0$.

\emph{Step 2 (state by state).} Let $j_1$ be the first element of $\prec$. Rivals who report a fixed band at every value are truthful on $\mathcal N_{\prec j_1}=\varnothing$, so $\mathcal F_{j_1}(\varnothing)$ contains the point mass at every $\mathbf k$, and \eqref{eq:interior} for $j_1$ holds state by state. Let $h=j_1+1$ if $j_1<n$ and $h=n-1$ if $j_1=n$, let $v_0$ be the common endpoint of $I_{j_1}$ and $I_h$, and let $v_1$ be the other endpoint of $I_{j_1}$. By the first two inequalities of \eqref{eq:interior} at point masses, $g_{j_1h}(\mathbf k,v_0)\ge0$ for every $\mathbf k$; by Step~1, $\sum_{\mathbf k}\ph(\mathbf k)g_{j_1h}(\mathbf k,v_0)=0$, since $g_{j_1h}(\cdot,v_0)=\pm g_{a,a+1}(\cdot,v_0)$ with $\{a,a+1\}=\{j_1,h\}$. As $\ph(\mathbf k)>0$, $g_{j_1h}(\mathbf k,v_0)=0$ for every $\mathbf k$. The third inequality at point masses then gives $g_{j_1h}(\mathbf k,v_1)\ge\eta(T_{j_1-1}-T_{j_1})$, and the slope identity yields the gap $p^r_a(\mathbf k)-p^r_{a+1}(\mathbf k)\ge\eta$ for every $\mathbf k$, because $v_1-v_0=T_{j_1-1}-T_{j_1}$ and $(a,a+1)=(j_1,h)$ if $h=j_1+1$, while $v_1-v_0=-(T_{n-1}-T_n)$ and $(a,a+1)=(h,j_1)$ if $j_1=n$.

\emph{Step 3 (contradiction).} By \eqref{eq:reserve}, $p^r_t(\mathbf k)=0$ if $t>r$ or $k_{t'}\ge1$ for some $t'<t$. If $a\ge2$, take $\mathbf k$ with $k_1\ge1$, possible as $m\ge2$: then $p^r_a(\mathbf k)=p^r_{a+1}(\mathbf k)=0$, contradicting the gap of Step~2. Otherwise $a=1$, so $j_1=1$ and $h=2$. If $r\ge2$, take $\mathbf k=(m-1)\mathbf e_n$: as $n\ge3$, reports $T_1$ and $T_2$ both face only lower rivals and are served, so $p^r_1(\mathbf k)=p^r_2(\mathbf k)=1$, contradicting the gap. If $r=1$, then $p^r_2\equiv p^r_3\equiv0$, so $g_{23}(\cdot,v)=u_2-u_3=-g_{32}(\cdot,v)$ for every $v$; as $\ph\in\mathcal F_2(\mathcal N_{\prec2})\cap\mathcal F_3(\mathcal N_{\prec3})$, the first inequality of \eqref{eq:interior} gives $\sum_{\mathbf k}\ph(\mathbf k)(u_2-u_3)(\mathbf k)\ge0$ and $\le0$, so it vanishes, whereas the third inequality for $(2,3)$ at $\ph$ requires $2\sum_{\mathbf k}\ph(\mathbf k)(u_2-u_3)(\mathbf k)\ge\eta(T_1-T_2)>0$. Each case is a contradiction, and adding \eqref{eq:cap} only shrinks the feasible set. \qed

%\subsection{Supporting Reductions and Allocation Bounds}\label{app:proofsstructure}
\subsubsection*{Proof of Proposition~\ref{prop:bandinterval}.}
\emph{Revenue bound.} Write $\ph(\cdot;\kappa)$ for the multinomial rival law with probabilities $\boldsymbol\pi(\kappa)$ and $\E_\Phi[f]=\sum_{\mathbf k}\Phi(\mathbf k)f(\mathbf k)$. At $T_j$, the adjacent gains $g_{j,j+1}(\mathbf k,T_j)$ and $g_{j+1,j}(\mathbf k,T_j)$ are opposites. By Theorem~\ref{thm:cvband}, both expected gains are nonnegative at this boundary under the same law, so $\mathrm{sl}_{j,j+1}(\ph(\cdot;\kappa),T_j)=0$ for $j=1,2$ and every $\kappa\in[0,\bar\kappa]$. This slack is a polynomial of degree at most $m-1$ in $\kappa$, as $\ph(\mathbf k;\kappa)$ is a product of $m-1$ affine factors times a multinomial coefficient and the contract is fixed. It vanishes identically and in particular at $\kappa=1$, where $\ph(\cdot;1)=\Phi$. Thus $\E_\Phi[u_j]=\E_\Phi[u_3]+\sum_{h>j}(T_{h-1}-T_h)\E_\Phi[p_h]$. Substitution gives
\[
 \operatorname{Rev}(\mathbf p,\mathbf u)
 =m\sum_jq_j b_j^{\mathrm{rat}}\E_\Phi[p_j]-m\E_\Phi[u_3]\le\mathcal R_B .
\]
Indeed, $\E_\Phi[u_3]\ge0$ and the first term is the expected allocated score. Profile by profile, \eqref{eq:FEA} bounds that score by the largest positive score present, or zero if there is none. The assumed score order makes $\mathbf p^2$ attain this bound.

\emph{Attainment and certificate.} Write $p_j=p_j^2$ and $d=T_1-T_2$ in this paragraph. By \eqref{eq:reserve}, $p_1\ge1/m$, $p_2=0$ if $k_1>0$ and $p_2\ge1/m$ if $k_1=0$, and $p_3=0$. Also $p_1\ge p_2$, so $0\le u_1=dp_2\le T_1p_1$. The adjacent equalities hold state by state and $u_3=0$. The six statewise gain functions are
\[
\begin{array}{c|l@{\qquad}c|l}
(j,h)&g_{jh}(\mathbf k,v)&(j,h)&g_{jh}(\mathbf k,v)\\
(1,3)&dp_2+(v-T_1)p_1 &(3,1)&(T_2-v)p_1+d(p_1-p_2)\\
(1,2)&(v-T_1)(p_1-p_2)& (2,1)&(T_1-v)(p_1-p_2)\\
(2,3)&(v-T_2)p_2 &(3,2)&(T_2-v)p_2.
\end{array}
\]
They are nonnegative on their respective closed bands. In the three rows, the endpoint sums divided by the corresponding band widths are bounded below by $p_1$, $p_1-p_2$, and $p_2$, respectively. For $(1,2)$, every band-1 rival can report only $1$ or $2$. On the event that all rivals are in band 1, either some report is $1$ and $p_1-p_2=p_1\ge1/m$, or all report $2$ and $p_1-p_2=1-1/m\ge1/m$. For $(2,1)$ the same event gives the bound because band-1 rivals then report truthfully. For $(2,3)$, all rivals in band 2 can report only $2$ or $3$, so when all are in band 2, $p_2\ge1/m$. For $(3,2)$, the same bound holds when none are in band 1 because both band-2 and band-3 rivals can report only $2$ or $3$. The respective event probabilities are $\pi_1(\kappa)^{m-1}$, $\pi_2(\kappa)^{m-1}$, and $[1-\pi_1(\kappa)]^{m-1}$. These bounds hold for each admissible map and hence for mixtures. Averaging gives \eqref{eq:interior} for the stated $\mathcal E$ and $\eta$. Truthful rivals are admissible at every deletion, so this certificate implies the truthfulness required for the upper bound, and the reserve reaches it. Proposition~\ref{prop:interior}'s report-by-report extension gives uniqueness almost everywhere at positive $\eta$. \qed

\subsubsection*{Proof of Theorem~\ref{thm:compression}.}
(i)--(ii) Fix $\mathbf p\in\cP$; let $c_{jh}=(T_j-T_h)P_{jh}(\mathbf p)$. \emph{Step 1 (cost bound).} As $\ph_j\le\Lambda_j\Phi$ and $u_j\ge0$, $\sum_{\mathbf k}\Phi(\mathbf k)u_j(\mathbf k)\ge\Shat_j/\Lambda_j$, with equality exactly when $u_j$ vanishes where $\Lambda_j(\mathbf k)<\Lambda_j$.

\emph{Step 2 (leakage bound).} As $\ph_j\ge\rh_{jh}\ph_h$ pointwise, $\sum_{\mathbf k}\ph_j(\mathbf k)u_h(\mathbf k)\ge\rh_{jh}\Shat_h$, so \eqref{eq:ic} implies
\begin{equation}\label{ec:levels}
 \Shat_j\ge\rh_{jh}\Shat_h+c_{jh}\quad(j\ne h),\qquad \Shat_j\ge0.
\end{equation}

\emph{Step 3 (attainment).} For any $\widehat{\boldsymbol U}$ satisfying \eqref{ec:levels}, set $u_h(\mathbf k)=\Shat_h\ind\{\mathbf k=\mathbf d_h\}/\ph_h(\mathbf d_h)$. Alignment gives $\sum_{\mathbf k}\ph_j(\mathbf k)u_h(\mathbf k)=\rh_{jh}\Shat_h$ for $j\ne h$ and equality in the cost bound. This schedule satisfies \eqref{eq:ic}--\eqref{eq:ir}, delivers $\widehat{\boldsymbol U}$, and attains its reduced objective. With Steps~1--2, this proves exact reduction.

\emph{Step 4 (the fixed point).} The map \eqref{eq:fixedpoint} is monotone and a contraction in the supremum norm with modulus $\bar\rh<1$, since taking a maximum is nonexpansive. Its iterates from zero converge to its unique fixed point $\widehat{\boldsymbol U}^*=\widehat{\boldsymbol U}^*(\mathbf p)$. A vector $\mathbf v$ satisfies \eqref{ec:levels} exactly when $\mathbf v\ge0$ and $\boldsymbol\Gamma_{\mathbf p}(\mathbf v)\le\mathbf v$. Monotonicity then bounds every iterate, and hence $\widehat{\boldsymbol U}^*$, by $\mathbf v$. Thus the fixed point is the componentwise smallest feasible vector and the unique reduced optimum because all weights $q_j/\Lambda_j$ are positive.

\emph{Step 5 (uniqueness).} For any optimal schedule, $m\sum_jq_j\sum_{\mathbf k}\Phi(\mathbf k)u_j(\mathbf k)\ge m\sum_j(q_j/\Lambda_j)\Shat_j\ge m\sum_j(q_j/\Lambda_j)\Shat_j^*$, and optimality makes both inequalities equalities. Positivity of the weights and $\widehat{\boldsymbol U}\ge\widehat{\boldsymbol U}^*$ give $\Shat_j=\Shat_j^*$ and equality in the cost bound for every $j$. Under strict-in-cost alignment, this forces the concentrated schedule of Step~3, and hence the unique payment rule.

\emph{Step 6 (endogenous allocation).} Steps~1--4 apply to every feasible allocation, making each implementable and identifying its minimum rent with \eqref{eq:reduced}. The constraints of \eqref{eq:reduced}, equivalent to $\widehat{\boldsymbol U}\ge\boldsymbol\Gamma_{\mathbf p}(\widehat{\boldsymbol U})$, are finitely many linear inequalities in $(\mathbf p,\widehat{\boldsymbol U})$, so maximizing revenue over $\cP$ proves (ii).

(iii) The concentrated optimum $\mathbf u^*$ of part~(i) satisfies \eqref{eq:cap} exactly when $\Shat_j^*\le\ph_j(\mathbf d_j)(T_j+\varsigma)p_j(\mathbf d_j)$ for all $j$. These inequalities therefore suffice for $\Rp^\varsigma(\mathbf p)=\Rp(\mathbf p)$. Under strict-in-cost alignment, equality of the two values makes every capped optimum an unrestricted optimum, necessarily $\mathbf u^*$ by Step~5, so the inequalities are also necessary. \qed

\subsubsection*{Proof of Corollary~\ref{cor:recursion}.}
Let $\mathbf p$ be monotone, write $P_{jh}=P_{jh}(\mathbf p)$ and $c_{jh}=(T_j-T_h)P_{jh}$, and set $\rh_{jj}=1$. The pointwise leakage bounds, their products, and allocation monotonicity imply, respectively, $\rh_{jh}P_{ht}\le P_{jt}$, $\rh_{jh}\rh_{ht}\le\rh_{jt}$, and $P_{jt}\ge P_{jt'}$ for $t<t'$. Define $v_n=0$ and $v_j=\max_{h>j}(\rh_{jh}v_h+c_{jh})$. These values are nonnegative and satisfy all downward constraints. To prove the upward constraint for $h<j$, induct on $j-h$ and choose $t>h$ attaining the recursion for $v_h$. The preceding inequalities give
 $\rh_{jh}v_h+c_{jh}
 \le\rh_{jt}v_t+(T_h-T_t)P_{jt}+(T_j-T_h)P_{jh}
 =\rh_{jt}v_t+(T_j-T_t)P_{jt}+(T_j-T_h)(P_{jh}-P_{jt})
 \le\rh_{jt}v_t+(T_j-T_t)P_{jt}\le v_j$.
The last inequality is equality for $t=j$, follows from the recursion for $t>j$, and from the induction hypothesis for $h<t<j$ (absent when $j-h=1$). Thus $\mathbf v$ satisfies \eqref{ec:levels}; conversely, downward induction makes every feasible vector at least $\mathbf v$. Step~4 of the proof of Theorem~\ref{thm:compression} gives $\mathbf v=\widehat{\boldsymbol U}^*(\mathbf p)$, which is \eqref{eq:downward}; one maximizing downward arc per positive-rent type gives an acyclic supporting graph, so Proposition~\ref{prop:settlement} applies.

For reserve $T_r$, $P_{jh}(\mathbf p^r)=P_{jh}(\mathbf p^n)\ind\{h\le r\}$. Downward induction gives zero rents for $j\ge r$ and \eqref{eq:truncated} for $j<r$; those rents are positive since $(T_j-T_r)P_{jr}(\mathbf p^n)>0$ by full support. Symmetry at profiles with highest value $T_h$ gives $mq_h\bar p_h=w_h$, hence $G(\mathbf p^r)=\sum_{h\le r}w_hT_h$ and the revenue formula in \eqref{eq:truncated}. Under subjective independence, the same count gives $P_{jh}(\mathbf p^n)=\{(\sum_{t\ge h}\pi_{jt})^m-(\sum_{t>h}\pi_{jt})^m\}/(m\pi_{jh})$, computable for all pairs in $O(n^2)$ operations using cumulative row sums. \qed

\subsubsection*{Proof of Corollary~\ref{cor:projection}.}
\emph{Step 1 (alignment and payments).} For $\alpha>0$, $\pi_{jt}/q_t$ equals $\pi_{jj}/q_j>1$ at $t=j$ and $\omega<1$ elsewhere, giving the unique cost peak $\mathbf D_j$ and $\Lambda_j=(\pi_{jj}/q_j)^{m-1}$. For $j\ne h$, the per-rival imitation ratio is uniquely minimized at $t=h$, with value $\omega q_h/\pi_{hh}$; its product gives $\rh_h$ at $\mathbf D_h$. This proves strict-in-cost alignment.

For $h>j$, every winning configuration of report $T_h$ has $k_j=0$, so $P_{jh}(\mathbf p^n)=\omega^{m-1}\bar p_h$ and $\rh_h\Lambda_h=\omega^{m-1}$. Put $\zeta_j=\Shat_j^r/\omega^{m-1}$, so $U_j^r=\rh_j\zeta_j$. Dividing \eqref{eq:truncated} by $\omega^{m-1}$ gives $\zeta_j=\max_{j<h\le r}\{U_h^r+(T_j-T_h)\bar p_h\}$ for $j<r$ and $\zeta_j=0$ for $j\ge r$. Theorem~\ref{thm:compression}, $\ph_j(\mathbf D_j)=\Lambda_jq_j^{m-1}$, and $p_j^r(\mathbf D_j)=1/m$ for served reports give \eqref{eq:projrec} and uniqueness.

\emph{Step 2 (rent and approximation bounds).} At $\alpha=0$, $\rh_j=1$ and the recursion gives
\begin{equation}\label{ec:rationalrents}
 U_j^{\mathrm{rat},r}=\sum_{h=j+1}^r(T_{h-1}-T_h)\bar p_h\quad(j<r),\qquad U_j^{\mathrm{rat},r}=0\quad(j\ge r),
\end{equation}
the rational rents, which by Theorem~\ref{thm:common}(ii), with $\Lambda_j\equiv1$ and $P_h(\mathbf p^r)=\bar p_h\ind\{h\le r\}$, are the optimal objective rents at $\mathbf p^r$ under correct beliefs; hence $S-\RE^{\mathrm{rat}}=m\sum_{j<n}q_jU_j^{\mathrm{rat},n}$. Downward induction in this recursion yields $0\le\zeta_j\le U_j^{\mathrm{rat},r}$, proving $U_j^r\le\rh_jU_j^{\mathrm{rat},r}$. Each $\rh_j$ is continuous and strictly decreasing from one to zero. Inductively, the maximum defining $\zeta_j$ is continuous, nonincreasing, and bounded below by $(T_j-T_r)\bar p_r>0$ for $j<r$; hence $U_j^r=\rh_j\zeta_j$ is continuous and strictly decreasing on $[0,1)$.

Since $U_n^n=0$, this bound and \eqref{ec:rationalrents} give $S-\RE=m\sum_{j<n}q_jU_j^n\le\bar\rh_Pm\sum_{j<n}q_jU_j^{\mathrm{rat},n}=\bar\rh_P(S-\RE^{\mathrm{rat}})$, and $\RE\le\RR\le\Rstar\le S$ gives the stated chain. As $\alpha\uparrow1$, $\bar\rh_P\to0$, so each rebate in \eqref{eq:projrec} with $r=n$ tends to zero. All payments are eventually nonnegative. \qed

\subsubsection*{Proof of Proposition~\ref{prop:settlement}.}
Write $\boldsymbol\Gamma=\boldsymbol\Gamma_{\mathbf{p}}$, $\widehat{\boldsymbol U}^*=\widehat{\boldsymbol U}^*(\mathbf p)$ for its unique fixed point, and $c_{jh}=(T_j-T_h)P_{jh}(\mathbf p)$. Full support gives $\rh_{jh}>0$.

\emph{Step 1 (the sweep).} Monotonicity and nonnegativity of $\boldsymbol\Gamma$ give $\mathbf0=\widehat{\boldsymbol U}^{(0)}\le\widehat{\boldsymbol U}^{(1)}\le\widehat{\boldsymbol U}^{(2)}\le\cdots\le\widehat{\boldsymbol U}^*$. Equal consecutive iterates are a fixed point, so $\widehat{\boldsymbol U}^{(n)}=\widehat{\boldsymbol U}^{(n-1)}$ is equivalent to reaching it by round $n-1$. Once a component reaches its limit it stays there. When the sweep settles, define $t_j=\min\{t:\Shat_j^{(t)}=\Shat_j^*\}$; then $t_j=0$ exactly for zero-rent types.

\emph{Step 2 (supporting arcs).} For a positive-rent type, choose $h\ne j$ attaining the maximum at its first settling round $t_j\ge1$. Then $\Shat_j^*=\rh_{jh}\Shat_h^{(t_j-1)}+c_{jh}\le\rh_{jh}\Shat_h^*+c_{jh}\le\Shat_j^*$. Equality and $\rh_{jh}>0$ imply that $j\to h$ is tight and $t_h<t_j$.

\emph{Step 3 (equivalence).} Given $\widehat{\boldsymbol U}^{(n)}=\widehat{\boldsymbol U}^{(n-1)}$ (that is, the sweep settles), select the arcs of Step~2 at positive-rent types and none at zero-rent types; settling rounds strictly decrease along them, proving the equivalent condition in Proposition \ref{prop:settlement}. Conversely, let $\mathcal G'$ satisfy it, and let $L(j)\le n-1$ be the longest path length from $j$. Its sinks have zero rent. Induct on $L(j)$: at sinks, Step~1 fixes rent at zero; otherwise a tight arc $j\to h$ has $L(h)<L(j)$, so for $t\ge L(j)$,
$\Shat_j^{(t)}\ge\rh_{jh}\Shat_h^{(t-1)}+c_{jh}=\Shat_j^*$,
with the reverse inequality from Step~1. Thus $\widehat{\boldsymbol U}^{(n-1)}=\widehat{\boldsymbol U}^*$, proving $\widehat{\boldsymbol U}^{(n)}=\widehat{\boldsymbol U}^{(n-1)}$. Monotone allocations satisfy the equivalent condition by the proof of Corollary~\ref{cor:recursion}.

\emph{Step 4 (the recursion).} $\widehat{\boldsymbol U}^{(n)}=\widehat{\boldsymbol U}^{(n-1)}$ gives $\widehat{\boldsymbol U}^*$ and the settling rounds of Step~1. The graph $\mathcal G$ is acyclic because $t_h<t_j$ along its arcs. Zero-rent types have $t_j=0$ and no outgoing arcs, while every positive-rent type has a tight arc in $\mathcal G$ by Step~2; hence the sinks of $\mathcal G$ are the zero-rent types. At each nonsink, every bracket in \eqref{eq:settlingrecursion} is at most $\Shat_j^*$ by the fixed-point property, and one is equal, so $\widehat{\boldsymbol U}^*$ solves \eqref{eq:settlingrecursion}; induction in increasing $t_j$ proves uniqueness, since all successors of $j$ in $\mathcal G$ settle earlier.

In the two-type example of Section~\ref{sec:settlement}, $P_{12}(\mathbf p)=1/m$ and $P_{21}(\mathbf p)=0$ make $\boldsymbol\Gamma$ injective on $\mathbb R^2_+$ with $\boldsymbol\Gamma(\mathbf0)\ne\mathbf0$, so $\widehat{\boldsymbol U}^{(t)}=\widehat{\boldsymbol U}^{(t-1)}$ would propagate back to $\boldsymbol\Gamma(\mathbf0)=\mathbf0$. \qed

%\emph{Step 5 (stopped iterates and cyclic dependence).} The triangle inequality and the contraction bound on $\boldsymbol\Gamma_{\mathbf p}(\mathbf v)-\boldsymbol\Gamma_{\mathbf p}(\widehat{\boldsymbol U}^*)$ give the residual bound $\norm{\mathbf v-\widehat{\boldsymbol U}^*}_\infty\le\norm{\mathbf v-\boldsymbol\Gamma_{\mathbf p}(\mathbf v)}_\infty/(1-\bar\rh)$. For the stopped iterate, let $\mathbf w=(v_j+d)_{j\in\mathcal N}$, with $d$ as in Section~\ref{sec:settlement}. As $\boldsymbol\Gamma_{\mathbf p}$ is monotone with slopes at most $\bar\rh$, each component of $\boldsymbol\Gamma_{\mathbf p}(\mathbf w)$ exceeds that of $\boldsymbol\Gamma_{\mathbf p}(\mathbf v)$ by at most $\bar\rh d$, so $\boldsymbol\Gamma_{\mathbf p}(\mathbf w)\le\mathbf w$: $\mathbf w\ge\mathbf0$ satisfies the constraints of \eqref{eq:reduced}, and its concentrated payments satisfy \eqref{eq:ic} by Theorem~\ref{thm:compression}. By the residual bound, $\widehat{\boldsymbol U}^*\ge\mathbf v-d$ componentwise, so $w_j-\Shat_j^*\le2d$, and the revenue loss $m\sum_jq_j(w_j-\Shat_j^*)/\Lambda_j$ is at most $2dm\sum_jq_j/\Lambda_j$. In the two-type example of Section~\ref{sec:settlement}, $P_{12}(\mathbf p)=1/m$ and $P_{21}(\mathbf p)=0$ make $\boldsymbol\Gamma$ injective on $\mathbb R^2_+$ with $\boldsymbol\Gamma(\mathbf0)\ne\mathbf0$, so $\widehat{\boldsymbol U}^{(t)}=\widehat{\boldsymbol U}^{(t-1)}$ would propagate back to $\boldsymbol\Gamma(\mathbf0)=\mathbf0$. \qed

\subsubsection*{Verification of Example~\ref{ex:settlement}.}
For each $j$, $\Lambda_j(\mathbf k)$ equals $3/2$ at $\mathbf k=\mathbf e_j$ and $3/4$ elsewhere, and $\ph_j/\ph_h$ ($j\ne h$) is minimized at $\mathbf e_h$ with value $1/2$. Feasibility follows from $p_j\le1/2$. In \eqref{eq:sweep}, $c_{jh}=(T_j-T_h)p_h$ gives the stated iterates: in rounds two and three, the positive maxima occur only on $1\to2$ and $2\to3$, with values $5/8$ and $1/4$. Theorem~\ref{thm:compression} gives rents $5/4$ and $1/2$ at $\mathbf e_1$ and $\mathbf e_2$ for reports 1 and 2, leaving payments $1/4$ and $1$ at those entries and $T_jp_j$ elsewhere. \qed

\subsubsection*{Proof of Theorem~\ref{thm:dual}.}
If $k_h+\ind\{h=j\}>0$ and $\mathbf k'=\mathbf k+\mathbf e_j-\mathbf e_h$, the multinomial formula gives $q_j\Phi(\mathbf k)(k_h+\ind\{h=j\})=q_h\Phi(\mathbf k')(k'_j+\ind\{j=h\})$. Multiplying by any $f_h(\mathbf k')$ and summing gives
\begin{equation}\label{ec:regroupsum}
 \sum_jq_j\sum_{\mathbf k}\Phi(\mathbf k)\sum_h(k_h+\ind\{h=j\})f_h(\mathbf k+\mathbf e_j-\mathbf e_h)
 =m\sum_hq_h\sum_{\mathbf k}\Phi(\mathbf k)f_h(\mathbf k),
\end{equation}
because $(j,\mathbf k,h)\mapsto(h,\mathbf k',j)$ is an involution and the profile contains $m$ buyers.
Attach $m\mu_{jh}\ge0$ to \eqref{eq:ic} and $q_j\Phi(\mathbf k)\chi_j(\mathbf k)\ge0$ to \eqref{eq:FEA}. Constraints describing the same full profile coincide, so take $\chi_j(\mathbf k)=\chi_h(\mathbf k+\mathbf e_j-\mathbf e_h)$ whenever $k_h+\ind\{h=j\}>0$. The dual of \eqref{eq:primal} can then be written as
\begin{equation}\label{eq:dualLP}
\begin{aligned}
\min_{\boldsymbol\mu\ge0,\,\boldsymbol\chi\ge0}\quad&\sum_jq_j\sum_{\mathbf{k}}\Phi(\mathbf{k})\chi_j(\mathbf{k})\\
 \text{subject to}\quad&\chi_j(\mathbf{k})\ge\psi_j(\mathbf{k};\boldsymbol\mu),\ \ \ell_j(\mathbf{k};\boldsymbol\mu)\le0,&&j\in\mathcal N,\ \mathbf{k}\in\cK,\\
 &\chi_j(\mathbf{k})=\chi_h(\mathbf{k}+\mathbf{e}_j-\mathbf{e}_h),&&j,h\in\mathcal N,\ \mathbf{k}\in\cK,\ k_h+\ind\{h=j\}>0.
\end{aligned}
\end{equation} With $\mathrm{sl}_{jh}=\mathrm{sl}_{jh}(\ph_j)$, every $(\mathbf p,\mathbf u)$ and $\boldsymbol\mu\ge0$ satisfy
\begin{equation}\label{eq:lagidentity}
 \frac{\operatorname{Rev}(\mathbf{p},\mathbf{u})}m+\sum_{j,h:\,j\ne h}\mu_{jh}\mathrm{sl}_{jh}
 =\sum_{h,\mathbf{k}}q_h\Phi(\mathbf{k})\psi_h(\mathbf{k};\boldsymbol\mu)p_h(\mathbf{k})
 +\sum_{h,\mathbf{k}}\Phi(\mathbf{k})\ell_h(\mathbf{k};\boldsymbol\mu)u_h(\mathbf{k}).
\end{equation}

\emph{Step 1 (the identity \eqref{eq:lagidentity}).} On its left, the coefficients of $p_h(\mathbf k)$ and $u_h(\mathbf k)$ are, respectively, $q_h\Phi(\mathbf k)T_h-\sum_{j\ne h}\mu_{jh}\ph_j(\mathbf k)(T_j-T_h)=q_h\Phi(\mathbf k)\psi_h(\mathbf k;\boldsymbol\mu)$ and $-q_h\Phi(\mathbf k)+\ph_h(\mathbf k)M_h-\sum_{j\ne h}\mu_{jh}\ph_j(\mathbf k)=\Phi(\mathbf k)\ell_h(\mathbf k;\boldsymbol\mu)$, proving the identity. \emph{Step 2 (an upper bound).} For $\boldsymbol\mu\in\cD$ and feasible $(\mathbf p,\mathbf u)$, nonnegative IC slacks, \eqref{eq:lagidentity}, and $\ell_h\le0$ give
\begin{equation}\label{ec:dualchain}
 \operatorname{Rev}(\mathbf p,\mathbf u)
 \le\operatorname{Rev}(\mathbf p,\mathbf u)+m\sum_{j\ne h}\mu_{jh}\mathrm{sl}_{jh}
 \le m\sum_{h,\mathbf k}q_h\Phi(\mathbf k)\psi_h(\mathbf k;\boldsymbol\mu)p_h(\mathbf k)
 \le H(\boldsymbol\mu).
\end{equation}
By \eqref{ec:regroupsum} with $f_h=p_h\psi_h$, $H(\boldsymbol\mu)$ minus the penultimate expression equals
\begin{equation}\label{eq:regret}
 \sum_jq_j\sum_{\mathbf{k}}\Phi(\mathbf{k})\Bigl[\chi_j(\mathbf{k};\boldsymbol\mu)-\sum_h(k_h+\ind\{h=j\})\,p_h(\mathbf{k}+\mathbf{e}_j-\mathbf{e}_h)\,\psi_h(\mathbf{k}+\mathbf{e}_j-\mathbf{e}_h;\boldsymbol\mu)\Bigr],
\end{equation}
with terms having $k_h+\ind\{h=j\}=0$ omitted. Each bracket is nonnegative: all scores are at most the profile maximum $\chi\ge0$, while their allocation weights are nonnegative and sum to at most one by \eqref{eq:FEA}. This proves the last inequality and $\Rstar\le H(\boldsymbol\mu)$.

\emph{Step 3 (the dual and strong duality).} By the multinomial identity, feasibility pricing subtracts $mq_h\Phi(\mathbf k)\bar\chi_h(\mathbf k)$ from the coefficient of $p_h(\mathbf k)$, where $\bar\chi_h(\mathbf k)=m^{-1}\sum_j(k_j+\ind\{j=h\})\chi_j(\mathbf k+\mathbf e_h-\mathbf e_j)$. This average depends only on the full profile; replacing $\boldsymbol\chi$ by $\bar{\boldsymbol\chi}$ preserves all variable coefficients and, by \eqref{ec:regroupsum}, the dual constant, while profile-constant arrays are fixed by averaging. Thus the equality constraints of \eqref{eq:dualLP} impose no loss. Step~1 gives coefficients $mq_h\Phi(\mathbf k)(\psi_h-\chi_h)$ for allocations and $m\Phi(\mathbf k)\ell_h$ for rents. Their nonpositivity is exactly dual feasibility; the dual objective is its constant term. Hence \eqref{eq:dualLP} is the dual LP and attains $\Rstar$, as the primal is feasible and bounded. For each $\boldsymbol\mu\in\cD$, the smallest feasible profile-constant $\boldsymbol\chi$ is the maximum in $\chi_j(\mathbf{k};\boldsymbol\mu)$. Eliminating $\boldsymbol\chi$ gives the attained minimum $\Rstar=\min_{\boldsymbol\mu\in\cD}H(\boldsymbol\mu)$, the uncapped part of~(i).

\emph{Step 4 (a fixed allocation).} The minimum-rent LP for implementable $\mathbf p$ is feasible and bounded below. Its dual maximizes $m\sum_{j\ne h}\mu_{jh}(T_j-T_h)P_{jh}(\mathbf p)$ over $\cD$, and $G(\mathbf p)-m\sum_{j\ne h}\mu_{jh}(T_j-T_h)P_{jh}(\mathbf p)=m\sum_hq_h\sum_{\mathbf k}\Phi(\mathbf k)p_h(\mathbf k)\psi_h(\mathbf k;\boldsymbol\mu)$, so LP duality gives the uncapped identity in \eqref{eq:capdual}. By complementary slackness, its nonempty set of minimizers is $\cD(\mathbf p)=\{\boldsymbol\mu\in\cD:\ \mu_{jh}\mathrm{sl}_{jh}(\ph_j)=0\ (j\ne h),\ \ell_h(\mathbf k;\boldsymbol\mu)u_h(\mathbf k)=0\ (h\in\mathcal N,\ \mathbf k\in\cK)\}$ for any optimal rent schedule $\mathbf u$, independently of its choice. At these multipliers the IC-slack and rent terms in \eqref{eq:lagidentity} vanish, yielding
\begin{equation}\label{eq:virtualidentity}
 \Rp(\mathbf{p})=m\sum_jq_j\sum_{\mathbf{k}}\Phi(\mathbf{k})p_j(\mathbf{k})\psi_j(\mathbf{k};\boldsymbol\mu),\qquad\boldsymbol\mu\in\cD(\mathbf p).
\end{equation}

\emph{Step 5 (subsidy caps).} Attach multipliers $m\Phi(\mathbf k)\nu_h(\mathbf k)\ge0$ to $(T_h+\varsigma)p_h(\mathbf k)-u_h(\mathbf k)\ge0$. In the normalized identity \eqref{eq:lagidentity}, the rent and allocation coefficients become
$\Phi(\mathbf k)(\ell_h-\nu_h)$ and
$q_h\Phi(\mathbf k)\{\psi_h+(T_h+\varsigma)\nu_h/q_h\}$.
Thus dual feasibility requires $\nu_h\ge\pos{\ell_h}$. Steps~2--3 give the global dual with these modified scores in $H$. As $T_h+\varsigma>0$, its objective is nondecreasing in every $\nu_h$, so the smallest admissible choice $\nu_h=\pos{\ell_h}$ is optimal, yielding the stated formula.

At a fixed $\mathbf p$ implementable under \eqref{eq:cap}, the rent-minimizing primal is feasible and bounded. Its dual objective is $m\sum_{j\ne h}\mu_{jh}(T_j-T_h)P_{jh}(\mathbf p)-m\sum_{h,\mathbf k}\Phi(\mathbf k)\nu_h(\mathbf k)(T_h+\varsigma)p_h(\mathbf k)$, with value $G(\mathbf p)-\Rp^\varsigma(\mathbf p)$. Lowering $\boldsymbol\nu$ to $\boldsymbol\nu(\cdot;\boldsymbol\mu)=\pos{\boldsymbol\ell}$ preserves feasibility and cannot lower this objective. At that choice the objective is $G(\mathbf p)$ minus the score sum in \eqref{eq:capdual}, proving the formula and attainment. By complementary slackness, the set $\cD^\varsigma(\mathbf p)$ of minimizers is nonempty and consists of the $\boldsymbol\mu\ge0$ with $\mu_{jh}\,\mathrm{sl}_{jh}(\ph_j)=0$, $\pos{\ell_h(\mathbf k;\boldsymbol\mu)}\{(T_h+\varsigma)p_h(\mathbf k)-u_h(\mathbf k)\}=0$, and $\pos{-\ell_h(\mathbf k;\boldsymbol\mu)}u_h(\mathbf k)=0$ for any optimal capped rent schedule $\mathbf u$, independently of its choice; for these $\boldsymbol\mu$, $\Rp^\varsigma(\mathbf p)=m\sum_hq_h\sum_{\mathbf k}\Phi(\mathbf k)p_h(\mathbf k)\psi^\varsigma_h(\mathbf k;\boldsymbol\mu)$.

\emph{Step 6 (certificates and optimality).} For $\boldsymbol\mu\ge0$, part~(i) gives $\mathcal R^{*,\varsigma}=\min H^\varsigma\le H^\varsigma(\boldsymbol\mu)$, with equality exactly at minimizers; subtracting $\Rp^\varsigma(\mathbf p)\le\mathcal R^{*,\varsigma}$ gives the bound in (iii). If $\Delta^\varsigma(\mathbf p;\boldsymbol\mu)=0$, then $\Rp^\varsigma(\mathbf p)=H^\varsigma(\boldsymbol\mu)\ge\mathcal R^{*,\varsigma}$, so $\mathbf p$ is globally optimal. Conversely, let $\mathbf p$ be globally optimal under \eqref{eq:cap}, with optimal rents $\mathbf u$, and let $\boldsymbol\mu^*$ minimize $H^\varsigma$. With $\boldsymbol\nu=\pos{\boldsymbol\ell(\cdot;\boldsymbol\mu^*)}$, Step~5 and the argument of \eqref{ec:dualchain} give
$\operatorname{Rev}(\mathbf p,\mathbf u)\le\operatorname{Rev}(\mathbf p,\mathbf u)+m\sum_{j\ne h}\mu^*_{jh}\mathrm{sl}_{jh}+m\sum_{h,\mathbf k}\Phi(\mathbf k)\nu_h(\mathbf k)\{(T_h+\varsigma)p_h(\mathbf k)-u_h(\mathbf k)\}
 \le m\sum_{h,\mathbf k}q_h\Phi(\mathbf k)\psi^\varsigma_h(\mathbf k;\boldsymbol\mu^*)p_h(\mathbf k)\le H^\varsigma(\boldsymbol\mu^*)$.

The two ends equal $\mathcal R^{*,\varsigma}$, so every inequality binds; as all complementary-slackness terms have fixed signs and $\Phi>0$, they vanish, so $\boldsymbol\mu^*\in\cD^\varsigma(\mathbf p)$ and $\Delta^\varsigma(\mathbf p;\boldsymbol\mu^*)=0$. The uncapped statements follow likewise from \eqref{ec:dualchain}, with $\boldsymbol\mu\in\cD$. On $\cD(\mathbf p)$, \eqref{eq:virtualidentity} and Step~2 show that $\Delta(\mathbf p;\boldsymbol\mu)=H(\boldsymbol\mu)-\Rp(\mathbf p)$ equals \eqref{eq:regret}; on $\cD^\varsigma(\mathbf p)$, the same holds with $\psi^\varsigma$ and $\chi^\varsigma$. 

\paragraph{Dual feasibility under correct beliefs.} With $\ph_j=\Phi$, take $\mu_{h-1,h}=\sum_{j<h}q_j$ for $h\ge2$ and all other multipliers zero. Then $\psi_h=b_h^{\mathrm{rat}}$, $\ell_h=0$ for $h<n$, and $\ell_n=-1$, proving dual feasibility. At every $\mathbf p^r$, the optimal objective rents \eqref{ec:rationalrents} make adjacent downward constraints bind and $u_n\equiv0$. Hence these multipliers belong to $\cD(\mathbf p^r)$.

\paragraph{Dual feasibility under alignment.} Under alignment, membership in $\cD$ reduces to $n$ inequalities: dividing $\ell_h(\mathbf k;\boldsymbol\mu)\le0$ by $\Lambda_h(\mathbf k)>0$ gives $M_h\le q_h/\Lambda_h(\mathbf k)+\sum_{j\ne h}\mu_{jh}\ph_j(\mathbf k)/\ph_h(\mathbf k)$, and alignment makes $\mathbf d_h$ minimize every term on the right, so $\boldsymbol\mu\in\cD$ if and only if
\begin{equation}\label{eq:flow}
 M_h\le q_h/\Lambda_h+\sum_{j\ne h}\mu_{jh}\rh_{jh},\qquad h=1,\ldots,n.
\end{equation} \qed

\subsubsection*{Proof of Corollary~\ref{cor:commondual}.}
With a common law, $\ell_h(\mathbf k;\boldsymbol\mu)=-q_h+\Lambda_h(\mathbf k)\bigl(M_h-\sum_{j\ne h}\mu_{jh}\bigr)$. For the stated multipliers, with $\mu_{01}=0$, $M_h-\sum_{j\ne h}\mu_{jh}=\mu_{h,h+1}-\mu_{h-1,h}=q_h\vartheta_h$ for $h<n$, and $M_n-\sum_{j\ne n}\mu_{jn}=-\mu_{n-1,n}$. Hence $\ell_h(\mathbf k;\boldsymbol\mu)=q_h\{\vartheta_h\Lambda_h(\mathbf k)-1\}$ for $h<n$ and $\ell_n(\mathbf k;\boldsymbol\mu)=-q_n-\mu_{n-1,n}\Lambda_n(\mathbf k)<0$. Since $\mu_{jh}>0$ requires $j=h-1$, $\psi_h(\mathbf k;\boldsymbol\mu)=T_h-\ind\{h\ge2\}\mu_{h-1,h}(T_{h-1}-T_h)\Lambda_h(\mathbf k)/q_h$, and $\nu_h=\pos{\ell_h}$ gives the stated $\psi_h^\varsigma$. Let $\mathbf u$ be optimal for $\mathbf p$. By Theorem~\ref{thm:common}(ii), $\Shat_j=\underline{\Shat}_j(\mathbf p)$ for all $j$, so the adjacent downward constraints bind, and $u_n\equiv0$ because $\underline{\Shat}_n=0$ and $\ph>0$. For $h<n$, $u_h$ solves the LP defining $U_h^\varsigma(\mathbf p)$, and $\vartheta_h$ with $\eta=\Phi\nu_h/q_h$ solves its dual (proof of Proposition~\ref{prop:knapsack}). Complementary slackness gives $u_h(\mathbf k)=(T_h+\varsigma)p_h(\mathbf k)$ where $\nu_h(\mathbf k)>0$, and $u_h(\mathbf k)=0$ where the reduced cost $\Phi(\mathbf k)\pos{-\ell_h(\mathbf k;\boldsymbol\mu)}/q_h$ is positive. These are the conditions defining $\cD^\varsigma(\mathbf p)$. Without a cap, $\vartheta_h=1/\Lambda_h$ gives $\ell_h\le0$, so $\boldsymbol\mu\in\cD$, and optimal $u_h$ vanishes off $\argmax_{\mathbf k}\Lambda_h(\mathbf k)$, where $\ell_h=0$; hence $\boldsymbol\mu\in\cD(\mathbf p)$, and $\nu_h=0$. \qed

\subsubsection*{Proof of Corollary~\ref{cor:settlingdual}.}
Write $\widehat{\boldsymbol U}^*=\widehat{\boldsymbol U}^*(\mathbf p)$ and concentrate rents as in Theorem~\ref{thm:compression}. By Step~2 of the proof of Proposition~\ref{prop:settlement}, every positive-rent type has a tight arc in $\mathcal G$, and every arc of $\mathcal G$ entering $j$ leaves a type with a larger settling round. Processing types in decreasing $t_j$ therefore fixes all $\mu_{hj}$ before type $j$ distributes its budget, and placing the budget on a tight arc of $\mathcal G$ attains the stated $M_j$; zero-rent types pass nothing. Hence \eqref{eq:flow} holds, with equality at positive-rent types, and $\boldsymbol\mu\in\cD$. Under concentrated rents, alignment gives $\sum_{\mathbf k}\ph_j(\mathbf k)u_h(\mathbf k)=\rh_{jh}\Shat_h^*$, so $\mathrm{sl}_{jh}(\ph_j)=\Shat_j^*-\rh_{jh}\Shat_h^*-(T_j-T_h)P_{jh}(\mathbf p)$ vanishes on tight arcs, which carry every positive multiplier. Positive rent occurs only at $\mathbf d_h$ of positive-rent types, where $\Lambda_j(\mathbf d_h)=\rh_{jh}\Lambda_h$ gives $\ell_h(\mathbf d_h;\boldsymbol\mu)=\Lambda_h(M_h-q_h/\Lambda_h-\sum_{j\ne h}\mu_{jh}\rh_{jh})=0$. Both complementary-slackness conditions of Step~4 of the proof of Theorem~\ref{thm:dual} hold, so $\boldsymbol\mu\in\cD(\mathbf p)$. 

%\begin{figure}[!t]
%\vspace{-0.1in}
%\centering
%\begin{tikzpicture}[
%  font=\small,
%  type/.style={circle, draw, thick, minimum size=10mm, inner sep=0pt},
%  zero/.style={type, fill=black!12},
%  tight/.style={-{Stealth[length=2.6mm]}, very thick},
%  slack/.style={-{Stealth[length=2.3mm]}, thick, dashed, black!60},
%  info/.style={align=center, inner sep=1pt, font=\footnotesize}
%]
%\node[type] (T1) at (0,0) {$T_1$};
%\node[type] (T2) at (5.7,0) {$T_2$};
%\node[zero] (T3) at (11.4,0) {$T_3$};
%\draw[tight] (T1) -- node[above=1pt]{tight: $\mu_{12}=\tfrac29$} (T2);
%\draw[tight] (T2) -- node[above=1pt]{tight: $\mu_{23}=\tfrac13$} (T3);
%\draw[slack] (T1.north east) .. controls +(3.2,1.9) and +(-3.2,1.9) .. node[above=1pt]{slack: $\mu_{13}=0$} (T3.north west);
%\node[info, below=3mm of T1] (I1) {$\Shat_1^*=\tfrac58$,\ \ $t_1=2$\\[2pt] budget $q_1/\Lambda_1=\tfrac29$\\[1pt] passed on in full};
%\node[info, below=3mm of T2] (I2) {$\Shat_2^*=\tfrac14$,\ \ $t_2=1$\\[2pt] budget $q_2/\Lambda_2+\rh_{12}\mu_{12}=\tfrac29+\tfrac19=\tfrac13$\\[1pt] passed on in full};
%\node[info, below=3mm of T3] (I3) {$\Shat_3^*=0$,\ \ $t_3=0$\\[2pt] budget $q_3/\Lambda_3+\rh_{23}\mu_{23}=\tfrac29+\tfrac16=\tfrac7{18}$\\[1pt] zero rent: nothing passed on};
%\end{tikzpicture}
%\caption{Multipliers from finite settlement (Corollary~\ref{cor:settlingdual}) in Example~\ref{ex:settlement}.}\label{fig:settlingdual}
%\vspace{-0.2in}
%\end{figure}

\paragraph{The illustration after Corollary~\ref{cor:settlingdual}.} With the beliefs of Example~\ref{ex:settlement}, $\Lambda_j(\mathbf e_j)=\tfrac32$ and $\Lambda_j(\mathbf e_h)=\tfrac34$ for $h\ne j$. Under $\mathbf p^n$, Corollary~\ref{cor:recursion} gives $\widehat{\boldsymbol U}^*(\mathbf p^n)=(\tfrac12,\tfrac14,0)$, with tight arcs $1\to2$ and $2\to3$ only, so Corollary~\ref{cor:settlingdual} gives the stated $\boldsymbol\mu$, and $\psi_1=4$, $\psi_2=3-\tfrac23\Lambda_1\ge2$, $\psi_3=1-2\Lambda_2$. By \eqref{eq:regret}, $\Delta(\mathbf p^n;\boldsymbol\mu)=\tfrac1{18}$; with $S=\tfrac{10}3$ and $\RE=3$, $H(\boldsymbol\mu)=\tfrac{55}{18}$, while $\Shat_1^*(\mathbf p^2)=\tfrac38$ and $\Shat_2^*(\mathbf p^2)=\Shat_3^*(\mathbf p^2)=0$ give $\Rp(\mathbf p^2)=\tfrac{55}{18}$. The replication package checks these values in exact arithmetic. \qed

\subsubsection*{Proof of Theorem~\ref{thm:primitive}.}
\emph{Step 1 (attached downward multipliers).} Full support and \eqref{eq:truncated} at $r=n$ give $\Shat_j^n>0$ for $j<n$ and $\Shat_n^n=0$. Set $\mu_{jh}=0$ unless $j<h$. Process $j=1,\ldots,n-1$, distributing the full budget $M_j=q_j/\Lambda_j+\sum_{t<j}\mu_{tj}\rh_{tj}$ among maximizers in \eqref{eq:truncated} with $r=n$. This construction satisfies \eqref{eq:flow}, and every positive multiplier prices a tight IC constraint under the concentrated optimal rents of Theorem~\ref{thm:compression}. At each cost peak carrying positive rent, the budget equality gives $\ell_j(\mathbf d_j;\boldsymbol\mu)=0$, as in the proof of Corollary~\ref{cor:settlingdual}. Complementary slackness therefore gives $\boldsymbol\mu\in\cD(\mathbf p^n)$.

\emph{Step 2 (budget charges and scores).} Let $\boldsymbol\mu\in\cD(\mathbf p^n)$ with $\mu_{jh}=0$ for $j>h$, such as the multipliers of Step~1. The concentrated rents of Theorem~\ref{thm:compression}(i) are optimal at $\mathbf p^n$ and positive at $\mathbf d_j$ for $j<n$, so complementary slackness (Step~4 of the proof of Theorem~\ref{thm:dual}) gives $\ell_j(\mathbf d_j;\boldsymbol\mu)=0$, that is, equality in \eqref{eq:flow}, for $j<n$; these are the budget equations of Step~1. Alignment gives $\rh_{tj}\Lambda_j=\Lambda_t(\mathbf d_j)\le\Lambda_t$, so the budget equations imply $\sum_{t>j}\mu_{jt}\Lambda_j\le q_j+\sum_{t<j}\mu_{tj}\Lambda_t$ for $j<n$. Fix $h\ge2$, multiply by $T_j-T_h$, and sum over $j<h$. For each internal edge $j\to t$ with $j<t<h$, its outgoing weight $T_j-T_h$ is at least its incoming weight $T_t-T_h$; edges with $t>h$ also contribute nonnegatively. Dropping these terms leaves $\sum_{j<h}\mu_{jh}\Lambda_j(T_j-T_h)\le\sum_{j<h}q_j(T_j-T_h)$. Whenever $p_h^n(\mathbf k)>0$, the definition of $\theta_h$ gives $\Lambda_j(\mathbf k)\le\theta_h\Lambda_j$ for $j<h$, so this inequality proves \eqref{eq:scorelower}. Since only downward multipliers are positive, $\psi_1=T_1$ and $\psi_t\le T_t$ for every report $t$.

\emph{Step 3 (regret and optimality).} At a full profile with highest report $T_h$, all buyers making that report face the same configuration and have the same score. Efficient allocation serves only them, with total probability one. Its bracket in \eqref{eq:regret} is zero if $h=1$ and otherwise at most $\pos{T_{h+1}-\psi_h(\mathbf k;\boldsymbol\mu)}\le\pos{\theta_h\sum_{j<h}(q_j/q_h)(T_j-T_h)-(T_h-T_{h+1})}$, where $T_{n+1}=0$ also covers the no-sale alternative at the all-low profile. Profiles with highest report $T_h$ have total probability $w_h$, so Theorem~\ref{thm:dual} gives $\Rstar-\RE\le\Delta(\mathbf p^n;\boldsymbol\mu)\le B_{\rm loc}$. The other revenue inequalities follow from $\RE\le\RR\le\Rstar$. Clearly, $B_{\rm loc}=0$ if $\theta_h\le K_h$ for all $h\in[2,n]$.

If the conditions are strict, \eqref{eq:scorelower} makes the highest report's score positive and strictly greater than every lower report's score at every profile. As $H(\boldsymbol\mu)=\RE=\Rstar$, any optimal mechanism attains equality in \eqref{ec:dualchain}, so its nonnegative regret brackets vanish at every profile, as each has positive objective probability. Positive winning scores force sale, and strictly lower scores exclude every lower report, so its allocation is $\mathbf p^n$ in the anonymous class. An asymmetric optimum anonymizes to this allocation by Lemma~\ref{lem:model}; nonnegativity forces its original no-sale and lower-report winning probabilities to vanish as well. Only the division among tied highest reports can differ.

If the concentrated efficient payments satisfy \eqref{eq:cap}, then $\RE^\varsigma=\RE$. Since $\mathcal R^{*,\varsigma}\le\Rstar$ and $\RR^\varsigma\ge\RE^\varsigma$, the same local bound holds with superscripts $\varsigma$. Whenever efficient allocation is uncapped optimal, it then attains the capped optimum as well; under the strict conditions, every capped optimum is also an uncapped optimum, so uniqueness persists.

Finally, under projection, $p_h^n(\mathbf k)>0$ implies $k_j=0$ for every $j<h$. Thus $\Lambda_j(\mathbf k)=\omega^{m-1}$ for those types, giving $\Lambda_j(\mathbf k)/\Lambda_j=\rh_j$ and $\theta_h=\max_{j<h}\rh_j\to0$. The constants $K_h$ are fixed and positive, so the conditions are eventually strict. The eventual nonnegative payments in Corollary~\ref{cor:projection} satisfy every cap, proving the projection conclusions. \qed

\subsubsection*{Proof of Proposition~\ref{prop:universal}.}
The cap restricts \eqref{eq:primal}, so $\mathcal R^{*,\varsigma}\le\Rstar$, and an unrestricted optimum satisfying the cap remains optimal after imposing it. For any monotone $\mathbf p$, set
\begin{equation}\label{ec:dsicrent}
 u_j^{\mathrm{rat}}(\mathbf k)=\sum_{h=j+1}^n(T_{h-1}-T_h)p_h(\mathbf k).
\end{equation}
For $j<h$, telescoping gives $u_j^{\mathrm{rat}}-u_h^{\mathrm{rat}}=\sum_{t=j+1}^h(T_{t-1}-T_t)p_t\ge(T_j-T_h)p_h$. For $j>h$, the difference is the negative sum from $h+1$ to $j$, whose allocations are all at most $p_h$, giving the same inequality. Thus $g_{jh}\ge0$ at every configuration, and IC holds under every law. Also $0\le u_j^{\mathrm{rat}}\le(T_j-T_n)p_j\le T_jp_j$, proving \eqref{eq:ir} and \eqref{eq:cap} for all $\varsigma\ge0$.

At $\mathbf p^n$, this schedule costs $m\sum_jq_j\sum_{h>j}(T_{h-1}-T_h)\bar p_h=S-\RE^{\mathrm{rat}}$ by \eqref{ec:rationalrents}, so $\RE,\RE^\varsigma\ge\RE^{\mathrm{rat}}$. Since every mechanism earns at most $S$ and $\RR^\varsigma\ge\RE^\varsigma$, this proves the first two inequalities of \eqref{eq:universal}. For $h\ge2$, winning requires no rival of type $T_1$, hence $\bar p_h\le(1-q_1)^{m-1}$. Therefore $\sum_{h>j}(T_{h-1}-T_h)\bar p_h\le(T_j-T_n)(1-q_1)^{m-1}$, proving the third inequality after weighting by $mq_j$ and summing. 

\paragraph{Smallest attached certificates.} By Theorem~\ref{thm:dual}(ii), $\cD^\varsigma(\mathbf p)$ is the sublevel set of the score sum in \eqref{eq:capdual} at its minimum $\Rp^\varsigma(\mathbf p)$. For $(\boldsymbol\mu,\boldsymbol\nu)\ge0$, let $\tilde\psi_h(\mathbf k)=\psi_h(\mathbf k;\boldsymbol\mu)+(T_h+\varsigma)\nu_h(\mathbf k)/q_h$, which is affine in $(\boldsymbol\mu,\boldsymbol\nu)$, and consider the LP
\begin{equation}\label{ec:mincert}
\begin{aligned}
 \min_{\boldsymbol\mu,\boldsymbol\nu,\boldsymbol\chi\ge0}\quad&\sum_jq_j\sum_{\mathbf k}\Phi(\mathbf k)\chi_j(\mathbf k)-\Rp^\varsigma(\mathbf p)\\
 \text{subject to}\quad&\nu_h(\mathbf k)\ge\ell_h(\mathbf k;\boldsymbol\mu),\quad\chi_j(\mathbf k)\ge\tilde\psi_h(\mathbf k+\mathbf e_j-\mathbf e_h)\ \text{ if }k_h+\ind\{h=j\}>0,\\
 &m\sum_hq_h\sum_{\mathbf k}\Phi(\mathbf k)p_h(\mathbf k)\tilde\psi_h(\mathbf k)\le\Rp^\varsigma(\mathbf p).
\end{aligned}
\end{equation}
Let $(\boldsymbol\mu,\boldsymbol\nu,\boldsymbol\chi)$ be feasible. As $\boldsymbol\nu\ge\pos{\boldsymbol\ell(\cdot;\boldsymbol\mu)}$ and $T_h+\varsigma>0$, $\tilde\psi_h\ge\psi_h^\varsigma(\cdot;\boldsymbol\mu)$. Since $\mathbf p\ge0$, the last constraint and \eqref{eq:capdual} give
$\Rp^\varsigma(\mathbf p)\ge m\sum_hq_h\sum_{\mathbf k}\Phi(\mathbf k)p_h(\mathbf k)\psi_h^\varsigma(\mathbf k;\boldsymbol\mu)\ge\Rp^\varsigma(\mathbf p)$,
so $\boldsymbol\mu\in\cD^\varsigma(\mathbf p)$. Moreover, $\chi_j(\mathbf k)\ge\chi_j^\varsigma(\mathbf k;\boldsymbol\mu)$, so the objective is at least $\Delta^\varsigma(\mathbf p;\boldsymbol\mu)$. Conversely, each $\boldsymbol\mu\in\cD^\varsigma(\mathbf p)$, with $\boldsymbol\nu=\pos{\boldsymbol\ell(\cdot;\boldsymbol\mu)}$ and $\boldsymbol\chi=\boldsymbol\chi^\varsigma(\cdot;\boldsymbol\mu)$, is feasible with objective $\Delta^\varsigma(\mathbf p;\boldsymbol\mu)$. Hence the value of \eqref{ec:mincert} is $\Delta^\varsigma_*(\mathbf p)$. The LP is feasible, as $\cD^\varsigma(\mathbf p)\ne\emptyset$, and bounded below by Theorem~\ref{thm:dual}(iii), so it has an optimal solution, whose $\boldsymbol\mu$ attains $\Delta^\varsigma_*(\mathbf p)$. By Theorem~\ref{thm:dual}(iii), $\Delta^\varsigma_*(\mathbf p)=\mathcal R^{*,\varsigma}-\Rp^\varsigma(\mathbf p)$ if and only if a minimizer of $H^\varsigma$ lies in $\cD^\varsigma(\mathbf p)$, as when $\mathbf p$ is globally optimal. Without \eqref{eq:cap}, the same LP with $\boldsymbol\nu=\mathbf0$, $\boldsymbol\mu\in\cD$, and $\Rp$ in place of $\Rp^\varsigma$ computes $\Delta_*(\mathbf p)$. \qed

\subsubsection*{Proof of Corollary~\ref{cor:stcert}.}
Write $P_h=P_h(\mathbf p^n)$. Report $T_h$ wins with probability $p_h^n(\mathbf k)=\ind\{k_t=0\text{ for }t<h\}/(k_h+1)$, so $p_1^n\ge p_2^n\ge p_3^n$ pointwise, $P_1-P_2\ge\ph(\mathbf D_1)/m>0$, and $P_2-P_3\ge(1-1/m)\ph(\mathbf D_3)>0$.

\emph{Step 1 (adjacent multipliers).} Let $\boldsymbol\mu\in\cD^\varsigma(\mathbf p^n)$. Optimal rents have $\Shat_j=\underline{\Shat}_j(\mathbf p^n)$ by Theorem~\ref{thm:common}(ii), so $\mathrm{sl}_{13}=(T_1-T_2)(P_2-P_3)$, $\mathrm{sl}_{21}=(T_1-T_2)(P_1-P_2)$, $\mathrm{sl}_{32}=(T_2-T_3)(P_2-P_3)$, and $\mathrm{sl}_{31}=(T_1-T_2)(P_1-P_2)+(T_2-T_3)(P_1-P_3)$ are positive. Complementary slackness leaves only $\mu_{12}$ and $\mu_{23}$, so $\ell_1(\mathbf k;\boldsymbol\mu)=-q_1+\Lambda_j(\mathbf k)\mu_{12}$ and $\ell_2(\mathbf k;\boldsymbol\mu)=-q_2+\Lambda_j(\mathbf k)(\mu_{23}-\mu_{12})$.

\emph{Step 2 (pinning).} As $\beta>\beta_c$ and $\lambda^\varsigma>0$, \eqref{eq:ST} gives $\underline{\Shat}_1(\mathbf p^n)<(T_1+\varsigma)\ph(\mathbf D_1)/m$, while $\underline{\Shat}_1(\mathbf p^n)\ge(T_1-T_2)P_2>0$. Since $\mathbf D_1$ uniquely maximizes $\Lambda_j(\cdot)$, Proposition~\ref{prop:knapsack} makes the optimal $u_1$ vanish off $\mathbf D_1$, for otherwise $u_1$ would fill $\mathbf D_1$ and exceed the floor; thus $0<u_1(\mathbf D_1)<(T_1+\varsigma)p_1^n(\mathbf D_1)$. Next, $0<\underline{\Shat}_2(\mathbf p^n)=(T_2-T_3)P_3<(T_2+\varsigma)P_2$, and $p_2^n$ vanishes unless $k_1=0$, where $\Lambda_j(\mathbf k)=Z_1^{1-m}$. Hence $u_2=c(T_2+\varsigma)p_2^n$, with $c=\underline{\Shat}_2(\mathbf p^n)/\{(T_2+\varsigma)P_2\}\in(0,1)$, and $u_3\equiv0$ complete an optimal schedule. As $\cD^\varsigma(\mathbf p^n)$ does not depend on the optimal schedule, use this one. Where $0<u_h(\mathbf k)<(T_h+\varsigma)p_h(\mathbf k)$, the conditions defining $\cD^\varsigma(\mathbf p^n)$ force $\nu_h(\mathbf k;\boldsymbol\mu)=\pos{-\ell_h(\mathbf k;\boldsymbol\mu)}=0$, that is, $\ell_h(\mathbf k;\boldsymbol\mu)=0$. At $(1,\mathbf D_1)$ and $(2,\mathbf D_3)$, Step~1 yields $\mu_{12}=q_1/\Lambda_1$ and $\mu_{23}=\mu_{12}+q_2Z_1^{m-1}$, the values in $\boldsymbol\mu^\circ$. Thus $\cD^\varsigma(\mathbf p^n)\subseteq\{\boldsymbol\mu^\circ\}$, with equality because $\cD^\varsigma(\mathbf p^n)\ne\emptyset$.

\emph{Step 3 (the bound).} By Step~6 of the proof of Theorem~\ref{thm:sharedtop}, $\boldsymbol\mu^\circ$ gives the scores of its Step~3 with $a=1$: $\psi_1^\varsigma(\cdot;\boldsymbol\mu^\circ)\equiv T_1$ and, by the definition of $\beta_c$, $\psi_2^\varsigma(\mathbf D_1;\boldsymbol\mu^\circ)=T_2+(T_2+\varsigma)(\beta^{m-1}-1)-(q_1/q_2)(T_1-T_2)=T_1+(T_2+\varsigma)(\beta^{m-1}-\beta_c^{m-1})$. As $\boldsymbol\mu^\circ$ attains \eqref{eq:capdual}, \eqref{ec:regroupsum} expresses $\Delta^\varsigma(\mathbf p^n;\boldsymbol\mu^\circ)$ as \eqref{eq:regret} with $\psi^\varsigma$ in place of $\psi$: an objective expectation, over full profiles, of brackets that are nonnegative by Step~2 of the proof of Theorem~\ref{thm:dual}. At $\mathbf D_1+\mathbf e_2$, of probability $mq_1^{m-1}q_2$, $\mathbf p^n$ serves only the $m-1$ top reports, so the bracket is at least $\psi_2^\varsigma(\mathbf D_1;\boldsymbol\mu^\circ)-T_1$. Since $\Delta^\varsigma_*(\mathbf p^n)=\Delta^\varsigma(\mathbf p^n;\boldsymbol\mu^\circ)$ by Step~2, this gives the inequality; Theorem~\ref{thm:sharedtop} and the definition of $\lambda^\varsigma$ give the equality.

\emph{Step 4 (remaining claims).} By Theorem~\ref{thm:common}(ii) and $G(\mathbf p^n)=S$, $S-\RE^\varsigma=m\sum_jq_jU_j^\varsigma(\mathbf p^n)$. As $\mathbf p^n$ is implementable under \eqref{eq:cap}, filling all capacity delivers each floor, so $U_j^\varsigma(\mathbf p^n)\le(T_j+\varsigma)\sum_{\mathbf k}\Phi(\mathbf k)p_j^n(\mathbf k)$; by \eqref{ec:regroupsum}, the resulting bound is the objective expectation of $T_h+\varsigma$ for the winning report $T_h$, at most $T_1+\varsigma$. As $\beta\to\infty$ with $\beta_c$ fixed, $\bar\psi_2\to T_2$, $\bar\psi_3\to T_3$, $\lambda^\varsigma\to0$, $\ph(\mathbf D_1)\to1$, and $\underline{\Shat}_1(\mathbf p^n)\le(T_1-T_3)P_2\le(T_1-T_3)\{(1-q_1)/Z_1\}^{m-1}\to0$, so \eqref{eq:ST} and $\beta>\beta_c$ hold for all large $\beta$. \qed

\subsubsection*{Proof of Proposition~\ref{thm:kink}.}
\emph{Step 1 (feasibility and monotonicity).} The altered profile has probabilities $1-\lambda,\lambda/2,\lambda/2$, summing to one. Against two lowest rivals the own-report probabilities are $1,1-\lambda,1/3$; against a middle and a lowest rival they are $1,1/2,\lambda/2$. All other comparisons are unchanged, so feasibility and monotonicity hold for $0\le\lambda\le2/3$.

\emph{Step 2 (rents).} Summing winning probabilities against the stated subjective marginals gives $P_{23},P_{12},P_{13}$. Theorem~\ref{thm:compression} and Corollary~\ref{cor:recursion} then give the stated optimal rents, concentrated at $\mathbf D_1,\mathbf D_2$, with all IC constraints satisfied. The slopes of the two affine functions $\Shat_1^{(2)}$ and $\Shat_1^{(3)}$ differ by
\[
 \partial_\lambda\Shat_1^{(3)}-\partial_\lambda\Shat_1^{(2)}
 =\omega^2q_3\Bigl\{q_2(T_1-T_3)-\frac{\omega q_2}{\pi_{22}}\,q_2(T_2-T_3)+q_3(T_1-T_2)\Bigr\}
 \ge\omega^2q_3(q_2+q_3)(T_1-T_2)>0,
\]
because $\rh_2\pi_{22}\omega q_3=\omega^2q_3(\omega q_2/\pi_{22})q_2$ and $\omega q_2\le \pi_{22}$. The denominator of $\bar\mu_{12}$ equals this difference divided by $q_2q_3^2$, and its numerator is positive, so $\bar\mu_{12}>0$. The two functions therefore cross exactly once, and under \eqref{eq:KR} they do so at $\lambda^*\in(0,2/3)$. Hence both downward constraints of $T_1$ bind at $\lambda^*$, only the one toward $T_2$ binds at $\lambda=0$, and the constraint from $T_2$ to $T_3$ binds throughout.

\emph{Step 3 (multipliers).} Let $\mu_{12}=\bar\mu_{12}$, $\mu_{13}=q_1/\Lambda_1-\bar\mu_{12}$, and $\mu_{23}=q_2/\Lambda_2+\rh_2\bar\mu_{12}$, with all other multipliers zero; they are positive under \eqref{eq:KR}. Then $M_1=q_1/\Lambda_1$, $M_2=q_2/\Lambda_2+\rh_2\mu_{12}$, and $M_3=0$, so \eqref{eq:flow} holds, with equality for $h=1,2$. Hence $\boldsymbol\mu\in\cD$ and $\ell_h(\mathbf{D}_h;\boldsymbol\mu)=0$ for $h=1,2$. The positive multipliers sit on constraints that bind at $\lambda^*$, so $\boldsymbol\mu\in\cD(\mathbf{p}(\lambda^*))$.

\emph{Step 4 (scores).} At profiles containing $T_1$, its score $\psi_1=T_1$ strictly exceeds all others. Otherwise $\Lambda_1(\mathbf k)=\omega^2$, so the middle report has constant score
$\psi_2(\mathbf{k};\boldsymbol\mu)=T_2-\bar\mu_{12}\omega^2(T_1-T_2)/q_2$,
which is positive under \eqref{eq:KR}. The score of a $T_3$ report at $(T_2,T_3,T_3)$ is
\[
 \psi_3(\mathbf{e}_2+\mathbf{e}_3;\boldsymbol\mu)=T_3-\Bigl(\frac{q_1}{\Lambda_1}-\bar\mu_{12}\Bigr)\frac{\omega^2(T_1-T_3)}{q_3}
 -\Bigl(\frac{q_2}{\Lambda_2}+\rh_2\bar\mu_{12}\Bigr)\frac{\pi_{22}\omega(T_2-T_3)}{q_2q_3},
\]
and equating it with the score of the $T_2$ report gives exactly the definition of $\bar\mu_{12}$. Replacing the lowest buyer's middle rival by a lowest rival reduces $\Lambda_2(\mathbf{k})$ from $\pi_{22}\omega/q_2$ to $\omega^2$, which strictly raises her score at $(T_3,T_3,T_3)$; replacing her lowest rival by a second middle rival raises $\Lambda_2(\mathbf{k})$ to $\pi_{22}^2/q_2^2$, which strictly lowers her score at $(T_2,T_2,T_3)$. Both comparisons are strict because $\pi_{22}>\omega q_2$ and $\mu_{23}>0$. Therefore, at every profile, $\mathbf{p}(\lambda^*)$ sells with probability one to reports with the largest score, which is positive. Thus $\Delta(\mathbf{p}(\lambda^*);\boldsymbol\mu)=0$, and Theorem~\ref{thm:dual} gives global optimality and $\Rstar=H(\boldsymbol\mu)$.

\emph{Step 5 (the gain).} The efficient allocation $\mathbf{p}(0)$ also selects largest scores at every profile, because the scores are tied at the only altered profile. At its optimal payments the rents sit at $\mathbf{D}_1$ and $\mathbf{D}_2$, where $\ell_h=0$, and the only positively weighted slack is that of the constraint from $T_1$ to $T_3$, equal to $\Shat_1^{(2)}(0)-\Shat_1^{(3)}(0)$ by Step~2. With $m=3$, \eqref{eq:lagidentity} therefore gives $H(\boldsymbol\mu)-\RE=3\mu_{13}\{\Shat_1^{(2)}(0)-\Shat_1^{(3)}(0)\}$, the first identity in the proposition. It is positive under \eqref{eq:KR}.

\emph{Step 6 (reserves, surplus, and payments).} For $r<3$, $\mathbf{p}^r$ withholds the all-$T_3$ profile, at which the largest score is positive; the last inequality of \eqref{ec:dualchain} is then strict and $\Rp_r<H(\boldsymbol\mu)=\Rstar$. The altered profile has probability $3q_2q_3^2$ and its surplus falls by $(T_2-T_3)\lambda^*$, which is the second identity in the proposition. Payments equal $T_jp_j(\mathbf{k})$ except at $\mathbf{D}_1$ and $\mathbf{D}_2$, where a truthful buyer wins with probability $1/3$ and $\ph_j(\mathbf{D}_j)=\pi_{jj}^2$; they are nonnegative exactly under the two stated conditions, and Proposition~\ref{prop:universal} then gives optimality with nonnegative payments. \qed

%\paragraph{The numerical example.} For values $(10,9,7)$, probabilities $(3/5,1/10,3/10)$, and $\alpha=1/10$, the two rent branches are $\Shat_1^{(2)}(\lambda)=80001/722000-9477\lambda/190000$ and $\Shat_1^{(3)}(\lambda)=729(1+\lambda)/10000$, so $\lambda^*$ and the multipliers of Step~3 are rational. In exact rational arithmetic, every constraint in \eqref{eq:ic} holds, all payments are nonnegative, and the scores of Step~4 certify the optimal revenue $151098763/15564800$ reported after Proposition~\ref{thm:kink}.

\subsubsection*{Proof of Proposition~\ref{prop:twocap}.}
For two types, $\Lambda_1(\mathbf k)=(\pi_{11}/q_1)^{k_1}\omega^{m-1-k_1}$ strictly increases with $k_1$, whereas $\ph_2(\mathbf k)/\ph_1(\mathbf k)=(\omega q_1/\pi_{11})^{k_1}\{\pi_{22}/(\omega q_2)\}^{m-1-k_1}$ strictly decreases.

\emph{Step 1 (relaxation).} Drop upward IC $(2,1)$. Lowering $u_2$ to zero preserves all remaining constraints and strictly raises revenue when any component was positive. Reallocating every profile with a high report equally among high reports then increases surplus unless unchanged, weakly lowers $P_{12}$, and weakly raises high-report rent caps. Hence every relaxed optimum has $u_2=0$, $p_1(\mathbf k)=1/(k_1+1)$, and $p_2(\mathbf k)=0$ off $\mathbf D_2$.

\emph{Step 2 (rent delivery).} Given $p_2(\mathbf D_2)$, minimize $mq_1\sum\Phi u_1$ subject to
$\sum\ph_1u_1\ge(T_1-T_2)\ph_1(\mathbf D_2)p_2(\mathbf D_2)$ and
$0\le u_1(\mathbf k)\le(T_1+\varsigma)/(k_1+1)$.
Perceived rent costs $mq_1/\Lambda_1(\mathbf k)$ per unit, so the unique fractional-knapsack solution fills caps in decreasing $k_1$ until IC binds. Feasibility follows because $\mathbf D_2$ alone has capacity $(T_1+\varsigma)\ph_1(\mathbf D_2)>(T_1-T_2)\ph_1(\mathbf D_2)/m$.

\emph{Step 3 (sale probability).} Revenue equals $(1-q_2^m)T_1+mq_2^mT_2p_2(\mathbf D_2)$ minus rent cost. While configuration $\mathbf k$ fills, its derivative with respect to $mp_2(\mathbf D_2)$ is
\[
 q_2^mT_2-\frac{q_1(T_1-T_2)\ph_1(\mathbf D_2)}{\Lambda_1(\mathbf k)}
 =q_1q_2^{m-1}(T_1-T_2)\left\{K_2-\frac{\Lambda_1(\mathbf D_2)}{\Lambda_1(\mathbf k)}\right\}.
\]
Successive configurations have decreasing $\Lambda_1$, so revenue is concave. Fill precisely those with positive derivative, or stop at full sale first, yielding \eqref{eq:capsale}. A zero derivative permits the interval obtained by also including that configuration. All capacities scale with $T_1+\varsigma$, proving the claimed monotonicity.

\emph{Step 4 (the dropped constraint).} The maximal ratio $\ph_2/\ph_1$ occurs at $\mathbf D_2$, so binding downward IC gives $\sum_{\mathbf k}\ph_2u_1\le\{\ph_2(\mathbf D_2)/\ph_1(\mathbf D_2)\}\sum_{\mathbf k}\ph_1u_1=(T_1-T_2)\ph_2(\mathbf D_2)p_2(\mathbf D_2)\le(T_1-T_2)P_{21}$, where $p_2(\mathbf D_2)\le1/m<1=p_1(\mathbf D_2)$. Thus all relaxed optima satisfy upward IC and coincide with the original optima, proving the structure and low winner payment $T_2$. Without a zero derivative, the sale probability is unique; if it lies in $(0,1)$, both reserve endpoints are strictly suboptimal. \qed

\section{Calibration and Computational Detail}\label{app:caldetail}
\subsection{Observation Process and Calibration}\label{app:calobs}
Of 1,000 listings, 863 are sales. Positive prices, at least two bidders, and the screen $y_a\in[r_a+0.02,3.00)$ retain 749, where $y_a$ and $r_a$ are price and reserve relative to book value. Let $F_a$ be the c.d.f.\ of the lognormal with log mean $\mathbf z_a^\top\boldsymbol\gamma$ and log variance $\sigma_X^2$, and $\tilde F_a(x)=(F_a(x)-F_a(r_a))/(1-F_a(r_a))$. Among $N_a$ independent draws from $\tilde F_a$, the second-highest has c.d.f.\ $H_a=\tilde F_a^{N_a}+N_a\tilde F_a^{N_a-1}(1-\tilde F_a)$. The selected log density is
\begin{equation}\label{ec:likelihood}
 \ell_a=\log H'_a(y_a)-\log\{H_a(3.00)-H_a(r_a+0.02)\}.
\end{equation}
The conditional and pooled fits give log likelihoods 236.40 and 139.19. At fixed covariates, bidder counts, and reserves, each of 9,999 parametric resamples redraws selected prices and refits. The selected-price probability integral transforms give Kolmogorov--Smirnov distances 0.03 and 0.04, with plus-one bootstrap $p$-values 0.10 and 0.01, respectively.

\begin{table}[!htbp]
\vspace{-0.1in}
\centering
\caption{Retained sample and conditional lognormal estimates}\label{tab:ECcoef}
\begin{footnotesize}\setlength{\tabcolsep}{5pt}
\begin{tabular*}{0.9\textwidth}{@{\extracolsep{\fill}}lrrrrr}\toprule
\textit{A. Retained listings} & Mean & SD & Median & Minimum & Maximum\\\midrule
Price-to-book ratio $y_a$ & 0.96 & 0.31 & 0.91 & 0.31 & 2.77 \\
Reserve-to-book ratio $r_a$ & 0.44 & 0.30 & 0.35 & 0.02 & 1.50 \\
Bidders $N_a$ & 4.69 & 2.02 & 5.00 & 2.00 & 12.00 \\
Book value (\$) & 33.10 & 31.52 & 24.50 & 7.50 & 175.00 \\
\midrule
\end{tabular*}
\begin{tabular*}{0.9\textwidth}{@{\extracolsep{\fill}}lrrr}
\textit{B. Conditional lognormal model} & Estimate & 95\% interval & Share with indicator $= 1$\\\midrule
Intercept & $-0.38$ & $[-0.44,-0.34]$ & ---\\
Minor blemish & $0.06$ & $[0.01,0.13]$ & 0.09 \\
Major blemish & $-0.40$ & $[-0.54,-0.25]$ & 0.04 \\
Power seller & $0.03$ & $[-0.03,0.07]$ & 0.47 \\
Verified seller identity & $0.00$ & $[-0.12,0.13]$ & 0.03 \\
Sealed package & $0.32$ & $[0.27,0.39]$ & 0.12 \\
Negative feedback above 1\% & $0.00$ & $[-0.06,0.06]$ & 0.09 \\
$\log(\text{book value}/22.50)$ & $-0.13$ & $[-0.16,-0.10]$ & ---\\
Standard deviation of $\log X$ & $0.42$ & $[0.39,0.45]$ & ---\\
\bottomrule\end{tabular*}\end{footnotesize}
\begin{minipage}{0.9\textwidth}\footnotesize\medskip
\textit{Notes.} Intervals give the central 95\% of paired refits in Appendix~\ref{app:calrefit}.
\end{minipage}
\vspace{-0.2in}
\end{table}

For fitted c.d.f.\ endpoints $a<b$ in the central 99\% at the modal profile, the representative value is $22.50\int_a^b F^{-1}(z)\,dz/(b-a)$. The source excludes bids whose sum with shipping is at most one-quarter of the smaller of price and book value \citep{Wegmann2011}. At $r_a>0.25$ no admissible bid can be removed. Refitting those 493 listings yields the check in Section~\ref{sec:datacal}.

\begin{table}[!t]
\vspace{-0.1in}
\centering
\caption{How service changes support a one-cent reporting certificate}\label{tab:policy}
\begin{small}\setlength{\tabcolsep}{4pt}
\begin{tabularx}{\linewidth}{@{}>{\raggedright\arraybackslash}p{0.50\linewidth}>{\centering\arraybackslash}X>{\centering\arraybackslash}X@{}}
\toprule
Policy component & Projection & Shared-top\\\midrule
First eliminated pair & $(3,2)$ & $(2,3)$\\
Middle report's winning probability against top rivals & $\delta/(T_2-T_3)\approx0.15\%$ & $\delta/T_2\approx0.06\%$\\
Winner's payment in those states & $T_2\approx\$18.16$ & \$0.00\\
Middle type's truthful rent there & \$0.00 & $\delta=\$0.01$\\
Binding first-step advantage & $g_{32}=(T_2-T_3)p_2=\delta$ & $g_{23}=T_2p_2=\delta$\\
Deletion schedule & \shortstack{$(3,2),(3,1),(2,1),$\\$(2,3),(1,3),(1,2)$} & \shortstack{$(2,3),(3,1),(3,2),$\\$(2,1),(1,3),(1,2)$} \\
Matched allocation gain (cents) & 1.48 & 3.32\\
\bottomrule
\end{tabularx}
\end{small}
\begin{minipage}{\linewidth}\footnotesize\medskip
\textit{Notes.} Primitives from Table~\ref{tab:main}. $(j,h)$ deletes $T_h$ for type $T_j$. Top–middle probabilities are per middle-report buyer; top reports get residual, low reports lose. Displayed states bind deletion, and six steps complete certification.
\end{minipage}
\vspace{-0.2in}
\end{table}

\subsection{Optimization and Verification}\label{app:calverify}
Report-by-report optimization searches all 720 schedules for joint design and each reserve. With four buyers, Theorem~\ref{thm:unique} uses at most $\binom83=56$ rank profiles per step, and whole-type steps use $|\cK|=10$ configurations. Enumerating initially $27^3=19{,}683$ ordered pure-map profiles verifies the selected one-cent margins within $10^{-14}$ dollars. The compact and unreduced common-law LPs agree to numerical precision.

\noindent\textbf{Allocation regimes.} At $\beta=1.30$, knapsack prices $(148877/162500,148877/125000,0)$ give a strictly positive highest-report score advantage of at least $720373091/150000000$, certifying unique efficiency. Figure~\ref{fig:regimes} uses extrema of every allocation coordinate and welfare on each optimal revenue face at 253 strengths. The largest welfare range is 1.93 cents. Allocation probabilities vary by less than $10^{-4}$ outside $\beta\in(1.41,1.83)$. At $\beta=1.50$, shifting service reduces top rent in two-top rival states from 1.48 to 1.24 and increases it in one-top rival states from 0.55 to 0.88 perceived dollars. At $\beta=3.00$, peak capacity 0.54 falls short of top rent 1.10. Both checks illustrate use of costlier tiers outside \eqref{eq:ST}.

\noindent\textbf{Belief intervals.} Endpoint LPs add violating interior stationary points of the cubic slacks (Proposition~\ref{prop:verify}). Cuts were needed for 6 and 44 of the 2,880 schedule--allocation pairs at $\bar\kappa=0.10$ and $0.20$. The point-contract check retains all four one-cent shared-top joint maximizing schedules. Maximizing minimum endpoint truthful-IC slack on their revenue-optimal faces (tolerance $10^{-9}$ dollars) gives best slacks of $-10.82$ and $-23.97$ cents at the two endpoints.

\subsection{Paired Refits and Fixed Contracts}\label{app:calrefit}
The 300 paired resamples of 749 listings use NumPy's SeedSequence with master seed 20,261,101. Each refits the model, reconstructs values at fixed shares, and reoptimizes both allocation classes over the relevant schedules. For fixed $(\mathbf p,\mathbf x)$, rents are recomputed at the refit values. Truthful IC holds in 104 projection and 116 shared-top refits. Participation holds in the 89 refits in which all representative values weakly rise. The one-cent certificate survives in 0 and 89 refits, respectively. Conditional-on-truth receipts stay fixed because shares and payments are fixed.

Near-withholding at the all-low profile occurs in six shared-top joint designs (2.00\% of refits, exact 95\% interval 0.74--4.30\%). In five, the best reserve excludes the low type and loses more welfare than the joint design. In the remaining refit, the reserve is efficient and the joint design loses 63.76 cents more. Within each of the six refits, welfare varies by less than 0.01 cent across all joint revenue-optimal schedule faces. Table~\ref{tab:ECboot} reports extrema to capture these rare outcomes.

\begin{table}[!htbp]
\vspace{-0.1in}
\centering
\caption{Paired refits: revenue gains, welfare levels, and incremental loss}\label{tab:ECboot}
\begin{footnotesize}\setlength{\tabcolsep}{4pt}
\begin{tabular*}{\textwidth}{@{\extracolsep{\fill}}lrrrrr}\toprule
 & Estimate & 2.50\% & 97.50\% & Minimum & Maximum\\\midrule
\multicolumn{6}{@{}l}{\textit{A. Allocation comparisons, shared-top unless stated (cents)}}\\
Nominal gain & 24.72 & 23.49 & 25.80 & 23.14 & 26.53 \\
Report-by-report matched gain, projection & 1.48 & 1.31 & 1.66 & 1.20 & 1.79 \\
Report-by-report matched gain & 3.32 & 3.28 & 3.37 & 3.28 & 4.33 \\
Whole-type conservative difference & 34.69 & 32.99 & 36.19 & 32.49 & 37.13 \\
Welfare loss, report-by-report joint, projection & 0.80 & 0.77 & 0.83 & 0.76 & 0.85 \\
Welfare loss, report-by-report joint & 0.30 & 0.27 & 0.33 & 0.26 & 64.89 \\
Welfare loss, report-by-report best reserve & 0.00 & 0.00 & 0.00 & 0.00 & 66.41 \\
% Shared-top refits_rbr.csv: summarize loss_returned - reserve_loss within each rep, then round.
Welfare loss, joint minus reserve & 0.30 & 0.27 & 0.33 & $-1.53$ & 63.76 \\
Welfare loss, whole-type joint & 2.32 & 2.14 & 2.51 & 2.07 & 2.66 \\
\multicolumn{6}{@{}l}{\textit{B. Payment comparisons under projection, nonnegative payments (dollars)}}\\
Gain over the best second-price auction & 3.87 & 3.64 & 4.11 & 3.57 & 4.28 \\
Premium over correct beliefs & 1.93 & 1.79 & 2.07 & 1.74 & 2.18 \\
Cost of nonnegative payments & 0.25 & 0.23 & 0.27 & 0.22 & 0.29 \\
\bottomrule\end{tabular*}\end{footnotesize}
\begin{minipage}{\textwidth}\footnotesize\medskip
\textit{Notes.} Conditional scale, $\alpha=0.15$, $\beta=4.00$, nonnegative payments, and $\delta=\$0.01$ for certificates. The whole-type difference uses the zero-margin reserve bound. All reported revenue comparisons are positive in every refit. Welfare losses use each refit's efficient allocation. Joint minus reserve is calculated within each refit.
\end{minipage}
\vspace{-0.1in}
\end{table}

\subsection{Reporting Resolution and Matched Band Certificates}\label{app:calresolution}
Table~\ref{tab:ECresolution} fixes the population and quadruples the top quintile's weight. Exact types and band welfare use group means. Band payments and participation use floors, and IC uses both endpoints. Whole-type certification searches 6 or 120 orders for three or five reports. Lemma~\ref{lem:bandreserve} gives the zero-coefficient band reserve bound.

\begin{table}[!htbp]
\vspace{-0.1in}
\centering
\caption{Matched reporting resolution: shared-top category, four buyers, nonnegative payments}\label{tab:ECresolution}
\begin{footnotesize}\setlength{\tabcolsep}{5pt}
\begin{tabular*}{\textwidth}{@{\extracolsep{\fill}}rlrrrrrr}\toprule
\multicolumn{2}{l}{\textit{A. Revenue (\$), differences (cents)}} & \multicolumn{3}{c}{Nominal} & \multicolumn{3}{c}{Whole-type comparison}\\\cmidrule(lr){3-5}\cmidrule(lr){6-8}
Reports & Interpretation & Joint & Reserve & Gain & \shortstack{Whole-type\\joint} & \shortstack{Zero-margin\\reserve bound} & \shortstack{Conservative\\difference}\\\midrule
3 & Exact types & 23.17 & 22.93 & 24.72 & 20.99 & 20.64 & 34.69 \\
 & Every-value bands & 18.43 & 18.12 & 31.18 & 16.66 & 16.67 & $-0.21$ \\
5 & Exact types & 23.40 & 22.88 & 52.69 & 21.14 & 20.13 & 100.86 \\
 & Every-value bands & 19.25 & 18.78 & 47.35 & 17.08 & 17.09 & $-0.54$ \\
10 & Exact types & 23.74 & 22.92 & 81.36 & --- & --- & --- \\
 & Every-value bands & 21.11 & 20.35 & 75.84 & --- & --- & --- \\
\midrule
\end{tabular*}
\begin{tabular*}{\textwidth}{@{\extracolsep{\fill}}rrlrrrr}
\multicolumn{3}{l}{\textit{B. Allocated-value loss (cents)}} & \multicolumn{2}{c}{Nominal} & \multicolumn{2}{c}{Whole-type comparison}\\\cmidrule(lr){4-5}\cmidrule(lr){6-7}
Reports & Interface & Interpretation & Joint & Reserve & \shortstack{Whole-type\\joint} & \shortstack{Zero-margin\\reserve bound}\\\midrule
3 & 92.99 & Exact types & 0.20 & 0.00 & 2.32 & 0.00 \\
 & & Every-value bands & 0.36 & 71.07 & 71.38 & 71.07 \\
5 & 67.20 & Exact types & 0.44 & 1.41 & 39.37 & 1.41 \\
 & & Every-value bands & 0.47 & 1.41 & 1.88 & 1.41 \\
10 & 19.83 & Exact types & 0.60 & 0.08 & --- & --- \\
 & & Every-value bands & 0.42 & 1.57 & --- & --- \\
\bottomrule\end{tabular*}\end{footnotesize}
\begin{minipage}{\textwidth}\footnotesize\medskip
\textit{Notes.} Whole-type joint designs use $\delta=\$0.01$ for exact types or $\eta=0.10\%$ for bands. The difference subtracts the zero-margin reserve bound from joint revenue. Interface loss is $S_V-G(\mathbf p^n)$ at $S_V=\$24.15$, and allocation loss is $G(\mathbf p^n)-G(\mathbf p)$.
\end{minipage}
\vspace{-0.2in}
\end{table}

At each deletion, Proposition~\ref{prop:interior} adds a midpoint inequality to the endpoint checks. The 720 schedules at $\eta=0$ and $0.10\%$, with joint allocation and each reserve, give 5,760 outcomes. Pure-map enumeration verifies endpoint violations below $10^{-15}$ dollars on the selected contracts and the joint coefficient of 0.10\%. Across programs, violations are below $10^{-12}$ and optimal-face welfare ranges are below $10^{-6}$ dollars. The positive-coefficient reserve uses floor \$15.33 and schedule $((1,3),\allowbreak(3,1),\allowbreak(2,3),\allowbreak(1,2),\allowbreak(3,2),\allowbreak(2,1))$. Tightening can make reserve schedules infeasible, so Section~\ref{sec:dataresolution} measures finite changes.

\noindent\textbf{Fixed-contract belief perturbation.} Fix the point-belief solutions at $\eta=0.10\%$ from \texttt{bandcert.py} for the reserve above and joint schedule $((1,3),\allowbreak(2,3),\allowbreak(1,2),\allowbreak(3,1),\allowbreak(3,2),\allowbreak(2,1))$. With truthful rivals at $\kappa=0.001$, maximize $100\pos{-\mathrm{sl}_{jh}(\ph(\cdot;\kappa),v)}$ over $j\ne h$ and the two endpoints of band $j$. Reporting band~2 at $v=T_1$ attains both maxima, 0.088 cents for joint and 0.086 for reserve.

\section{Verification of the Motivating Examples}\label{app:examples}
\subsection{Allocation Creates Rent Capacity}\label{app:examplecap}
Let $m=2$, $(T_1,T_2,T_3)=(30,20,12)$, $\mathbf q=(\tfrac13,\tfrac13,\tfrac13)$, $\varsigma=0$, and $\boldsymbol\pi_j=(\tfrac23,\tfrac16,\tfrac16)$ for every type, the shared-top beliefs \eqref{eq:sharedcat} with $\beta=4$. The threshold of Theorem~\ref{thm:sharedtop} is $\beta_c=2$, the scores are $\bar\psi_2=17.50>\bar\psi_3=6>0$, and the top rent floor $3.50$ fits within capacity $10$, so \eqref{eq:ST} holds and the theorem gives $\lambda^\varsigma=\tfrac1{20}$, a best reserve revenue $\tfrac{409}{18}$, and joint revenue $\tfrac{139}6$. Table~\ref{tab:introexample} lists the policies; prices are conditional on winning, and losers pay zero. With rows indexing own reports and columns rival reports, their truthful rents, and the joint mechanism's subjective advantages of truth, are
\[
 \mathbf u^{\rm res}=\begin{pmatrix}\tfrac{19}4&0&0\\0&0&4\\0&0&0\end{pmatrix},\qquad
 \mathbf u^{\rm joint}=\begin{pmatrix}\tfrac{21}4&0&0\\1&0&0\\0&0&0\end{pmatrix},\qquad
 \bigl(\mathrm{sl}_{jh}(\ph_j)\bigr)_{j,h}=\begin{pmatrix}0&0&2\\\tfrac{15}4&0&0\\\tfrac{167}{20}&\tfrac85&0\end{pmatrix}.
\]
Both satisfy \eqref{eq:ir} and nonnegative payments; the last matrix verifies \eqref{eq:ic} for the joint mechanism.

\begin{table}[!htbp]
\vspace{-0.1in}
\centering
\caption{Complete policies for the first motivating example}\label{tab:introexample}
\begin{footnotesize}\setlength{\tabcolsep}{5pt}
\begin{tabularx}{\textwidth}{@{}l>{\raggedright\arraybackslash}X>{\raggedright\arraybackslash}X@{}}\toprule
Reports & Best reserve allocation with optimized payments & Joint allocation and payment design\\\midrule
$(30,30)$ & Uniform winner, price \$20.50 & Uniform winner, price \$19.50\\
$(30,20)$ & \$30 report wins, price \$30 & \$30 report wins with probability $\tfrac{19}{20}$, price \$30; \$20 report wins with probability $\tfrac1{20}$, price \$0\\
$(30,12)$ & \$30 report wins, price \$30 & Same\\
$(20,20)$ & Uniform winner, price \$20 & Same\\
$(20,12)$ & \$20 report wins, price \$16 & \$20 report wins, price \$20\\
$(12,12)$ & Uniform winner, price \$12 & Same\\
\bottomrule\end{tabularx}\end{footnotesize}
\vspace{-0.2in}
\end{table}

The middle type's perceived rent is $4\cdot\tfrac16=1\cdot\tfrac23=\tfrac23$ in both mechanisms, while her objective conditional rent falls from $\tfrac43$ to $\tfrac13$; the top type's rises from $\tfrac{19}{12}$ to $\tfrac74$. The total objective rent saving is thus $2\cdot\tfrac13(1-\tfrac16)=\tfrac59$. The lottery changes service only at the high--middle profile, of probability $\tfrac29$, with welfare loss $\tfrac29\cdot\tfrac1{20}(30-20)=\tfrac19$, so revenue rises by $\tfrac59-\tfrac19=\tfrac49$. Under correct beliefs, the rational optimum is \$20 by Lemma~\ref{lem:rational}. One implementing dominant-strategy rule uses reserve \$20, charges a high winner \$25 against a middle report and \$20 against a low report, and charges the winning report at tied admitted reports; this optimized discrete-type benchmark is distinct from a second-price auction with the same reserve.

\subsection{Nominal Optimality and a Pooling Equilibrium}\label{app:examplepool}
Let $m=2$, $(T_1,T_2)=(10,1)$, $\mathbf q=(\tfrac15,\tfrac45)$, and $\alpha=\tfrac3{10}$, so the subjective high-rival probabilities are $\pi_{11}=\tfrac{11}{25}$ and $\pi_{21}=\tfrac7{50}$. With nonnegative payments, Proposition~\ref{prop:twocap} gives, with rows indexing own reports and columns rival reports, $\mathbf p=\bigl(\begin{smallmatrix}1/2&1\\0&55/126\end{smallmatrix}\bigr)$ and $\mathbf u=\bigl(\begin{smallmatrix}5&0\\0&0\end{smallmatrix}\bigr)$. Thus two high reports receive a free uniformly allocated object, a lone high report pays \$10, and two low reports trigger sale with probability $\tfrac{55}{63}$ at price \$1. The truthful revenue is $2\cdot\tfrac15\cdot\tfrac45\cdot10+(\tfrac45)^2\tfrac{55}{63}=\tfrac{1184}{315}$. The high type's truthful perceived utility is $\tfrac{11}{25}\cdot5=\tfrac{11}5$, and reporting low gives $\tfrac{14}{25}\cdot9\cdot\tfrac{55}{126}=\tfrac{11}5$; the low type's false-high payoff is $\tfrac7{50}\cdot\tfrac12+\tfrac{43}{50}(1-10)=-\tfrac{767}{100}<0$. Hence \eqref{eq:ic} holds. If everyone instead always reports high, a type of value $v$ obtains $v/2>0$, and a unilateral low report obtains zero. Pooling is therefore a strict equilibrium that yields revenue zero and is strictly preferred to truthful reporting by both types, the implementation issue diagnosed by Proposition~\ref{prop:pool}.

\end{APPENDICES}

\end{document}